\documentclass[twocolumn,table]{aastex62}

\usepackage{amsmath,amssymb,amstext}
\usepackage{mathtools}
\usepackage{gensymb}
\usepackage{comment}
\usepackage{todonotes}

\usepackage{graphicx}
\usepackage{multirow}
\usepackage{chngcntr}
\usepackage{booktabs}
\usepackage{xcolor}
\usepackage{soul}

\newcommand{\iso}[2]{\ensuremath{^{#2}{\rm #1}}}
\newcommand{\nickel}{\ensuremath{^{56}{\rm Ni}}}
\newcommand{\mej}{\ensuremath{M_{\rm ej}}}
\newcommand{\ekin}{\ensuremath{E_{\rm k}}}
\newcommand{\mni}{\ensuremath{M_{56}}}
\newcommand{\mrp}{\ensuremath{M_{r \rm p}}}
\newcommand{\msun}{\ensuremath{M_{\odot}}}
\newcommand{\mmix}{\ensuremath{M_{\rm mix}}}
\newcommand{\xmix}{\ensuremath{\psi_{\rm mix}}}
\newcommand{\per}[1]{#1\ensuremath{^{-1}}}
\newcommand{\dd}[2]{\ensuremath{\frac{{\rm d}#1}{{\rm d}#2}}}
\newcommand{\rp}{\emph{r}-process}

\newcommand{\Trp}{\ensuremath{T_{r \rm p}}}
\newcommand{\Mr}{\ensuremath{M_{\rm R}}}
\newcommand{\tR}{\ensuremath{t_{\rm R,pk}}}

\newcommand{\taurp}{\ensuremath{\tau_{r \rm p}}}
\newcommand{\tautr}{\ensuremath{\tau_{\rm tr}}}
\newcommand{\Rl}{\ensuremath{\mathcal{R}_{\rm L}}}

\newcommand{\bej}{\ensuremath{\beta_{\rm ej}}}
\newcommand{\drx}{\ensuremath{\Delta(R{-}X)}}
\newcommand{\Lrp}{\ensuremath{L_{\rm neb}^{r \rm p}}}
\newcommand{\Lsn}{\ensuremath{L_{\rm neb}^{\rm sn}}}
\newcommand{\xmin}{\ensuremath{\psi_{\rm min}}}
\newcommand{\rccsne}{\emph{r}CCSNe}
\newcommand{\rccsn}{\emph{r}CCSN}
\newcommand{\RJ}{\ensuremath{R{-}J}}
\newcommand{\RH}{\ensuremath{R{-}H}}
\newcommand{\RK}{\ensuremath{R{-}K}}

\defcitealias{Taddia.ea_2019.AandA_IcBL.iPTF.survey}{T2019}
\defcitealias{Barbarino.ea_AandA2021_reg.Ic.iPTF.survey}{B2021}
\defcitealias{Siegel.Barnes.Metzger_2019.Nature_rp.collapsar}{SBM19}

\shorttitle{\emph{r}-process collapsars} 
\shortauthors{Barnes \& Metzger}

\begin{document}

\title{Signatures of \emph{R}-process Enrichment in Supernovae from Collapsars}

\author[0000-0003-3340-4784]{Jennifer Barnes}
\affil{Kavli Institute for Theoretical Physics, Kohn Hall, University of California, Santa Barbara, CA 93106 }
\email{jlbarnes@kitp.ucsb.edu}
\author[0000-0002-4670-7509]{Brian D. Metzger}
\affil{Department of Physics and Columbia Astrophysics Laboratory, Columbia University, New York, NY 10027, USA}
\affil{Center for Computational Astrophysics, Flatiron Institute, 162 5th Ave, New York, NY 10010, USA} 

\begin{abstract}
Despite recent progress, the astrophysical channels responsible for rapid neutron capture (\rp) nucleosynthesis remain an unsettled question.
Observations of kilonovae following gravitational wave-detected neutron star mergers established mergers as one site of the \rp{}, but additional sources may be needed to fully explain \rp{} enrichment in the Universe.
One intriguing possibility is that rapidly rotating massive stars undergoing core collapse launch \rp-rich outflows off the accretion disks formed from their infalling matter.
In this scenario, \rp{} winds comprise one component of the supernova (SN) ejecta produced by ``collapsar’’ explosions.
We present the first systematic study of the effects of \rp{} enrichment on the emission from collapsar-generated SNe.
We semi-analytically model \rp{} SN emission from explosion out to late times, and determine its distinguishing features.
The ease with which \rp{} SNe can be identified depends on 
how effectively wind material mixes into the initially \rp-free outer layers of the ejecta.
In many cases, enrichment produces a near infrared (NIR) excess that can be detected within ${\sim}75$ days of explosion.
We also discuss optimal targets and observing strategies for testing the \rp{} collapsar theory, and find that frequent monitoring of optical and NIR emission from high-velocity SNe in the first few months after explosion offers a reasonable chance of success while respecting finite observing resources.
Such early identification of \rp{} collapsar candidates also lays the foundation for nebular-phase spectroscopic follow-up in the near- and mid-infrared, for example with the \emph{James Webb Space Telescope}.
\end{abstract}

\section{Introduction} \label{sec:intro}

The astrophysical site---or sites---of rapid neutron capture (\rp) nucleosynthesis, which produces roughly half of all elements more massive than iron \citep{Burbidge2.Howler.Foyle_1957.RMP_rprocess,Cameron_1957.AJ_rprocess}, remains a major outstanding question in astrophysics (see \citealt{Horowitz.ea_2019.JPG_rproc.frib.connect}, \citealt{Thielemann.ea_2020.JPCS_rproc.sites.review}, and \citealt{Cowan.ea_2021.RMP_rprocess.origins} for recent reviews).

The association of a radioactively powered kilonova \citep[henceforth kn170817;][]{Abbott.ea_2017ApJL_gw.170817.multimess,Aracavi.ea_2017Natur_gw170817.lco.emcp.disc,Coulter.ea_2017Sci_gw.170817.emcp.disc,Drout.ea_2017Sci_gw.170817.emcp.disc,Evans.ea_2017Sci_gw.170817.em.blue.spec,Kasliwal.ea_2022.MNRAS_spitzer.late.obs.kn170817,Kilpatrick.ea_2017.Sci_gw.170817.spectrum.opt.nir,McCully.ea_2017ApJ_gw.170817.blue.spec,Nicholl.ea_2017ApJ_gw.170817.blue.spec, Shappee.ea_2017.Science_kn170817,Smartt.ea_2017Natur_gw170817.empc.disc, SoaresSantos.ea_2017ApJ_gw.170817.empc.decam.disc} with the gravitational wave-detected neutron star merger (NSM) GW170817 \citep{Abbott.ea_2017.PRL_gw170817.disc} represented a watershed in the understanding of \rp{} origins.
In addition to demonstrating the long-theorized viability of NSMs as \rp{} sites \citep[e.g.,][]{Lattimer.Schramm_1974.ApJL_rproc.nsbh.merger,Symbalisty.Schramm_1982.ApL_rproc.ns.merger,Eichler_1989.nat_sgrb.nsm.rproc,Freiburghaus.ea_1999.ApJL_rproc.nsm}, kn170817 provided an unprecedentedly detailed picture of the various environments in which the \rp{} may occur following a merger \citep[][among others]{Drout.ea_2017Sci_gw.170817.emcp.disc,Cowperthwaite+17,Kilpatrick.ea_2017.Sci_gw.170817.spectrum.opt.nir,Tanvir.ea+2017.ApJL_gw170817.emcp.disc}.
In particular, spectral analysis (e.g., \citealt{Chornock.ea_2017ApJ_gw.170817.em.red.spec}; \citealt{Kasen.ea_2017Natur_gw.170817.knova.theory}; and \citealt{Tanaka.ea_2017PASJ_gw.170817.knova.interp}; see \citet{Siegel_2019.EPJA_knova.review} and \citet{Barnes_2020.FrontPhys_knova.review} for reviews) pointed to accretion disk outflows 
\citep{Metzger.ea_2008.mnras_accretion.disk.co.merg.tdep,Fernandez.Metzger_2013.mnras_disk.outlfows.nsm,Perego.ea_2014.mnras_neutrino.winds.nsm,Just.ea_2015.arxiv_nsm.torus.nucleosyn,Siegel.Metzger_2017.PRL_nsm.accretion.disks.rproc,Fujibayashi.ea_2018.ApJ_mass.ejection.nsm.remnant.disks,Fernandez.ea_2019.mnras_grmhd.sims.nsm.disks}
as the locus of the heaviest element production.
(However, see \citet{Waxman.ea_2018.mnras_gw170817.emcp.ejecta.model} for an alternative interpretation.)

Despite this success, the idea that NSMs are the sole \rp{} sources in the Universe may be in tension with lines of evidence that call for \emph{r}-production in events with comparatively short delay times relative to star formation \citep{Cote.ea_2019.ApJ_other.rproc.sources,Siegel.Barnes.Metzger_2019.Nature_rp.collapsar,Zevin.ea_2019.ApJ_nsm.rproc.glob.cluster,vandeVoort.ea_2020_rproc.enrich.nsm.rare.ccsne,Jeon.ea_2021.mnras_rproc.stars.ufdwarf.gal,Molero.ea_2021.mnras_evol.rproc.dwarf.gal,delosReyes.ea_2022.apj_sfh.nucleosyn.sculptor} 
or kick velocities lower than the escape velocities of their sometimes diminutive host galaxies \citep[e.g.,][]{Ji.ea_2016.Nature_rproc.single.event.retII}.
Some recent Galactic chemical evolution studies \citep{Tsujimoto.ea_2021.ApJL_two.proc.sites,Naidu.ea_2022.ApJ_disrupted.halos.rproc.sources} have argued for two distinct \rp{} sources (though see \citealt{Beniamini.ea_2018.mnras_rproc.retention.dwarf.gal,Duggan.ea_2018.ApJ_nsm.main.rproc.src.dwarf,Macias.RamirezRuiz_2019.ApJL_rccsn.constraints.stellar.abund,Bartos.Marka_2019.ApJL_rproc.solar.rccsn.constraints,Schonrich.Weinberg_2019.mnras_rproc.nsm.ism} and \citealt{Fraser.Schonrich_2022.mnras_no.rccsne.MW.metallicity}).  
These clues hint that core-collapse supernovae (CCSNe) may contribute as an additional \rp{} site.

Though CCSNe were nominated as \rp{} sites as soon as the nuclear physics of the \rp{} was understood \citep{Burbidge2.Howler.Foyle_1957.RMP_rprocess}, decades of incremental progress culminated in the finding \citep[e.g.,][]{Qian.Woosley_1996.ApJ_neutrino.winds.ccsne.nucleosyn,Meyer.Brown_1997.ApJS_rproc.mods.survey,Hoffman.eq_1997.ApJ_neutrino.winds.nucleosyn,Thompson.ea_2001.ApJ_proto.ns.winds.rproc}
that matter ablated from a newly formed neutron star (NS) cannot in most cases achieve the combination of entropy and neutron-richness required for a successful \rp.
(Such proto-NS winds may still generate lighter neutron- or proton-rich nuclei; e.g., \citealt{Frohlich.ea_2006.ApJ_ccsne.inner.ejecta.comp,Arcones.Montes_2011.ApJ_light.elem.prod.neutrino.winds}).

More recent explorations have focused on rarer SN subtypes, such as explosions that leave behind very rapidly spinning \citep{Desai.ea_2022.arXiv_neutrino.winds.proto.ns} and/or highly magnetized NSs
\citep{Thompson.ea_2004.ApJ_magnetar.sne.grb,Metzger.ea_2007.ApJ_proto.ns.winds.bfields.rot}
and their concomitant ``jet-driven'' SN explosions 
\citep{Winteler.ea_2012.ApJL_magnetorot.sne.rproc,Mosta.ea_2014.ApJL_magnetorot.sne.3d,Mosta.ea_2018.ApJ_rproc.3d.magnetorot.sne,Kuroda.ea_2020.ApJ_magnetorot.sn.neutrino.gr3d}, in which the prompt advection of neutron-rich material away from the proto-NS surface avoids the problem of charged current interactions---neutrino capture and e$^+$/e$^-$ pair creation and capture---that in the absence of rapid expansion thwart the \rp{} by protonizing the outflowing matter.
However, 3D magnetohydrodynamic (MHD) simulations of the explosions \citep{Halevi.Mosta_2018.mnras_jet.sn.rproc.3d} cast doubt as to whether sufficiently rapid expansion can in fact be realized. 

Regardless, MHD SNe and the energetic processes that occur in their immediate aftermath may nevertheless be important for \emph{r}-production.  Broad-lined Type Ic (Ic-BL) SNe are the most obvious byproducts of MHD-driven explosions, with kinetic energies 
\citep[$\ekin \sim 10^{52}$ erg;][]{Maeda.ea_2002.ApJ_nucleosyn.hypernovae.1998bw.spec,Maeda03ApJ_twoCompNi_blic,Mazzali.ea_2002ApJL_snic.2002ap,Mazzali.ea_2003.ApjL_icbl.sn.2003dh} far exceeding what can be supplied by the standard neutrino mechanism 
(\citealt{Scheck.ea_2006/AAP_multiD.sn.sim.neutrino.transport,Bruenn.ea_2016.ApJ_ccsn.sims,Muller.ea_2017.mnras_sn.sim.3d.prog}; see \citet{Janka.ea_2016.arnps_ccsn.physics.3d} for a review).  
Some (possibly all) SNe Ic-BL occur in conjunction with long gamma-ray bursts 
\citep[GRBs;][]{Galama98_bwDisc,Bloom.ea_2002.ApJL_sn.grb.assoc.011121,Stanek.ea_2003.ApJL_sn2003dh.data,Hjorth.ea_2003_2003dh.disc.nature,WoosleyBloom06_snegrbRev}.
These ultra-relativistic jets may be powered by accretion onto a central compact object
\citep{Aloy.ea_2000.ApJL_grb.jets.collapsars,Bromberg.Tchekhovskoy_2016.mnras_rel.mhd.ccsn.grb,Gottlieb.ea_2022.mnras_bh_3d.grmhd.ccsn.jets}, and serve as indirect evidence for accretion flows with properties akin to those that give rise to the shorter-duration (but otherwise similar) GRBs 
\citep{Nakar_2007.PRep_sgrbs,Berger_2014.aara_sgrbs} associated with NSMs \citep{Abbott.ea_2017ApJ_gw170817.grb,Goldstein.ea_2017.ApJ_grb.gw170817}.

The term ``collapsar'' \citep[e.g.,][]{MacFadyenWoosley99_collapsar} refers to a
model for the production of long GRBs/SNe Ic-BL in which the large angular momentum in the outer layers of a rapidly rotating massive star allows material from those layers to circularize and form an accretion disk as the star undergoes core collapse.
Recent simulations of collapsar disks (\citealt{Siegel.Barnes.Metzger_2019.Nature_rp.collapsar}; henceforth \citetalias{Siegel.Barnes.Metzger_2019.Nature_rp.collapsar}) found that material in the disk becomes neutron-rich through weak interactions \citep{Beloborodov_2003.ApJ_grb.nuclear.comp}, and that winds launched from the disk retain a sufficiently low electron fraction to support an \rp.
(However, studies adopting different neutrino transport methods do not always find production of the heaviest \rp{} nuclei, at least not during epochs in the disk evolution when neutrino self-irradiation of the disk outflows is most important. See, e.g., \citet{Miller.ea_2020.ApJ_rproc.collapsar.blue} and \citet{Just.ea_2022.mnras_nuetrino.cooled.bh.acc.disks}.)

The potential for \emph{r}-production in energetic CCSNe motivates direct searches for its presence in SN light curves and spectra.
The signatures of \rp{} enrichment in SNe will depend sensitively on the quantity of \rp{} material synthesized and its distribution within the ejecta. 
Outflows of \rp-rich matter from a collapsar accretion disk may occur at a delay relative to the initial explosion that unbinds the star's outer layers if, for example, accretion rates onto the disk are initially too low to support an \rp{} \citepalias{Siegel.Barnes.Metzger_2019.Nature_rp.collapsar}.
This scenario implies an inner core of \rp{} products deposited behind an outer layer composed of ordinary stellar material (e.g., carbon and oxygen) and radioactive \nickel{} synthesized in the explosion.
Radiation transport simulations by \citetalias{Siegel.Barnes.Metzger_2019.Nature_rp.collapsar} found that such an ejecta structure produced light curves and spectra fairly consistent with observed SNe Ic-BL.

However, even if collapsar disks release an \rp{} wind into an already expanding SN ejecta, various processes, such as hydrodynamic instabilities at the wind-ejecta interface, may mix \rp{} elements out to higher mass coordinates.
While \citetalias{Siegel.Barnes.Metzger_2019.Nature_rp.collapsar} found that a fully mixed model could not reproduce emission from SNe Ic-BL, they did not consider intermediate levels of mixing, in which some but not all of the initially \rp-free ejecta from the prompt explosion becomes enriched with \rp{} material.

In the present work, we use analytic reasoning and semi-analytic modeling to improve on \citetalias{Siegel.Barnes.Metzger_2019.Nature_rp.collapsar}, extending their analysis to a much broader region of parameter space and investigating the possibility that signs of \emph{r}-production can be detected directly in the emission of \rp-enriched core-collapse SNe (\rccsne).
In \S\ref{sec:analytics}, we apply simple analytic arguments to establish baseline expectations for the strength of the \rp{} signal and the timescales on which it may appear.
A more detailed \rccsn{} emission model is developed in Section \ref{sec:analytic_mod}.
We validate the model against SNe Ic with typical ($\mathcal{O}(10^{51})$ erg) energies and---we assume---negligible \rp{} production, before extending it to enriched cases and exploring how the addition of \rp{} material impacts the SN's light-curve and color evolution.
In \S\ref{sec:results}, we consider a broad suite of models and discuss the prospects for constraining collapsar \rp{} production as a function of observational SN properties.
We present our conclusions in \S\ref{sec:conclusions}.

\section{Analytic considerations}\label{sec:analytics}

Before proceeding to more detailed SN emission models (\S\ref{sec:analytic_mod}), we present simple analytic arguments to build intuition and illustrate the key factors that determine how easily signs of \rp{} enrichment can be observed.
Here, as in later sections, we model the \rccsn{} ejecta as a spherical outflow of total mass \mej{} consisting of an \rp-enriched core and an \rp-free envelope.
We do not assume the core is composed purely of \rp{} products; rather, we mix some mass \mrp{} of \rp{} material into a central region of mass $\mmix = \xmix \mej$, where the mixing coordinate $\xmix \leq 1$. 

Motivated by models of both regular \citep{Yoon.ea_2019.ApJ_ni56.mix.color.ev} and broad-lined \citep{Taddia.ea_2019.AandA_IcBL.iPTF.survey} SNe Ic that find evidence of \nickel{} mixing out to high velocities, we assume that \nickel{} is distributed evenly throughout the ejecta.
In \S\ref{subsec:neb_model} and \S\ref{subsec:photolytics}, we ignore \rp{} decay and treat \iso{Ni/Co}{56} as the sole source of radioactive heating.
(We retire this simplification in later sections, since the fraction of energy due to \rp{} decay increases with time, and can be non-negligible depending on the relative masses of \nickel{} and \rp{} elements;  \citetalias{Siegel.Barnes.Metzger_2019.Nature_rp.collapsar}.)

We first consider \rp{} signatures during the nebular phase, before moving on to the earlier photospheric phase.

\subsection{Nebular phase}\label{subsec:neb_model}

Very late-time observations have been suggested \citepalias{Siegel.Barnes.Metzger_2019.Nature_rp.collapsar} as a key strategy for testing the \rp{} collapsar hypothesis, since by the nebular phase the ejecta is transparent and lines of sight extend into the inner regions where, for core-envelope models of \rccsne, the \rp{} material resides. 
However, even in the nebular phase, the strength of the signal depends on the degree to which the nebular spectrum of the \rp-rich layers diverges from that of the \rp-free envelope, as well as on the brightness of each component.

In the case of a completely optically thin ejecta heated by uniformly distributed radioactive \iso{Ni/Co}{56}, the ratio of luminosities from the \rp-rich and -free layers is simply $\Lrp/\Lsn = \xmix/(1-\xmix)$.
(Since we are ignoring heating from \rp{} decay, this estimate is a conservative lower limit.)

We assign to the \rp-free envelope a spectral energy distribution (SED) derived from late-time photometry of the Type Ic SN 2007gr \citep{Hunter.ea_2009.AandA_sn2007gr.phot,Bianco.Modjaz.ea_2014.ApJS_sesn.ccsne.lcs}, one of the few SNe Ic observed well past peak in both optical and near infrared (NIR) bands.
We assume that the \rp-enriched layers shine like a blackbody, and leave the effective temperature \Trp{} as a free parameter, deferring a more detailed discussion of our treatment of emission from optically thin material (whether \rp-enriched or not) to \S\ref{sec:analytic_mod}.

With these simplifications, we can determine how the spectrum of a totally optically thin \rccsn{} differs from the \rp-free case, for a given \xmix{} and \Trp.
For the range of \Trp{} we consider, which are broadly consistent with (admittedly limited) constraints from both theory \citep{Hotokezaka.ea_2021.MNRAS_kn.nebular.model} and observation \citep{Kasliwal.ea_2022.MNRAS_spitzer.late.obs.kn170817}, the impacts of the \rp{} are most visible in the NIR.
We therefore characterize the signal strength in terms of \drx{}, the change in $R{-}X$ color relative to the \rp-free SN case (\mrp = 0), where $X \in \{J,H,K\}$.
All magnitudes are calculated using the AB system and generic Bessel filters.

\begin{figure}\includegraphics[width=\columnwidth]{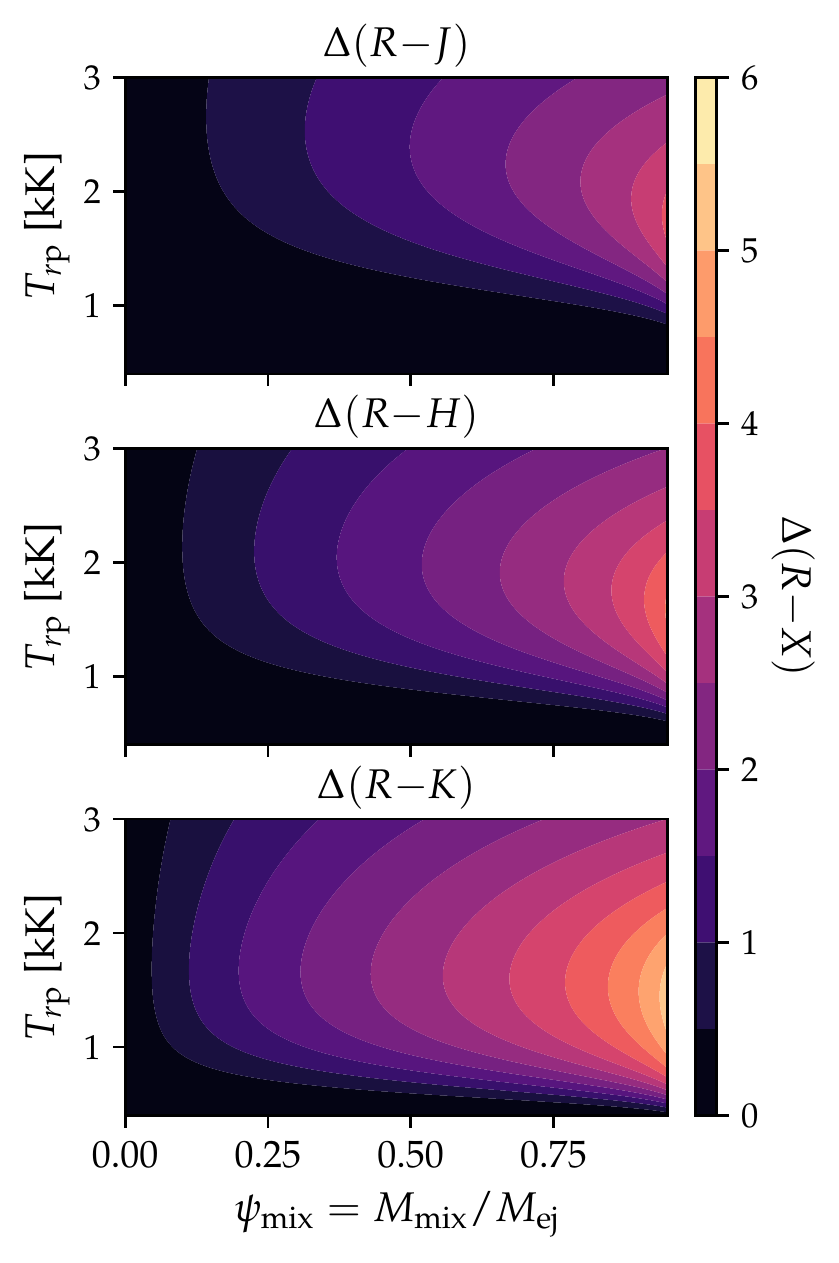}
    \caption{The addition of \rp{} material alters the optical-NIR colors of the nebular phase compared to SNe with no \rp{} enrichment.
    The magnitude of the effect depends on the mixing coordinate \xmix{} and on the nebular-phase \rp{} SED, which we model here as a blackbody of effective temperature \Trp.
    The signal is most apparent for high \xmix{} and for $1000 \text{ K} \lesssim \Trp \lesssim 2000 \text{ K}$.
    }
    \label{fig:rx_neb}
\end{figure}

Figure~\ref{fig:rx_neb} shows how each color changes for $400 \text{ K} \leq \Trp \leq 3000 \text{ K}$ and $0 \leq \xmix \leq 0.95$.
The signal strength increases with the mixing coordinate \xmix, and for a given \xmix{} is maximal for \Trp{} with blackbody functions peaking at wavelengths within the NIR band under consideration.
Regardless of \Trp{} and \xmix, the signal becomes easier to observe at longer wavelengths.
However, for cases of low to moderate mixing ($\xmix \lesssim 0.3$) the color difference even in $R{-}K$ is ${\lesssim}1.5$ mag for all \Trp. 

While this difference may still seem substantial, it is important to bear in mind that models of nebular-phase emission are not well constrained in either the standard or \rp-enriched case.
Relying exclusively on nebular observations may make it difficult to evaluate collapsars as \rp{} sites, particularly if the \rp{} matter is centrally concentrated (low \xmix).  We therefore turn our attention to the pre-nebular (photospheric) phase.

\subsection{Photospheric Phase}\label{subsec:photolytics}

Observations in the photospheric phase (before the ejecta is fully transparent) can sidestep some of the complications inherent in the acquisition and analysis of nebular-phase data.
First, \rp{} emission in the photospheric phase is better understood, thanks both to extensive theoretical studies of \rp{} elements' atomic structures, which enable descriptions of their behavior in local thermodynamic equilibrium \citep[e.g.,][]{Kasen_2013_AS,Tanaka.ea_2020.MNRAS_knova.kappas, Fontes.ea_2020.MNRAS_knova.opacs.lanl}, and to observations of kn170817 \citep{Aracavi.ea_2017Natur_gw170817.lco.emcp.disc,Coulter.ea_2017Sci_gw.170817.emcp.disc,Drout.ea_2017Sci_gw.170817.emcp.disc,Evans.ea_2017Sci_gw.170817.em.blue.spec,Kasliwal.ea_2022.MNRAS_spitzer.late.obs.kn170817,Kilpatrick.ea_2017.Sci_gw.170817.spectrum.opt.nir,McCully.ea_2017ApJ_gw.170817.blue.spec,Nicholl.ea_2017ApJ_gw.170817.blue.spec, Shappee.ea_2017.Science_kn170817,Smartt.ea_2017Natur_gw170817.empc.disc, SoaresSantos.ea_2017ApJ_gw.170817.empc.decam.disc}.

Second, the SN is brighter during the photospheric phase, and it is therefore easier to obtain high signal-to-noise photometry across a range of wavelengths. 
Finally, early observations do not preclude late-time follow-up.
To the contrary, they may be useful for filtering out the events most worthy of further attention during nebular epochs. 

Despite these advantages, the photospheric phase presents its own set of challenges.
While the ejecta remains optically thick, the \rp{} signal may be less clean than it is during the nebular phase.
This is of particular concern early on, when the \rp-free envelope is opaque and obscures emission from the enriched core underneath it.
The position of the photosphere (the surface that divides optically thick from optically thin material) is therefore a good indicator of how observable an \rp{} signature is at a given time.

\begin{figure}\includegraphics[width=\columnwidth]{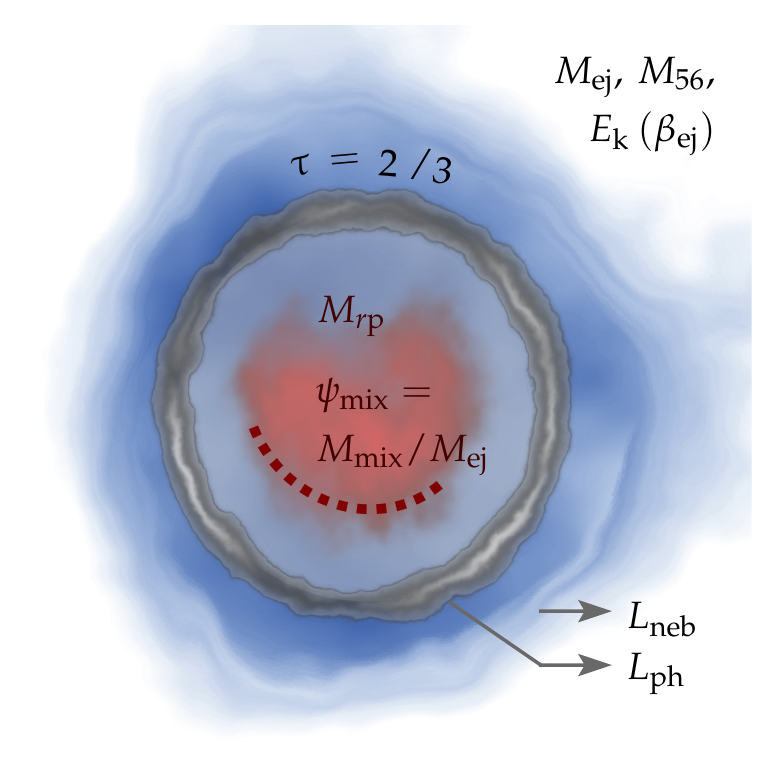}
\caption{A schematic illustration of our \rp{} collapsar model, with the main model parameters and emission components highlighted.
All models are defined by their total mass, \mej, and kinetic energy, \ekin (which together define a characteristic velocity \bej), as well as a \nickel{} mass, \mni.
For the \rp-enriched models, some amount \mrp{} of \rp{} matter is mixed into a central core (colored red above) of mass $\mmix > \mrp$.
The mixing coordinate \xmix{} is the ratio of \mmix{} to \mej.
The photosphere, defined as the surface at which $\tau = 2/3$, separates the optically thick and optically thin regions. 
Emission from the optically thick layers inside the photosphere takes the form of a blackbody, in contrast to emission from the optically thin (nebular) layers.
}\label{fig:illustration}
\end{figure}

To model the photospheric phase, we build on the simple ejecta structure introduced in \S\ref{subsec:neb_model}.
In addition to \mej{}, we now describe the ejecta in terms of its kinetic energy \ekin{}, and explicitly consider the \rp{} mass fraction in the enriched layers, $\zeta = \mrp/\mmix$. 
An illustration of the model is provided in Fig.~\ref{fig:illustration} and a summary of its parameters can be found in Table~\ref{tab:params}.

\begin{table}
\begingroup
\centering
\caption{Parameters of the \rp{} collapsar model}\label{tab:params}
    \begin{tabular}{p{0.33\columnwidth}>{\raggedright\arraybackslash}p{0.56\columnwidth}}
    \toprule
     \emph{Symbol} & \centering \emph{Definition}   \tabularnewline
    \hline
    \mej{} & The total mass of the ejecta \\
    \ekin{} & The kinetic energy of the ejecta \\
    \multirow{3}{*}{\bej{}} & The ejecta's average expansion velocity, normalized to $c$ ($\ekin = \mej \bej^2 c^2/2$) \\
    \multirow{2}{*}{\mni{}} & The mass of radioactive \nickel{} produced in the explosion \\
    \multirow{2}{*}{\mrp{}} & The mass of \rp{} material in the ejecta \\
    \multirow{2}{*}{\mmix{}} & The  mass of the ejecta enriched with \rp{} material (${\neq}\mrp$) \\
    \multirow{2}{*}{\xmix{}} & The fraction of \mej{} that is enriched (${=}\mmix/\mej$)  \\
    \multirow{2}{*}{$\zeta$} & The \rp{} mass fraction in the enriched layers (${=}\mrp/\mmix$) \\
    \multirow{2}{*}{$\kappa_{\rm sn} \; (=0.05$ cm$^2$ \per{g})} & The gray opacity of SN ejecta containing no \rp{} material \\
    \multirow{2}{*}{$\kappa_{r \rm p} \; (=10$ cm$^2$ \per{g})}  & The gray opacity of a pure \rp{} composition \\
    \bottomrule
    \end{tabular}
    \endgroup
\end{table}

As in \S\ref{subsec:neb_model}, our analysis here rests on the distinct optical properties of \rp{} elements.
We assume that the high opacity of the enriched regions causes them to shine in the NIR, significantly redder than both photospheric and nebular-phase emission from the \rp-free layers.
(The assumption of a NIR-dominated SED is a good one for cores composed primarily of \rp{} elements; however, with increased mixing both dilution and enhanced ionization due to energy from \iso{Ni/Co}{56} decay may reduce the opacity, resulting in somewhat bluer emission; e.g., \citealt{Barnes.ea_2021.ApJ_knova.nuc.landscape}.)

While the effect of the high-opacity core will be subtle during the light curve's early stages, it will become more apparent as the photosphere, which forms at ever-lower mass coordinates, sinks into the \rp{} layers.
At this point, the enriched core becomes ``visible,'' and its higher opacity exerts a greater influence on the overall SED of the \rccsn.

This argument suggests that \rp{} enrichment will be easier to detect in explosions with larger \xmix{} and \mrp{}.  Simple one-zone light-curve models allow us to map out this dependence.  We focus here on times after the outer layers have become transparent. 
During this period, radiation from the photosphere originates entirely from \rp-enriched material, while optically thin emission comes predominantly from the \rp-free envelope.
Both enriched and unenriched regions contain radioactive material, and both continue to radiate after becoming optically thin.
However, the recession of the photosphere slows dramatically after it reaches the inner high-opacity \rp{} layers, ensuring that nebular emission from the envelope dominates nebular emission from the core ($L_{\rm neb}^{\rm sn} \gg L_{\rm neb}^{r \rm p}$) out to fairly late times.
We therefore ignore for the time being the contribution of $L_{\rm neb}^{r \rm p}$, and assume $L_{\rm neb} = L_{\rm neb}^{\rm sn}$.

We define the ratio of the photospheric (\rp) and nebular (\rp-free) luminosities to be 
\begin{equation}
    \mathcal{R}_{\rm L} \equiv L_{\rm ph}^{r \rm p}/L_{\rm neb}^{\rm sn},\label{eq:Rl}
\end{equation}
and adopt it as a rough indicator of the detectability of an \rp-enrichment signature.

We estimate (see \S\ref{sec:analytic_mod} for more detail) that optically thin \rp-free material emits ${\sim}55$\% of its energy at NIR wavelengths ($1 \: \mu {\rm m} \leq \lambda \leq 2.5 \: \mu \rm{m}$).
The strength of the NIR excess from the photosphere can then be approximated as ${\approx}L_{\rm ph}^{r  \rm p}/0.55 \Lsn$.
To produce a NIR signal ${\sim}50$\% stronger than expected from \rp-free nebular emission alone requires $\mathcal{R}_{\rm L} \gtrsim 0.3$.

The \rp{}-detectability metric \Rl{} depends on a few fundamental timescales.
For a constant-density ejecta of mass \mej{} and kinetic energy \ekin, the \rp-free envelope becomes transparent at 

\begin{gather}
    \tau_{\rm tr} = 53 \text{ days} \times \frac{\mej/\msun} {(\ekin/1 \text{ foe})^{1/2}}\sqrt{1-\xmix^{1/3}},\nonumber \\
    \text{or }\left(\frac{\tau_{\rm tr}}{\tau_{\rm pk}}\right)^2 = 1.2 \frac{(1-\xmix^{1/3})}{\beta_{\rm ej}}, \label{eq:ttr_tpk}
\end{gather}
where 1 foe $=10^{51}$ erg, and \bej{} is the characteristic velocity in units of $c$ ($\ekin = \mej \bej^2 c^2 /2$).  
In the second line, we have normalized to the standard light-curve peak time, 
$\tau_{\rm pk} = 9 \text{ days} (\mej/\msun)^{3/4} (\ekin/1 \text{ foe})^{-1/4}$, which, like the transparency time $\tau_{\rm tr}$, was calculated
assuming an ejecta opacity of $\kappa_{\rm sn} = 0.05$ cm$^2$ \per{g}, much lower than the opacity of \rp{} compositions.  
This  choice presumes that the early light-curve evolution is driven by the unenriched layers, an assumption that becomes less reliable with increased outwards mixing (higher \xmix).

Once the unenriched layers are optically thin, their luminosity reflects the rate at which the material in those layers produces energy via radioactivity.
Thus, for $t > \tautr$, \Lsn{} declines as \iso{Ni/Co}{56} decay away.
In contrast, emission from the \rp{} layers evolves on a distinct timescale set by their opacity and mass.
It may be rising, maximal, or in decline at \tautr.
A second key timescale is therefore $\tau_{r \rm p}$, the time over which $L_{\rm ph}^{r \rm p}$ rises to a maximum.
This can be estimated as the time-to-peak for a transient consisting solely of the inner enriched layers,
\begin{gather}
    \tau_{r \rm p} = 9 \text{ days } \times \frac{ (\mej/\msun)^{3/4} \xmix^{1/3}}{(\ekin/1 \text{ foe})^{1/4}}\left[\zeta(\mathcal{K}-1) + 1 \right]^{1/2}, \nonumber \\
    \text{ or }
    \left( \frac{\tau_{r \rm p}}{\tau_{\rm tr}}\right)^2
    = 0.9 \frac{ \bej \xmix^{2/3}}{(1-\xmix^{1/3})} \left[ \zeta(\mathcal{K}-1)+1\right], \label{eq:trp_ttr}
\end{gather}
where $\zeta = \mrp/\mmix$ is the \rp{} mass fraction of the enriched layers and $\mathcal{K} = \kappa_{r \rm p}/\kappa_{\rm sn}$ is the ratio of the \rp{} opacity to the opacity of the (unenriched) SN ejecta.
In normalizing to \tautr{}, we have again assumed $\kappa_{\rm sn} = 0.05$  cm$^2$ \per{g}.  Since heavy \rp{} compositions have $\kappa_{r \rm p} \approx 10$ cm$^2$ \per{g}, this makes $\mathcal{K} \approx 200$.

If \tautr{} represents the \emph{first} chance to observe emission from the \rp-enriched layers, \taurp{} is a proxy for the \emph{best} chance---the point at which that emission component glows brightest.
Thus, systems for which $\taurp \approx \tau_{\rm tr}$ are close to ideal from an observability standpoint; the \rp{} layers are shining most strongly around the time they first come into view, when the SN overall is still fairly bright.
If instead $\taurp \ll \tautr$, the luminosity from the inner \rp{} core begins its decline before it is even visible.
(At the other extreme, for $\taurp \gg \tau_{\rm tr}$, there is a danger that $L_{\rm ph}^{r \rm p}$ will climb to its peak only after the \rccsn{} overall has grown faint.)
Thus, as we will see, the observability parameter \Rl{} is sensitive to $\taurp/\tautr$, and a successful observation is more likely when this ratio ${\gtrsim}1$.

\begin{figure}\includegraphics[width=\columnwidth]{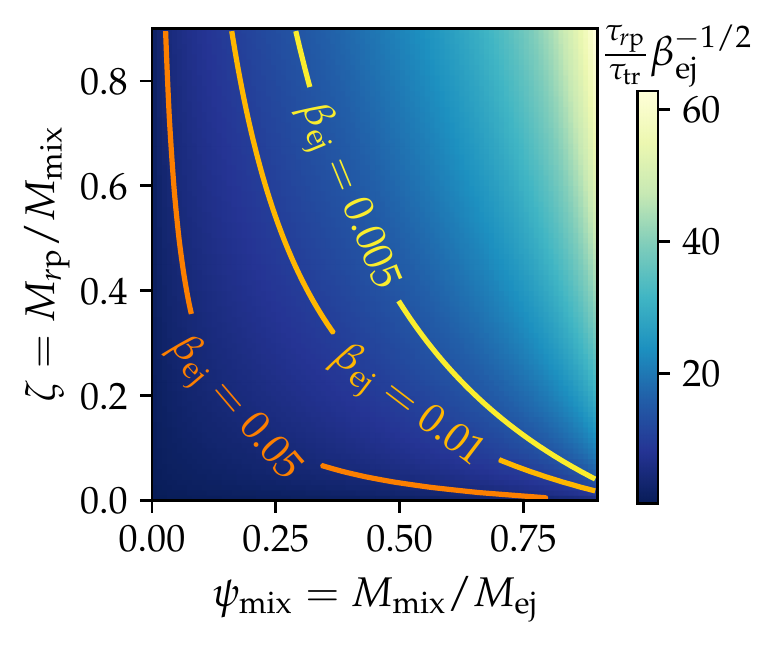}
    \caption{\emph{R}-process signals will be easier to see if the peak time of the enriched inner layers (\taurp) is close to the time at which the outer layers become transparent ($\tau_{\rm tr}$).
    This is the case only for certain combinations of the mixing coordinate \xmix{} (${=}\mmix/\mej)$, the \rp{} mass fraction $\zeta$ (${=}\mrp/\mmix)$, and the velocity \bej. 
    Shading denotes the quantity $(\tau_{r \rm p}/\tau_{\rm tr})\bej^{-1/2}$ for a range of \xmix{} and $\zeta$, with solid contours tracing, for different values of \bej, the points where $\tau_{r \rm p} = \tau_{\rm tr}$.  
    For higher \bej{}, $\taurp \approx \tautr$ even for low \xmix~and/or low $\zeta$.
    In contrast, for lower \bej{}, the signal is weaker absent significant mixing. 
    }
    \label{fig:bxz-plane}
\end{figure}

The value of $\tau_{r \rm p}/\tau_{\rm tr}$ depends on the interplay between $\xmix$, $\zeta$, and $\beta_{\rm ej}$ (Eq.~\ref{eq:trp_ttr}), as illustrated in Figure~\ref{fig:bxz-plane}.
Not surprisingly, for constant \bej, $\taurp/\tautr$ increases with both the mixing coordinate ($\xmix$) and the \rp{} mass fraction in the core ($\zeta$).
However, if \bej{} is sufficiently high, $\taurp/\tautr$ can approach unity even for low \xmix{} and $\zeta$.
This suggests that higher-velocity outflows will offer more opportunities to observe \rp{} enrichment for a greater variety of enrichment parameters (\xmix{} and $\zeta$).

However, even if the timescales are favorable ($\taurp \approx \tautr$), the \rp-visibility parameter $\Rl$ is limited by \xmix; only energy \emph{deposited} behind the photosphere can be \emph{emitted} from behind the photosphere.  
Our models have uniformly distributed \nickel{}, which establishes $\xmix/(\xmix-1)$ as a fundamental scale for \Rl. 
According to Arnett's Law \citep{Arnett_1980,Arnett_1982_Sne}, \Rl{} reaches a maximum of  $\xmix/(1-\xmix)$ at $t = \taurp$.

We estimate $L_{\rm ph}^{r \rm p}$ at $t \sim \taurp$ by Taylor-expanding the analytic light-curve solution to a one-zone model of a radioactively powered transient \citep{Chatzopoulos.ea_2012.ApJ_sn.lc.analytic},
\footnote{The equation of energy conservation in a homologously expanding, diffusive, homogeneous medium heated by radioactive decay gives an expression for the time-dependent emerging luminosity,
\begin{equation*}
    L(t) = \exp\left(\frac{-t^2}{2\tau_{\rm lc}^2}\right)\left[\int\limits_0^t \dot{Q}_{\rm rad}(t')\left(\frac{t'}{\tau_{\rm lc}^2}\right)\exp\left(\frac{t'}{2\tau^2_{\rm lc}}\right)\mathrm{d}t' \right],
    \end{equation*}
    where $\tau_{\rm lc}$ is a characteristic light-curve timescale defined in the same way as $\tau_{\rm pk}$ and \taurp.
    Expanding this solution about $t = \tau_{\rm lc}$, with the specification that the radioactive heating $\dot{Q}_{\rm rad}$ is due solely to \iso{Ni/Co}{56} decay (reasonable for $\mrp \lesssim \mni$, especially at early times), yields Eq.~\ref{eq:Lph_taylor}.
}
\begin{align}
    L_{\rm ph}^{r \rm p}(t \sim \taurp)
    &\approx \xmix \mni \biggl[ \dot{\epsilon}_{56}(\taurp) - \frac{1}{2} \frac{(t-\taurp)^2} {\taurp \tau_{\rm Ni}}  \nonumber \\
     \frac{}{} &\times \left( \dot{\epsilon}_{56}(\taurp) - \dot{\epsilon}_{\rm Co,eq}(\taurp) \right) \biggr], \label{eq:Lph_taylor}\\
     \intertext{where } \dot{\epsilon}_{\rm Co,eq}(t) &= \frac{q_{\rm Co}}{\tau_{\rm Co}}\exp\left[\frac{-t}{\tau_{\rm Co}}\right]. \nonumber
\end{align}
In Eq.~\ref{eq:Lph_taylor}, \mni{} is the \nickel{} mass, $\dot{\epsilon}_{56}$ the rate of specific energy production by \iso{Ni}{56} and \iso{Co}{56} decay, and $\tau_{\rm Ni}$ ($\tau_{\rm Co}$) is the \nickel{} (\iso{Co}{56}) lifetime. 
The quantity $q_{\rm Co}$ is the decay energy of \iso{Co}{56} divided by its mass.
Since we approximate the nebular luminosity from the \rp{} free layers as 
\begin{equation*}
L_{\rm neb}^{\rm sn} = (1-\xmix) \mni \dot{\epsilon}_{56},
\end{equation*}
Eq.~\ref{eq:Lph_taylor} enables a straightforward determination of \Rl{}, once all model parameters have been specified.

\begin{figure}\includegraphics[width=\columnwidth]{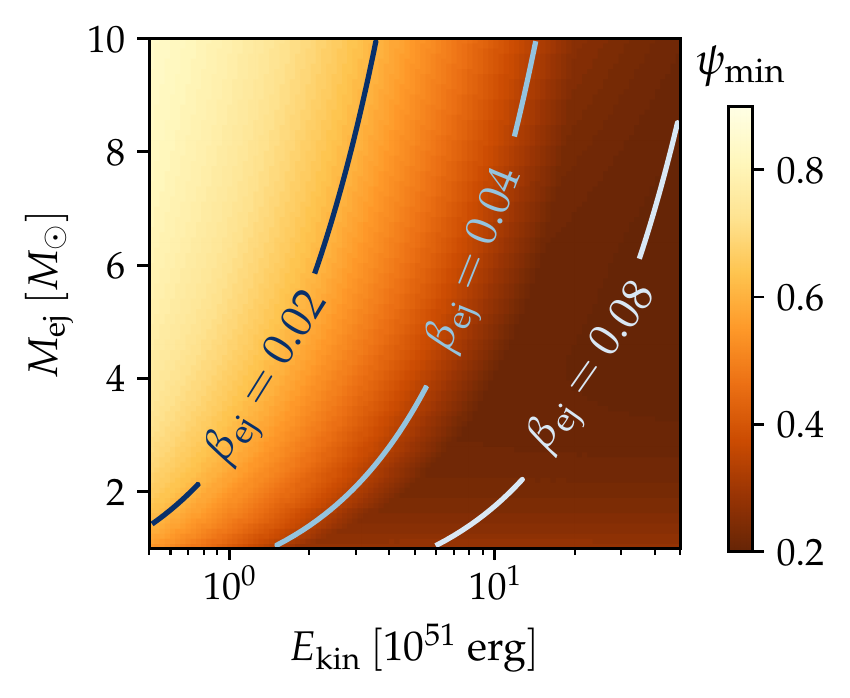}
    \caption{\emph{R}-process signatures may be detectable in the photospheric phase if ejecta velocities are high---even for low mixing coordinates \xmix.
    Colored fill indicates the minimum mixing coordinate (\xmin) for which $\mrp = 0.05 \msun$ produces a detectable signal, as a function of \mej{} and \ekin.
    Lines of constant \bej{} are over-plotted.
    As described in the text, to be observable, a signal must fulfill both a) $\tautr \leq 3\tau_{\rm pk}$, and b) $\Rl \geq 0.3$ at any time $\tautr \leq t \leq 3\tau_{\rm pk}$.  A deeply buried \rp{} core (small \xmix{}) is difficult to observe except in cases of low \mej{} and/or high \ekin{}.  For typical SN Ib/c parameters ($\mej/\msun \sim \text{ a few}, \: \bej \sim 0.03$), \rp{} detection will be challenging unless mixing is extensive.}
    \label{fig:xmin_mvplane}
\end{figure}

This framework allows us to estimate, given \mej{} and \bej{}, the minimum \xmix{} for which some \rp{} mass \mrp{} can produce an ``observable'' signal---i.e., one that produces a strong NIR excess on a timescale not too delayed relative to the light-curve peak.
This estimate is shown in Fig.~\ref{fig:xmin_mvplane} for $\mrp = 0.05 \msun$, an \rp{} mass comparable to what was produced in GW170817 \citep{Abbott.ea_2017ApJL_gw.170817.multimess,Kasen.ea_2017Natur_gw.170817.knova.theory,Kasliwal.ea_2017Sci_gw.170817.em.interp,Tanaka.ea_2017PASJ_gw.170817.knova.interp}.
In calculating \xmin, we defined an observable signal as one for which a) $\tautr \leq 3\tau_{\rm pk}$, and
b) $\Rl$ reaches a maximum ${\geq}0.3$ at some point in the interval $\tautr \leq t \leq 3\tau_{\rm pk}$.
The first criterion addresses whether the \rp{} signal will appear before the light curve has dimmed significantly, while the second requires the signal to be strong enough to appreciably alter the SED from the optically thin \rp-free ejecta.

Applying these conditions, we find (for $\mrp = 0.05 \msun$) that SNe with low mixing coordinates ($\xmix \lesssim 0.4$)
generate a visible \rp{} signal only for $\bej \gtrsim 0.04c$, somewhat larger than typical SNe Ib/c velocities.
Fortuitously, however, Fig.~\ref{fig:xmin_mvplane} suggests that
the high-velocity SNe Ic-BL connected to long GRBs---the same SNe most likely to be associated with collapsars, and hence significant \emph{r}-production---are the same events for which \rp{} enrichment is detectable even for low \xmix.

Despite the simplifications invoked in the preceding analysis, these trends suggest that rCCSNe search strategies need not be limited to the nebular phase. 
Indeed, they motivate a more rigorous consideration of the full evolution of \rccsn{} light curves.

\section{Semi-analytic light-curve modeling}\label{sec:analytic_mod}

We use a semi-analytic framework to model the emission from \rp{}-enriched and unenriched SNe from the initial explosion through to the nebular phase.

\subsection{Basic Structure}\label{subsec:model_basics}
We begin with an ejecta comprised of concentric, homologously expanding spherical shells.  
The density and composition of each shell are parameters of the model, and the frequency-independent (gray) opacity is a function of the local temperature and composition. 

The internal energy of a shell $i$ evolves as
\begin{align}
   \dd{E_{\rm int,i}}{t} = \dot{Q}_{\rm rad,i} - \frac{E_{\rm int,i}}{t} - L_{\rm i}, 
   \label{eq:dEdt}
\end{align}
where $\dot{Q}_{\rm rad,i}$ is the power injected by radioactive decays (in this case, of \iso{Ni}{56}, \iso{Co}{56}, and \rp{} nuclei), and $E_{\rm int,i}/t$ accounts for adiabatic losses.  The radiated luminosity $L_{\rm i}$ depends on the local diffusion ($t_{\rm diff}$) and light-crossing ($t_{\rm cross}$) timescales,
\begin{align*}
    L_{\rm i} &= \frac{E_{\rm int,i}}{t_{\rm diff,i}+t_{\rm cross,i}},\\
    \intertext{with}
    t_{\rm diff,i} &= \frac{2}{c}\sum\limits_{j \geq i} \rho_{\rm j} \kappa_{\rm j} r_{\rm j} \Delta r_{\rm j} \\
    \intertext{and} 
    t_{\rm cross,i} &= \frac{r_{\rm i}}{c},
\end{align*}
where $\rho_{\rm i}$ is the mass density of shell $i$, $\kappa_{\rm i}$ its opacity, $r_{i}$ its current radius, and $\Delta r_{\rm i}$ its radial width.  The emerging bolometric luminosity at any time $t$ is a sum over $L_{\rm i}$,
\begin{align*}
    L_{\rm bol}(t) = \sum\limits_i L_{\rm i}.
\end{align*}

In an optically thick system, radiation tends towards a blackbody distribution.
For a gray-opacity medium like ours, the photosphere is well-defined and the relationship between the luminosity and the SED is straightforward,
\begin{align}
    L(t) &= 4 \pi r_{\rm ph}^2 \sigma_{\rm SB} T_{\rm eff}^4, \nonumber \\
    \intertext{and}
    L_{\nu} &= 4 \pi^2 r_{\rm ph}^2 B_{\nu}(T_{\rm eff}),\label{eq:sed_bb}
\end{align}
where $\sigma_{\rm SB}$ is the Stefan-Boltzmann constant, $r_{\rm ph}$ is the radius of the photosphere, $T_{\rm eff}$ is the effective temperature, and $B_\nu$ is the Planck function.

As the ejecta becomes increasingly transparent, the blackbody approximation becomes less and less reliable.  We are interested in modeling the emission of the SN as it transitions from optically thick to optically thin, which motivates a modification of the blackbody prescription of Eq.~\ref{eq:sed_bb}.
As in \S\ref{sec:analytics}, we categorize the emerging luminosity as \emph{photospheric} if it originates interior to the radius $r_{\rm ph}$ defined by $\tau = \int\limits_{r_{\rm ph}}^\infty \rho(r')\kappa(r')\mathrm{d}r' = 2/3$, and \emph{nebular} otherwise. 

The photospheric component, $L_{\rm ph}$, is translated to a SED via Eq.~\ref{eq:sed_bb}.  The rigorous numerical modeling required to predict the emission of the nebular component is beyond the scope of this work.
Instead, we assume that the SED of radiation from optically thin regions depends only on the composition of the zone where it originates.
Building on the discussions in \S\ref{subsec:neb_model} and \S\ref{subsec:photolytics}, we associate one characteristic nebular SED with \rp{}-free SN ejecta, and a second with the \rp.
The net SED from an optically thin zone is a scaled sum of the two, as explained in \S\ref{subsec:model_imp}.

\subsection{Implementation of the Model}\label{subsec:model_imp}

We apply this blueprint to regular and \rccsn{} models characterized by the same parameters as in \S\ref{sec:analytics}: ejecta mass (\mej) and velocity (\bej), \nickel{} mass (\mni), the \rp{} mass \mrp{} (equal to zero for unenriched SNe), and, for $\mrp > 0$, the mixing coordinate \xmix.
All models have a uniform distribution of \nickel{} and a broken-power law mass density profile, $\rho \propto (v/v_{\rm tr})^{-\alpha}$, where $\alpha = 1 \:(10)$ for $v < \: (\geq) v_{\rm tr}$, and the transition velocity $v_{\rm tr}$ is chosen to ensure $\rho(v)$ integrates to the desired \mej{} and \ekin{} ($=\mej \bej^2 c^2/2$).

The decays of \iso{Ni/Co}{56} and, if present, \rp{} nuclei supply the energy ultimately radiated by the SN.
We assume $\gamma$-rays from \iso{Ni}{56} and \iso{Co}{56}, which constitute most of the energy from that decay chain, are deposited in the zone that produced them with an efficiency $f_{\rm dep,\gamma}$ that depends on $\tau_\gamma$, the ejecta's global optical depth to $\gamma$-rays.
We adopt the functional form of \citet{Colgate.ea_1980.ApJ_sn.grays.deposition.lum} for $f_{\rm dep,\gamma}(\tau_{\gamma})$, and calculate $\tau_\gamma$ using their suggested $\gamma$-ray opacity $\kappa_\gamma = 1.0/35.5$ cm$^{2}$ g$^{-1}$.
The fast positrons from the $\beta^+$-decay of \iso{Co}{56} are assumed to thermalize locally and instantaneously.

For the \mrp{} and \mni{} we consider, the energy from \rp{} decay is subdominant to that from \nickel{} on the timescales of interest \citepalias{Siegel.Barnes.Metzger_2019.Nature_rp.collapsar}.
Thus, while we include energy from \rp{} radioactivity in $\dot{Q}_{\rm rad, i}$ (Eq.~\ref{eq:dEdt}), we forgo complex treatments of the decay phase \citep[e.g.][]{Barnes.ea_2021.ApJ_knova.nuc.landscape} in favor of a power-law model \citep{Metzger_2010,Korobkin_NSM_rp},
\begin{align*}
    \dot{\epsilon}_{\rm rp} &= 3 \times 10^{10} \; \left(\frac{t}{1 \text{ day}}\right)^{-1.3} \:\: \text{erg \per{s} \per{g}}.
\end{align*}

We assume that 40\% of this energy is in $\gamma$-rays, which thermalize in the same way as $\gamma$-rays from \iso{Ni}{56} and \iso{Co}{56}. 
We divide the remaining energy between $\beta$-particles (35\%), which thermalize with perfect efficiency, and neutrinos, which do not thermalize at all.
This simplified treatment is justified by the sub-dominance of $\dot{\epsilon}_{\rm rp}$, and by the high densities in the enriched regions compared to the densities expected in kilonovae \citep{Bauswein_2013,Hotokezaka_2013_massEj,Kyutoku_2015_massEj,Bovard.ea_2017.PhRvD_rproc.nsm.mass.ej,Radice.ea_2018.ApJ_nsm.mass.eject}, which supports efficient thermalization \citep{Barnes_etal_2016}.

We model only emission derived from radioactivity.
In the case of a GRB-SN, the GRB afterglow could contribute to, or even dominate, optical and NIR emission at some epochs.
However, as we will show, the timescales of interest for \rp{} detection are generally much longer than the timescales on which the afterglow fades away, and contamination is not a major concern.

Opacity in our model is wavelength-independent, but varies with temperature and composition.
Ejecta free of both \nickel{} and \rp{} elements is assigned a baseline opacity $\kappa_{\rm sn} = 0.05$ cm$^2$ \per{g}, consistent with \S\ref{subsec:photolytics}.
The effects on opacity of \nickel{} and its daughter products  are accounted for with a simplified scheme, in which 
\begin{align*}
    \kappa_{56}(T) &= \begin{cases}
    \displaystyle
    \kappa_{56,0} &\text{ for } T < T_{\kappa} \\
    \displaystyle
    \kappa_{56,0} \left(\frac{T}{T_{\kappa}}\right)^{4/3} &\text{ for } T_\kappa \leq T \leq 2.5 \times 10^4 \text{ K} \\
    \displaystyle
    0.1 \text{ cm}^2\per{\text{g}} &\text{ for } T > 2 \times 10^4 \text{ K}, \\
    \end{cases}
\end{align*}
where $T_{\rm \kappa} = 3500$ K, and the lower limit $\kappa_{56,0} = 0.01$ cm$^2$ \per{g} reflects the dominance of electron-scattering opacity at low temperatures with a limited number of bound-bound transitions. 
This approximation is based on Planck mean opacities calculated for a mixture of \iso{Ni}{56}, \iso{Co}{56}, and \iso{Fe}{56} \citep[e.g.][]{Kasen_2013_AS}.

The total opacity in a zone is then given by
\begin{equation}
    \kappa_{\rm i} = \kappa_{\rm sn}(1 - X_{r \rm p,i} - X_{\rm 56}) + \kappa_{r \rm p}X_{r \rm p,i} + \kappa_{56}(T_{\rm i})X_{56}, \label{eq:ktot}
\end{equation}
where $\kappa_{r \rm p} = 10$ cm$^2$ g$^{-1}$  is the opacity of a pure \rp{} composition \citep{Kasen_2013_AS, Tanaka_Hotok_rpOps, Grossman_2014_kNe}, the \rp{} mass fraction $X_{r \rm p,i}$ is $\zeta$ within the enriched core and zero elsewhere, and the \nickel{} mass fraction $X_{56}$ equals $\mni/\mej$ in all zones.
The zone's temperature $T_{\rm i}$ is a function of its internal energy density.

Eq.~\ref{eq:ktot} allows the determination of the photosphere and the demarcation of the optically thin region.
As alluded to in \S\ref{subsec:model_basics}, emission from optically thin zones is modeled as the linear combination of two distinct SEDs associated with \rp{} and \rp{}-free material.
The SEDs are empirically derived and independent of time, and therefore elide the complex physics of nebular-phase spectral formation and evolution \citep[e.g.,][]{Jerkstrand.A_2017.hsn.book_neb.phase.sne}.
Nevertheless, \rp{} modeling \citep{Hotokezaka.ea_2021.MNRAS_kn.nebular.model} and SN observations \citep{Gomez.Lopez_2002.AJ_neb.spec.sne.ic,Tomita.ea_2006.ApJ_2002ap.opt.nir.500.days,Taubenberger.ea_2009.MNRAS_sne.ibc.emiss.lines.aspher} suggest that emission in the nebular phase may be fairly uniform in time and across different events, at least at the level of photometry.
Furthermore, we find this approach reproduces the photometry of SNe Ic with reasonable fidelity.

\begin{figure}
    \includegraphics[width=\columnwidth]{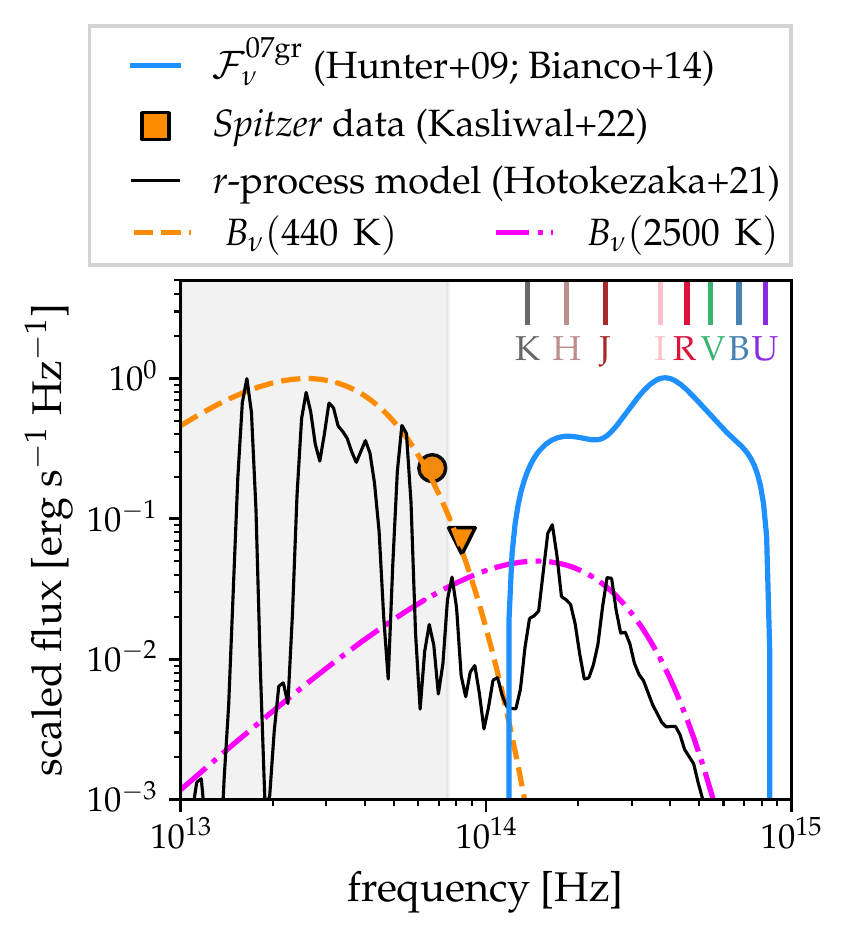}
    \caption{Our optically thin emission models and the observations and/or calculations that inform them. 
    Optically thin layers free of \rp{} material emit a spectrum $\mathcal{F}_{\nu}^{\rm 07gr}$ (blue curve) constructed from the photometry of the Type Ic SN 2007gr \citep{Hunter.ea_2009.AandA_sn2007gr.phot,Bianco.Modjaz.ea_2014.ApJS_sesn.ccsne.lcs} 120 days after $B$-band maximum.
    Observations of kn170817 43 days after the merger \citep[][orange markers]{Kasliwal.ea_2022.MNRAS_spitzer.late.obs.kn170817} are consistent with a blackbody at $T \leq 440$ K (dashed orange curve) or with the more complex spectrum (solid black line) predicted by \citet{Hotokezaka.ea_2021.MNRAS_kn.nebular.model} for \rp{} compositions in the nebular phase.
    We adopt a scaled $2500$ K blackbody SED (dot-dashed pink curve) to approximate optically thin emission associated with \rp{} material.
    The frequencies of the UVOIR and NIR bands are shown on the top axis.
    The shaded region indicates the frequencies that will be accessible to the \emph{James Webb Space Telescope} (\emph{JWST}).
    }
    \label{fig:rp-neb-mod}
\end{figure}

As mentioned in \S\ref{sec:analytics}, we construct the \rp{}-free SED from the late-time $B$- through $K$-band photometry of SN2007gr \citep{Hunter.ea_2009.AandA_sn2007gr.phot,Bianco.Modjaz.ea_2014.ApJS_sesn.ccsne.lcs}, accessed via the Open Supernova Catalog\footnote{R.I.P.} \citep[OSC;][]{Guillochon.ea_2017.ApJ_open.sn.cat}.
We consider data from $t \approx 120$ days after $B$-band maximum and perform a spline fit to convert photometry-derived monochromatic luminosities to a continuous SED, $\mathcal{F}_\nu^{\rm 07gr}$, as shown in Fig.~\ref{fig:rp-neb-mod}.
To improve the agreement between our model and SN Ic/Ic-BL observations, we assume that 30\% of the energy in this SED falls blueward of $U$ or redward of $K$. 

The SED associated with optically thin \rp{} compositions is highly uncertain.
\emph{Spitzer} observations \citep{Kasliwal.ea_2022.MNRAS_spitzer.late.obs.kn170817,Villar.ea_2018.ApJ_kn170817.mir.spitzer} of kn170817 (the only definitive \rp{} transient detected to date) at 43 days post-explosion yielded one mid-IR (MIR) photometric point and one upper limit.
If the kilonova's spectrum is a blackbody, these measurements constrain its temperature to ${\lesssim}440$ K.

On the other hand, numerical modeling by \citet{Hotokezaka.ea_2021.MNRAS_kn.nebular.model} of the emission from a nebula with properties similar to a collapsar disk wind and composed purely of Neodymium (a high-opacity element synthesized by the \rp{}) predicted a spectrum with far more complexity than a blackbody.
While their results showed MIR emission consistent with the observations of \citet{Kasliwal.ea_2022.MNRAS_spitzer.late.obs.kn170817}, they also found significant flux at lower wavelengths. 
We found that a blackbody at $\Trp = 2500$ K captures the features of the lower-wavelength emission, while the MIR component, which accounts for ${\sim}60$\% of the emitted energy, is too red to impact the photometry.
Fig.~\ref{fig:rp-neb-mod} shows both the  observations and the numerical model, as well as blackbody spectra for select temperatures, which we present for comparison.

The calculation of \citet{Hotokezaka.ea_2021.MNRAS_kn.nebular.model} relies on a simplified model of pure \rp{} ejecta; neither the assumed composition nor the heating rate (due exclusively to \rp{} decay) map directly onto the collapsar context. 
However, the argument for a bimodal spectrum is supported by simple arguments about \rp{} elements' atomic structures, which may be robust against increasing model complexity. 
In the absence of additional data, we 
therefore approximate the \rp-associated SED as a blackbody at $2500$ K, which we scale to account for the out-of-band emission. 

Though some of our zones are \rp-free, none are purely \rp{};
at a minimum, each zone contains a mixture of \rp{} elements and \nickel.
The appropriate way to model nebular emission from zones with complex compositions is an additional uncertainty.
Nebular spectra are dominated by the species that cool most efficiently, which may be distinct from the most abundant elements.

Here, we move away from the simpler approach of \S\ref{subsec:neb_model} and allocate the luminosity of an optically thin zone according to the fraction of the total optical depth a given component provides across that zone. 
For \rp{} material, that fraction is 
\begin{equation*}
    f_{r \rm p,i} = \frac{\kappa_{r \rm p} X_{r \rm p,i}}{\kappa_{\rm i}},
\end{equation*}
with $\kappa_{\rm i}$ defined by Eq.~\ref{eq:ktot}.
Thus, the luminosity $L_{\rm neb,i}$ from a zone outside the photosphere is converted to a SED following
\begin{align}
    L_{\rm neb,i,\nu} &= L_{\rm neb,i,\nu}^{r \rm p} + L_{\rm neb,i,\nu}^{\rm sn}, \nonumber \\
    \intertext{where}
    L_{\rm neb,i,\nu}^{r \rm p} &= 0.41 \times f_{r \rm p,i} \; \frac{ \pi L_{\rm neb,i} }{\sigma_{\rm SB}\Trp^4 } \; B_\nu(\Trp), \nonumber \\
    \Trp &= 2500 \:\text{K}, \nonumber \\
    \intertext{and}
    L_{\rm neb,i,\nu}^{\rm sn} &= 0.7 \times (1-f_{r \rm p,i}) \; L_{\rm neb,i} \; \mathcal{F}_\nu^{\rm 07gr}.
\end{align}

\subsection{Model validation}\label{subsec:ibc_obs}

\begin{figure}
    \includegraphics[width=\columnwidth]{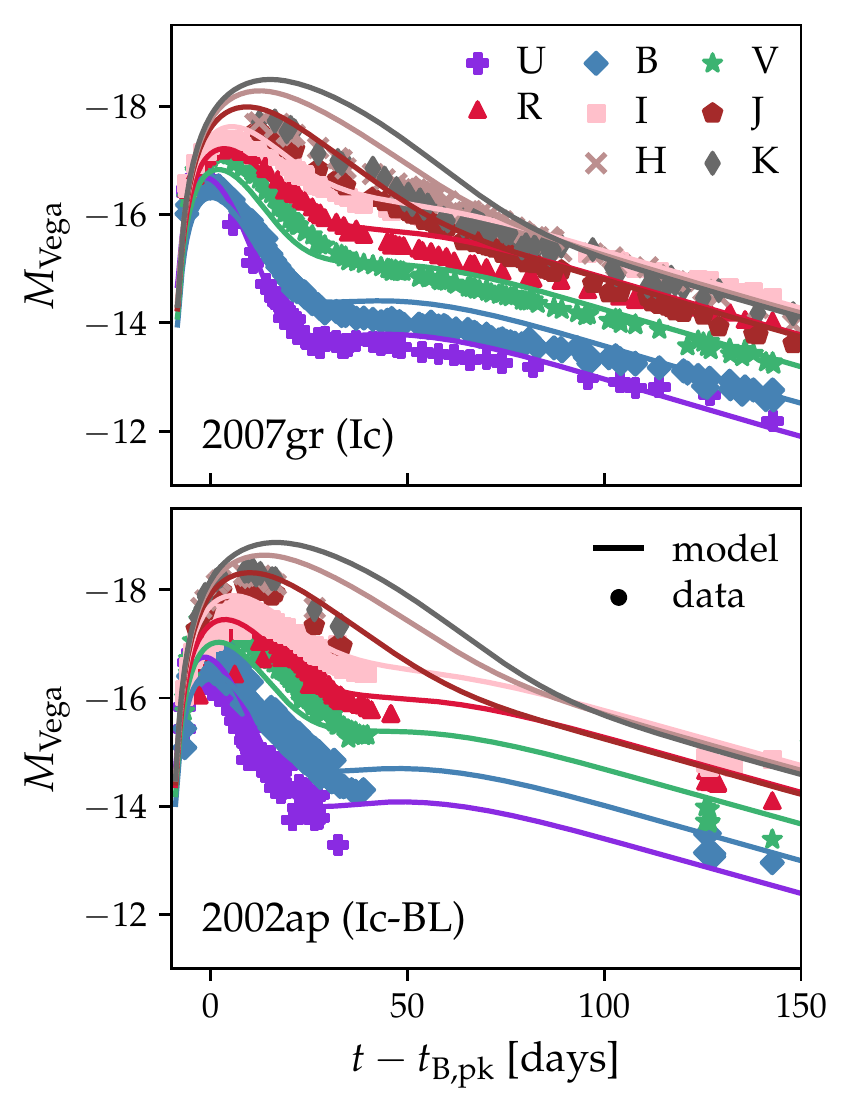}
    \caption{A comparison of model photometry (solid lines) and observations (markers) shows that our approach is well-suited for modeling SNe Type Ic at a range of kinetic energies.  SN 2007gr (\textit{top panel}) is classified as an ordinary SN Ic \citep{Madison.Li_2007.CBET_sn.2007gr.disc}, while SN 2002ap (\textit{bottom panel}) is a SN Ic-BL \citep{Meikle.ea_2002.IAUC_sn2002ap.disc.icbl.id,Filippenko.Chornock_2002.IAUC_sn2002ap.disc.icbl.id,GalYam.ea._2002_02ap.data}. The model parameters for each SN are given in Table~\ref{tab:f1_mods}.
    Photometry is from \citet{Bianco.Modjaz.ea_2014.ApJS_sesn.ccsne.lcs}, courtesy of the OSC \citep{Guillochon.ea_2017.ApJ_open.sn.cat}, and from \citet{Yoshii.ea_2003.ApJ_opt.nir.phot.sn2002ap} and references therein.
    To improve readability, in these plots only we calculate magnitudes using the Vega system.
    }
    \label{fig:data_comp}
\end{figure}

We validated our semi-analytic model against a handful of  SNe Ic/Ic-BL with late-time multi-band photometry.
In Fig.~\ref{fig:data_comp}, we show our predictions alongside observations for one ordinary and one broad-lined SN Ic.
The model parameters we used, as well as inferred ejecta properties reported in the literature, are recorded in Table~\ref{tab:f1_mods}.
Despite its simplifications, our model reproduces the basic features of the SNe, suggesting that we are accounting for the most important physical processes driving the light curve's evolution. 

Around maximum light, our model is most accurate in ultraviolet and optical bands. 
Though a lack of data makes it harder to gauge model performance in the NIR, the data that are available suggest we may over-predict $J$, $H$, and $K$ magnitudes in the light curve's early phases.
However, given our focus on optical-NIR color \emph{difference} as an \rp{} signature (e.g. \S\ref{subsec:neb_model}, \S\ref{sec:results}), slightly overestimating the NIR emission of \rp-free SNe near peak will merely make our calculation more conservative; the color differences we predict are likely to be underestimates, meaning the actual \rp{} signal may be stronger than we forecast.

We also found that our model is slightly less successful at reproducing the photometry of SNe Ic-BL. 
This is perhaps not surprising considering that the extreme kinetic energies (${\sim}10^{52}$ erg) of SNe Ic-BL, as well as the GRBs sometimes observed in conjunction with them, point to a nonstandard (e.g., ``engine-driven'') explosion mechanism that may induce ejecta asymmetries or unusual density profiles \citep[e.g.][]{Maeda03ApJ_twoCompNi_blic,Tanaka.ea_2008.ApJ_sn.2002ap.asphericity.rt,Barnes.Duffell.ea_2018.ApJ_grb.icbl.engine}.
They are also the SNe most closely linked, theoretically, to the collapsar explosion model \citep[e.g.,][]{WoosleyBloom06_snegrbRev}, and therefore the most likely to produce emission-altering \rp{} elements.

\begin{table}
\begingroup
\centering
\caption{Model parameters adopted in Figure~\ref{fig:data_comp}}\label{tab:f1_mods}
    \begin{tabular}{p{0.15\columnwidth}>{\centering}p{0.2\columnwidth}>{\centering}p{0.22\columnwidth}>{\centering\arraybackslash}p{0.23\columnwidth}}
    \toprule
    \emph{SN} & $\mej/\msun$ & $\ekin/10^{51}$ erg & \mni/\msun  \\
    \hline
    \multirow{2}{*}{2007gr$^{\rm a}$} & 2.0 ($2.0 - 3.5$) & 3.2 ($1.0-4.0$) & 0.08 ($0.076^{+0.02}_{-0.02}$) \\
    \multirow{2}{*}{2002ap$^{\rm b}$} & 3.2 ($2.5 - 5.0$) & 4.4 ($4.0-10.0$)  & \multirow{2}{*}{\shortstack{ 0.11 \\ (0.07) }} \rule{0pt}{3ex} \\
    \hline
    \end{tabular}
    \endgroup
    \vspace{0.5 mm}
    \newline \footnotesize{$^a$Inferred ejecta properties for SN 2007gr (in parentheses) from \citet{Hunter.ea_2009.AandA_sn2007gr.phot}.}
    \newline \footnotesize{$^b$Inferred ejecta properties for SN 2002ap (in parentheses) from \cite{Deng.ea_2003.NuPhA_2002ap.mods}.}
\end{table}

\subsection{Construction of a model suite}

We construct multi-band light curves extending to $t=200$ days after explosion for \rp{} enriched and unenriched SNe with a range of explosion and \rp-enrichment properties. 
The parameter values used in the model suite are presented in Table~\ref{tab:suite}.
We require that radioactive material not dominate the total ejecta mass (i.e., we enforce $\mni \leq 0.5\mej$ and $\mrp \leq 0.5(\mej - \mni)$), and we do not consider models with kinetic energies $\ekin > 5 \times 10^{52}$ erg.
Beyond these constraints, all parameter combinations are explored.
The \rp-free models reproduce the range of luminosities and timescales of observed SNe Ic and Ic-BL \citep{Drout.ea_2011.ApJ_sn.type.ibc.stats,Perley.ea_2020.ApJ_ZTF.sn.sample}, which validates both our modeling framework and the range of parameter values we adopt.

We do not consider here the case of complete mixing ($\xmix=1$). 
For fully mixed models, the high opacity of the \rp{} elements would affect the SN emission from the explosion onward, resulting in a transient very distinct from ordinary SNe at all phases of its evolution.
The question of detecting fully mixed \rccsne{} (or of recognizing one should it turn up in a blind search) is therefore better deferred to a separate work (though it has been at least partially addressed by \citet{Siegel.ea_2021.arXiv_super.kilonovae}, who discuss electromagnetic counterparts from very massive, uniformly mixed, \rp-enhanced ``superkilonovae'').

\begin{table}
\begingroup
\centering
\caption{Parameters of the model suite}\label{tab:suite}
    \begin{tabular}{p{0.15\columnwidth}>{\centering}p{0.18\columnwidth}>{\centering}p{0.18\columnwidth}>{\centering}p{0.1\columnwidth}>{\raggedleft\arraybackslash}r}
    \toprule
    Quantity & Minimum & Maximum & $N^\star$ & Spacing \\
    \hline 
    \mej{} & $0.5\msun$ & $12\msun$ & 24 & logarithmic \\
    \bej{} & 0.01 & 0.25 & 25 & logarithmic \\
    \mni{} & $0.05\msun$ & $1.0\msun$ & 20 & logarithmic \\
    \xmix{} & $0.1^{\dagger}$ & 0.9 & ${\leq}10$ & linear \\
    \arrayrulecolor{black!20}\hline
    \mrp{} &\multicolumn{4}{r}{ $(0.0\msun)$, $0.01\msun, \: 0.03 \msun, \: 0.08 \msun, \: 0.15 \msun$} \\
    \arrayrulecolor{black}
    \hline
    \end{tabular}
    \endgroup
    \vspace{0.5 mm}
    \newline \footnotesize{$^\star$The number of distinct values considered for each quantity. }
    \newline \footnotesize{$^\dagger$For every choice of \mej, \mni, and \mrp{} there is a minimum \xmix{} for which the enriched core contains only \nickel{} and \rp{} elements. Values of \xmix{} for each model include this minimum, $\psi_{0}$, and all \xmix{} indicated above for which $\xmix > \psi_0$. }
\end{table}

\section{Results}\label{sec:results}

We first explain how the addition of \rp{} material influences the evolution of SNe with specified explosion parameters (\mej, \bej, and \mni).
We then adopt a wider lens, and consider how signs of \rp{} enrichment may manifest in SNe with different observational properties.
Finally, we consider how our analysis could inform \rccsn{} search strategies.

\subsection{Effects of r-process enrichment on SN emission}\label{subsec:rpfx}

Enriching SN ejecta with \rp{} elements
extends the photospheric phase and alters the SED from the optically thin layers.
Each of these effects shifts the emitted spectrum from the optical toward the NIR; however, the strength of this shift, and the timescale at which it occurs, depend on the mass of \rp{} material (\mrp) and how extensively it is mixed in the ejecta (\xmix).  

\begin{figure*}
    \includegraphics[width=\textwidth]{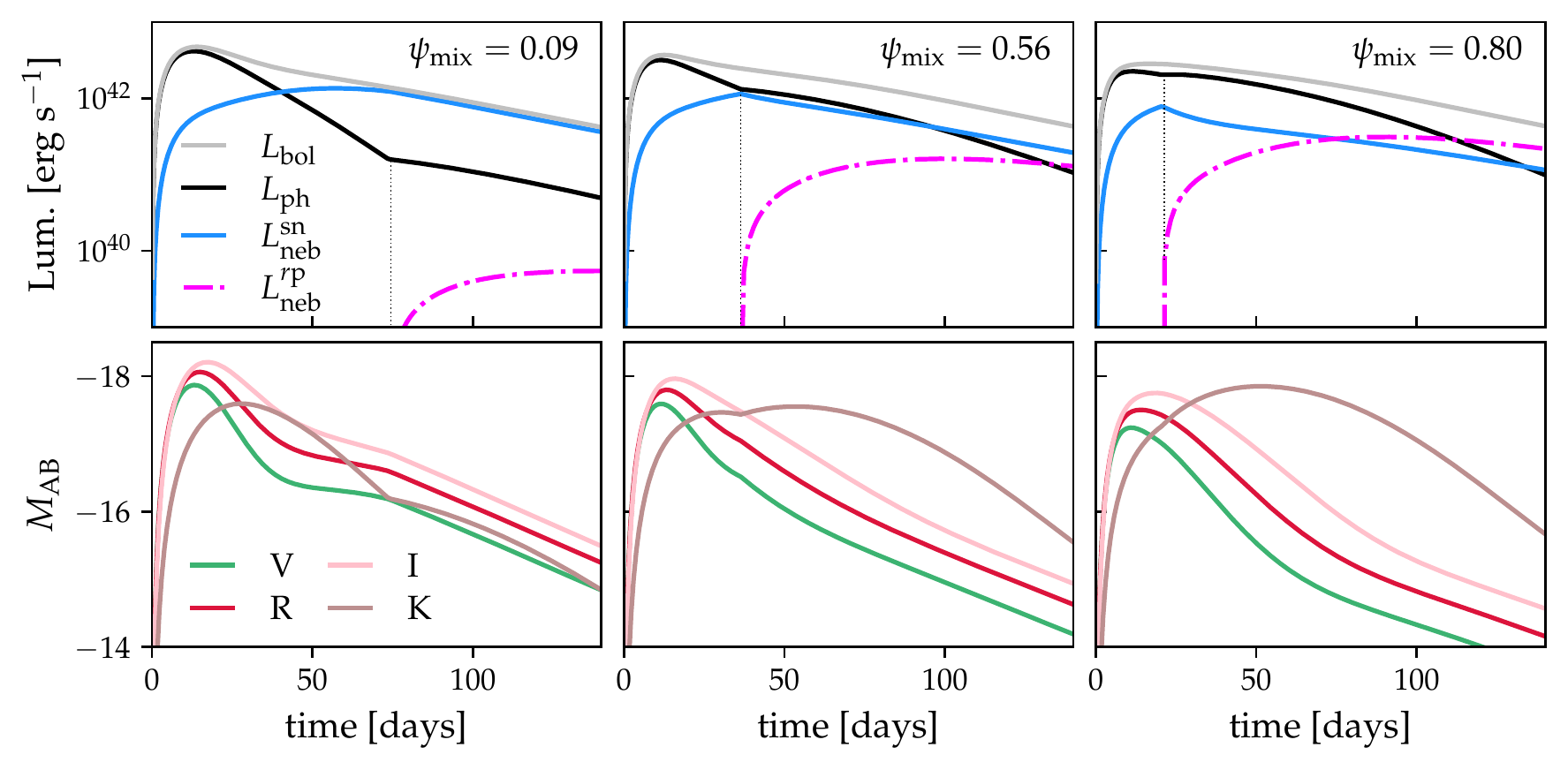}
    \caption{Increased mixing enhances emission in the NIR relative to the optical.
    All models above have $\mej = 4.0 \msun$, $\bej = 0.04$, $\mni = 0.25 \msun$, and $\mrp = 0.08\msun$.
    \textit{Top panels:} Mixing redistributes energy among luminosity components, with higher \xmix{} favoring $L_{\rm ph}$ and $L_{\rm neb}^{r \rm p}$ over $L_{\rm neb}^{\rm sn}$.
    (Note that $L_{\rm neb}^{\rm sn}$ and $L_{\rm neb}^{r \rm p}$ refer to luminosity components with distinct SEDs, rather than components emitted from \rp-rich or -free regions. 
    As described in \S\ref{subsec:model_imp}, enriched ejecta contributes energy to both $L_{\rm neb}^{\rm sn}$ and $L_{\rm neb}^{r \rm p}$.)
    The dotted black lines indicate $t=\tau_{\rm tr}$, the time at which the outer \rp-free layers become transparent. The evolution of $L_{\rm ph}$ slows at this point in response to the higher opacity of the core.
    \textit{Bottom panel:} Select broadband light curves showing the redistribution of energy from bluer to redder wavelengths. 
    }
    \label{fig:mix_fx}
\end{figure*}

The response of bolometric and broadband light curves to \xmix{} is illustrated in Fig.~\ref{fig:mix_fx}.
When the \rp{} is concentrated in the ejecta's center, its influence is minimal, since only a negligible fraction of the radiation originates in the enriched layers.
At higher \xmix, the effects are more visible.
The extended high-opacity core limits diffusion from the interior, producing lower $L_{\rm bol}$ and $L_{\rm ph}$ near peak.
After the outer layers reach transparency, the opaque core slows---or even reverses---the recession of the photosphere, sustaining a higher $L_{\rm ph}$ at the expense of $L_{\rm neb}^{\rm sn}$.
As the enriched layers (slowly) become transparent, their nebular emission begins to contribute to $L_{\rm neb}^{r \rm p}$, and
for high enough \xmix{} or late enough epochs, $L_{\rm neb}^{r \rm p}$ can overpower $L_{\rm neb}^{\rm sn}$.
The long-lived photosphere and the \rp{} nebular component each provide a luminous source of low-temperature emission that impacts the evolution of the SED and, therefore, the broadband light curves.

As seen in Fig.~\ref{fig:mix_fx}, mixing also affects light-curve shapes, though not always in a straightforward way.
In most cases, the opacity of the core is high enough that emission from the \rp-rich and \rp-free layers effectively becomes decoupled, rising to distinct peaks on distinct timescales (e.g., \S\ref{subsec:photolytics}).
Unless \xmix{} is very high, diffusion from the core is suppressed to the degree that the peak of $L_{\rm ph}$ is driven mainly by the \rp-free ejecta.
Increasing \xmix{} reduces the mass and increases the average velocity of this ejecta component, producing a narrower peak in $L_{\rm ph}$ and sharper light curves in optical bands, despite the increasing spatial extent of the high-opacity region.

To better understand how \rp{}-enriched SNe may be distinguished from their \rp-free counterparts, we expand the parameter space of Fig.~\ref{fig:mix_fx}, enriching a single explosion model $(\mej, \bej,\mni) = (4.0\msun, 0.04, 0.25\msun$) with a range of \mrp{} at various \xmix{}.
We calculate the broadband evolution for each combination $(\mrp,\xmix)$ and compare the colors to those of an \rp-free SN with the same \mej, \bej, and \mni. 

Because the effects of enrichment are seen primarily in the NIR, we use $R{-}X$ color as a proxy for the \rp{} signal strength, with $X \in \{J, H, K\}$.
As in \S\ref{subsec:neb_model}, we focus on color \emph{difference}: we determine the time, $t_\Delta$, at which the colors of each \rccsn{} model differ from those of the unenriched model by at least one magnitude. 
Since this divergence occurs at a range of $R{-}X$, the focus on color difference rather than absolute color is useful for making comparisons across a diverse set of models.
(However, we provide more concrete predictions of color itself in \S\ref{subsub:sn_egs} and \S\ref{subsub:cols_ts}.)
To demonstrate the importance of the NIR bands, we perform the same calculation for the optical color $V{-}R$, and find no models for which $\Delta(V{-}R)$ is ever greater than 1 mag.

\begin{figure}\includegraphics[width=\columnwidth]{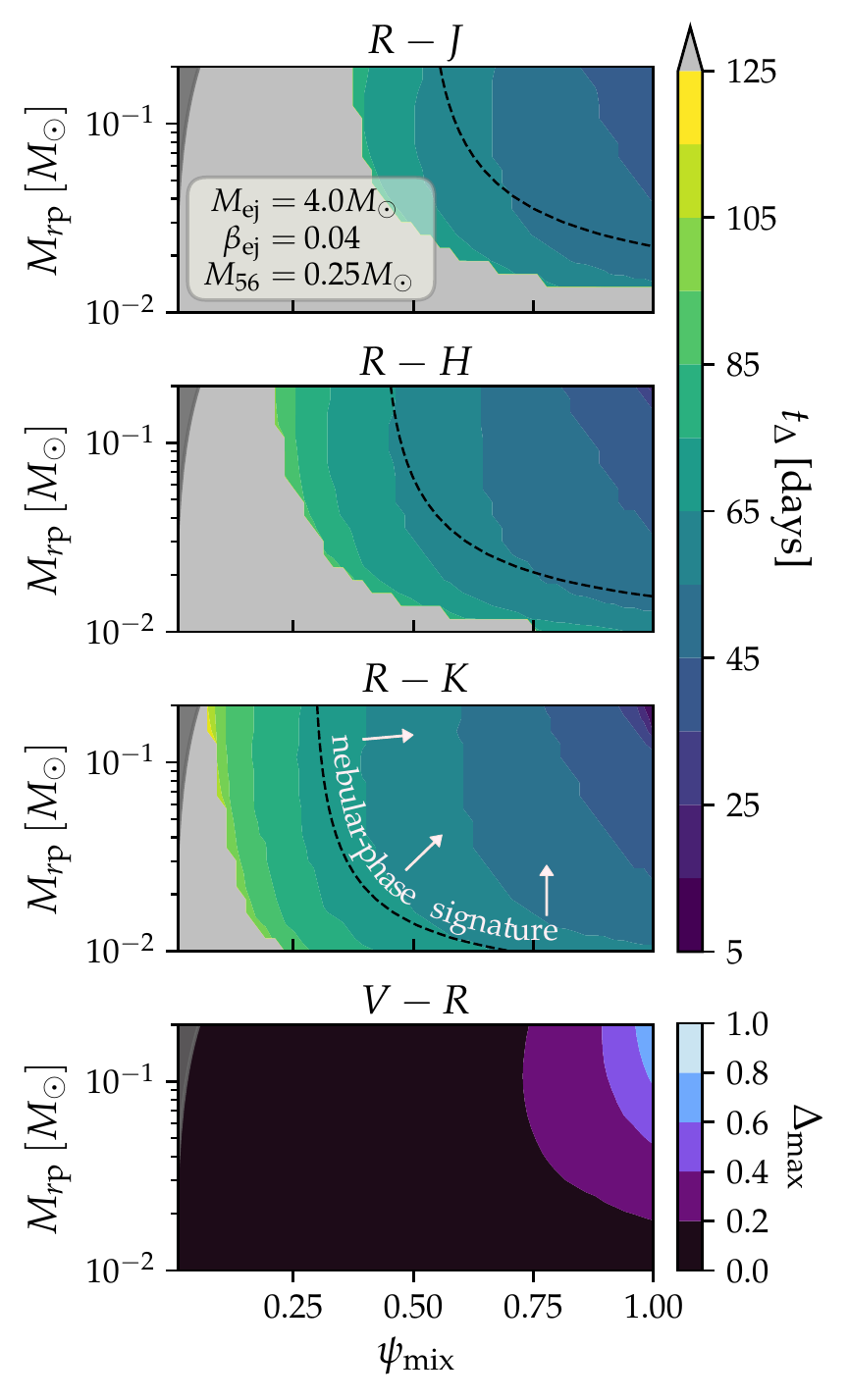}
    \caption{The effects of enrichment are strongest at redder wavelengths and  when the \rp{} mass and/or the degree of mixing is high.
    \emph{Top three panels:} The times\ ($t_\Delta$) at which select colors of \rccsne{} with variable \mrp{} and \xmix{} but uniform \mej, \bej, and \mni{} ($=4.0 \msun, \: 0.04$, and $0.25\msun$, respectively) first differ meaningfully ($\Delta \geq 1$ mag) from an unenriched SN with the same explosion parameters.
    The divergence occurs earlier for higher \mrp{} and \xmix, and for a larger fraction of the models as redder bands are considered.
    Models to the left of the dashed black lines in each panel are found, under the framework of \S\ref{subsec:model_imp}, to have $\Delta(R{-}X) < 1$ mag once the ejecta is fully transparent, and therefore meet our detection threshold only in the photospheric phase.
    \emph{Bottom panel:} The maximum difference (with respect to time) in $V{-}R$ of the \rccsne{} compared to the unenriched model.
    The effect is small in the optical bands; for no model does $\Delta(V{-}R)$ ever exceed 1 mag.
    \emph{All panels:} Dark gray shading marks parameter combinations disallowed by the requirement that \rp-enriched core also contain \nickel{} ($\xmix\mej \geq \xmix\mni + \mrp$).
    }
    \label{fig:tdiv_onemod}
\end{figure}

The results of this analysis, which we present in Fig.~\ref{fig:tdiv_onemod}, confirm that the emission of \rccsne{} is most distinct for high \xmix{} and high \mrp, consistent with earlier analytic arguments (e.g. Fig.~\ref{fig:bxz-plane}), and more obvious for \RK{} than for \RH{} or \RJ{}.
(Crucially, however, \emph{any} choice of $R{-}X$ is a more reliable \rp{} indicator than optical colors, as seen in the bottom panel, which shows the maximum difference in $V{-}R$ for each model.)

Fig.~\ref{fig:tdiv_onemod} also hints at when signs of more modest enrichment may appear; for $\xmix \leq 0.3$, the color difference for all $R{-}X$ is ${<}1$ mag until ${\sim}2$--3 months after explosion. 
In fact, for many $(\mrp, \xmix)$, \drx{} does not exceed our threshold within the time frame of the simulation, or exceeds it for only one of the three colors considered.

Still, Fig.~\ref{fig:tdiv_onemod} reinforces the potential of pre-nebular phase observations for \rp{} detection.
The dashed black curves in the top three panels show which \rccsne{}, once their ejecta were fully transparent, would have $\drx = 1$ mag. 
(Here, in contrast to Fig.~\ref{fig:rx_neb}, we have determined the nebular SED following the prescription of \S\ref{subsec:model_imp}.)
The swaths of parameter space that lie between the light gray regions and the dashed lines contain models whose enrichment is more visible late in the photospheric phase, when at least the enriched core remains opaque, than during the nebular phase, after the emission has achieved its asymptotic colors.

\subsubsection{Supernova case studies}\label{subsub:sn_egs}

We will argue later that, for questions of \rp{} detectability, classifying models based on \mej, \bej, and \mni{} is of limited utility.
Regardless, before proceeding to a more observationally motivated schema, we present detailed color-evolution predictions for a handful of models based on analyses of stripped-envelope SN demographics by \citet{Barbarino.ea_AandA2021_reg.Ic.iPTF.survey} and  \citet{Taddia.ea_2019.AandA_IcBL.iPTF.survey}.
(We note that many groups \citep{Drout.ea_2011.ApJ_sn.type.ibc.stats,Prentice.ea_2019.MNRAS_sesne.properties.progens,Perley.ea_2020.ApJ_ZTF.sn.sample} have contributed to efforts to uncover the distributiona of SNe Ic/Ic-BL properties, and that the characteristics of an ``average'' SN in a given category remain uncertain.)

In addition to cases representing typical SNe Ic/Ic-BL, we consider models based on individual SNe Ic-BL with inferred ejecta masses much higher and much lower than average, in order to explore how the signal may vary within the SN Ic-BL population. 
The explosion properties of our four models, along with the event or analysis on which each is based, can be found in Table~\ref{tab:rpfree}.
We enrich each of these models with 0.03\msun{} of \rp{} material, spread out to varying mixing coordinates \xmix.

\begin{table}
\caption{Explosion properties of the models of Fig.~\ref{fig:sn_casestudy}}\label{tab:rpfree}
    \begin{tabular}{p{0.175
    \columnwidth}>{\centering}p{0.16\columnwidth}>{\centering}p{0.08\columnwidth}>{\centering}p{0.08\columnwidth}>{\raggedleft\arraybackslash}p{0.28\columnwidth}}
    \toprule
    Type & \mej{} $[\msun]$ & \bej{} & \mni{} & Reference \\
    \hline
    Typical Ic-BL & \multirow{2}{*}{3.97} &  \multirow{2}{*}{0.044} &  \multirow{2}{*}{0.33} & \citetalias{Taddia.ea_2019.AandA_IcBL.iPTF.survey}$^\star$ \mbox{(average values)}  \\
    High-mass Ic-BL & \multirow{2}{*}{10.45} & \multirow{2}{*}{0.029} & \multirow{2}{*}{0.85} & \citetalias{Taddia.ea_2019.AandA_IcBL.iPTF.survey} (PTF10ysd) \\
    Low-mass Ic-BL & \multirow{2}{*}{1.51}  & \multirow{2}{*}{0.050} & \multirow{2}{*}{0.21} &\citetalias{Taddia.ea_2019.AandA_IcBL.iPTF.survey} (PTF10tqv) \\
    \multirow{2}{*}{Typical Ic}  & \multirow{2}{*}{3.97} & \multirow{2}{*}{0.020} & \multirow{2}{*}{0.21} &  \citetalias{Barbarino.ea_AandA2021_reg.Ic.iPTF.survey}$^\dagger$ \mbox{(average values)} \\
    \hline
    \end{tabular}
    \vspace{0.5 mm}
    \newline \footnotesize{$^\star$\citetalias{Taddia.ea_2019.AandA_IcBL.iPTF.survey}: \citet{Taddia.ea_2019.AandA_IcBL.iPTF.survey} }
    \newline \footnotesize{$^\dagger$\citetalias{Barbarino.ea_AandA2021_reg.Ic.iPTF.survey}: \citet{Barbarino.ea_AandA2021_reg.Ic.iPTF.survey} }
\end{table}

\begin{figure*}
    \includegraphics[width=\textwidth]{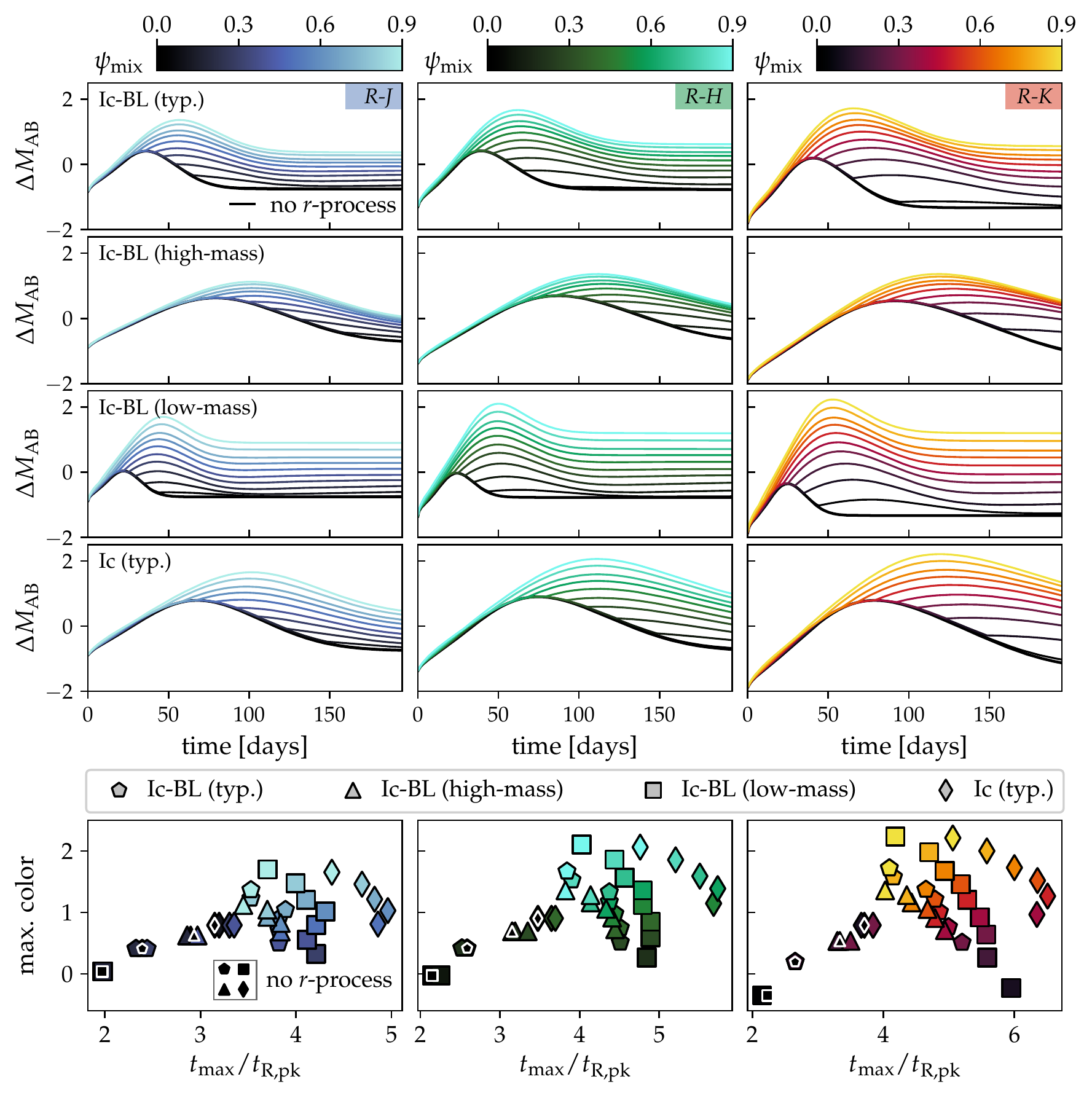}
    \caption{The effect of \rp{} enrichment ($\mrp = 0.03\msun$, variable \xmix) on $R{-}X$ for four explosion models with \mej, \bej, and \mni{} corresponding either to literature-reported averages, or to particular events representing extrema in the distribution of inferred SN Ic-BL properties
    (see Table~\ref{tab:rpfree} for details).
    We show the full color evolution for each model (\emph{top panels}) as well as the maximum color and the time at which occurs (\emph{bottom panels}).
    In all panels, we plot predictions for \rp-free models with the same explosion properties for comparison. 
    For all but the lowest \xmix, the presence of \rp{} material enhances $R{-}X$, producing either a secondary maximum, or a delayed global maximum relative to the unenriched cases.
    Significantly, the strongest enhancement can be transient in nature and may occur well before the \rccsn{} reaches its asymptotic colors.
    The impact is strongest for the highest velocity models (the typical and low-mass SNe Ic-BL), but can be significant even for a typical SN Ic if \xmix{} is high.
    }
    \label{fig:sn_casestudy}
\end{figure*}

Fig.~\ref{fig:sn_casestudy} shows the $R{-}X$ color evolution for each set of explosion parameters as a function of \xmix.
We display for comparison the colors of a SN with the same explosion properties but no \rp{} enrichment.
To better summarize the data, and to situate the \rp-induced changes in color more directly in the context of observational SN properties, we also plot in the bottom row of Fig.~\ref{fig:sn_casestudy} the maximum color for each model and the time at which that maximum occurs, normalized to the $R$-band rise time, $t_{\rm R,pk}$.

Regardless of \xmix, the colors of the \rccsn{} models track those of their unenriched counterparts at early times, becoming noticeably redder only later on (around $t \gtrsim 25$ days for this set of transients).
The extent of the reddening depends on \xmix{} and on \mej, \bej, and \mni; nevertheless, certain general trends are apparent.

While the effects of the \rp{} are present for all colors considered, they increase in prominence with the wavelengths of the NIR band.
This theme is also apparent in Fig.~\ref{fig:sn_casestudy}'s bottom panels.
For all colors, models with at least moderate \xmix{} occupy a distinct region of parameter space compared to poorly mixed or unenriched models.
The separation is strongest for $\RK$. 

More than color, the signal strength depends on the explosion parameters, particularly velocity. 
\emph{R}-process-induced changes in color are far more noticeable for the typical and low-mass SN Ic-BL models (with $\bej = 0.044$ and $0.050$, respectively) than for the high-mass SN Ic-BL or the typical SN Ic model (with $\bej < 0.03$).
This can be seen both in the full color evolution, and the position of each model in parameter space of the bottom panels.

However, consistent with Fig.~\ref{fig:tdiv_onemod}, the color change is most sensitive to the mixing coordinate \xmix. 
For nearly all models with $\xmix \geq 0.1$, \rp{} enrichment produces color extrema redder than the asymptotic colors of the corresponding unenriched SN. 
For some models, this is a global maximum occurring at a delay relative to the unenriched SN.
This is the case for $\xmix \gtrsim 0.2$--0.3 for the typical and low-mass Ic-BL models, and for $\xmix \gtrsim 0.4$ for the high-mass Ic-BL and the typical Ic model. 
(Even in the case of strong mixing though, the difference between the enriched and unenriched SN colors are much smaller for the latter two cases than for the former two.)

At lower levels of mixing, the colors of the \rccsne{} track those of their unenriched SN counterparts for a longer period of time, and rise only to a local maximum after diverging.
In these cases, as the bottom panels indicate, the maximum values of $R{-}X$ would not be sufficient to distinguish enriched from unenriched SNe; longer-term precision photometry would be required.

Nevertheless, it is reassuring that only the slowest and/or most poorly mixed models considered here fail entirely to form a distinct $R{-}X$ peak prior to the nebular phase.
For typical SN Ic-BL explosion parameters, \rccsne{} with $\xmix \geq 0.1$ exhibit either a global maximum $\RK$ color ${>}0$  or a secondary maximum ${<}0$ at $50 \lesssim t \lesssim 100$ days.
In contrast, without enrichment we predict a peak value of $\RK = 0.2$ at $t = 40$ days.
This again points to the value of pre-nebular phase observations for evaluating collapsars as sites of \emph{r}-production, particularly for $\xmix \geq 0.2$.

\subsection{Prospects for detection}\label{subsec:snvrpsn}

While the discussion of \S\ref{subsec:rpfx} and \S\ref{subsub:sn_egs} is useful for illustrating trends, it may overstate the differences between \rp-rich and \rp-free SNe.
Emission from models with large \mrp{} and \xmix---i.e., the models Figs.~\ref{fig:tdiv_onemod} and \ref{fig:sn_casestudy} suggest should be easiest to identify---is likely to be so impacted by enrichment that it bears little resemblance to the emission from an \rp-free counterpart of the same \mej, \bej, and \mni.
Since observers have no way of knowing a priori the physical properties of a SN explosion, a more appropriate reference case for an \rccsn---particularly if it is highly enriched or very well-mixed---is an \rp-free SN with similar \emph{observed} properties.

For the following analysis, we therefore categorize our models in terms of observable, rather than physical, parameters. 
Specifically, we classify them according to their $R$-band rise time, \tR;  peak $R$-band magnitude, \Mr; and velocity, \bej.
(While \bej{} is not an observed property in the same sense as \tR{} and \Mr, measurements of absorption features in SN spectra can provide  estimates of average ejecta velocities.)
SNe (whether \rp{}-enriched or not) with comparable \tR, \Mr, and \bej{} will not evolve in perfect synchronicity; still, this procedure allows us to at least compare models with similar behavior near peak light, when most observations are obtained.

\begin{figure}
\includegraphics[width=\columnwidth]{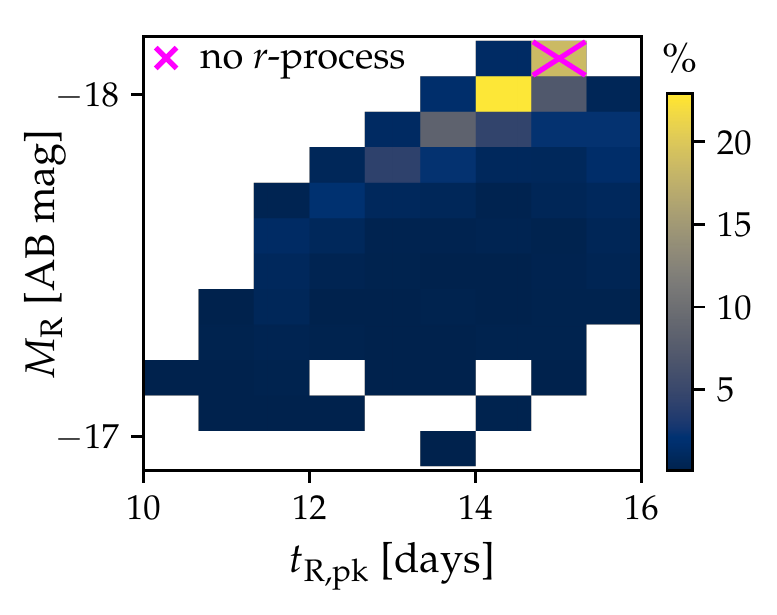}
\caption{
The presence of \rp{} material alters the SN light curves even near optical peak. 
As a result, \rp{}-free models with matching explosion parameters are not necessarily ideal points of comparison for \rccsne.
Above, we show how models with fixed $(\mej, \bej, \mni) = (4.0 \msun, 0.04, 0.25\msun)$ are distributed in  \tR{} and \Mr{}, where \tR{} (\Mr{}) is the rise time (peak magnitude) in $R$-band.
The histogram cell containing the \rp-free model with the same \mej, \bej, and \mni{} is marked with a fuchsia ``$\times$''. 
It contains only a minority of the enriched models.
}\label{fig:binhist}
\end{figure}

Fig.~\ref{fig:binhist} shows the advantage of this approach. 
As the histogram makes clear, models with identical explosion parameters nonetheless exhibit a range of \tR{} and \Mr{} when \rp{} enrichment varies.
More than 30\% of the models in Fig.~\ref{fig:binhist} have a \tR{} (\Mr) that differs from the \rp-free case by more than 1 day (0.5 mag).
This highlights the risk of assuming that \mej, \bej, and \mni{} can be extracted from light-curve data independent of the amount (or existence) of \rp{} enrichment.

However, a challenge of this framework is that there is no way to identify a single \rp-free model to which an \rccsn{} model should be compared.
(Unlike \mej, \bej, and \mni, we cannot force ordinary SNe and \rccsne{} to have the same \tR{} and \Mr{}.)
As in earlier sections, we focus on color as a diagnostic, and continue to use $\drx \geq 1$ mag as the criterion for detectability, with $X \in \{J,H,K\}$.
Now, however, instead of making one-to-one comparisons, we sort our models into bins of size $\Delta \tR = 2$ days, $\Delta \Mr = $0.25 mag, and $\Delta \bej/\bej = 0.37$, then contrast the colors of individual \rccsne{} with an average color evolution constructed from the \rp-free SNe in the same bin.

\begin{figure}\includegraphics[width=\columnwidth]{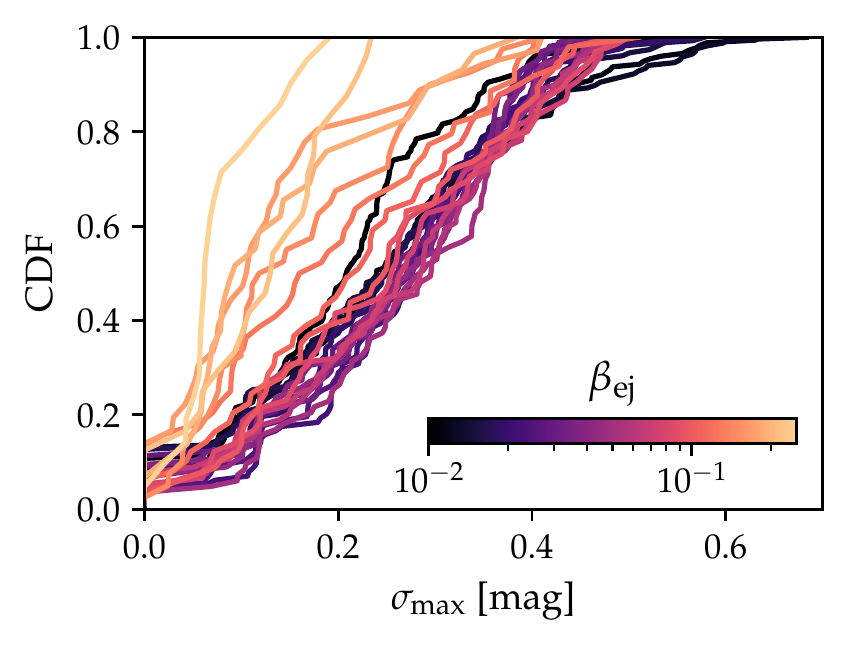}
    \caption{SNe with similar observed properties near peak do not exhibit perfectly uniform color evolution.
    Having binned the \rp-free SNe according to  \tR, \Mr, and \bej{} and averaged their $R{-}X$ color evolution, we determine $\sigma_{\rm max}$, the maximum standard deviation in each bin for any $R{-}X$, with respect to time. 
    The cumulative distribution function of $\sigma_{\rm max}$ is shown above for bins of constant central \bej.
    For typical SN Ic/Ic-BL velocities ($0.03 \lesssim \bej \lesssim 0.1$), most bins have $\sigma_{\rm R{-}X} \lesssim 0.4$ mag at all times, less than our \rp{} detection threshold $\drx \geq 1$ mag. 
    Furthermore, $\sigma_{\rm max}$ is not necessarily indicative of $\sigma_{\rm R{-}X}$ when a detection occurs (i.e., when $\drx \geq 1$ mag).
    In practice, the uncertainty at these times is often much less than the maximum value.}\label{fig:sigma_cdf}
\end{figure}

This introduces some uncertainty into the comparison, as unenriched models in a given bin do not produce identical SNe.
Indeed, the time-dependent standard deviation of the $R{-}X$ colors for \rp-free SNe, $\sigma_{\rm R{-}X}(t)$, can reach ${\sim}0.7$ mag at certain times for certain bins.
However, $\sigma_{\rm R{-}X}$ depends on velocity. 
This can be seen in Fig.~\ref{fig:sigma_cdf}, which presents the cumulative distribution function of the maximum $\sigma_{\rm R{-}X}$ for bins with a particular \bej.
For bins corresponding to typical SN Ic/Ic-BL velocities ($0.03 \lesssim \bej \lesssim 0.1$), $\sigma_{\rm R{-}X} < 0.5$ mag at all times for all $X$ considered.
Moreover, in this same velocity range, the average bin has at all times $\sigma_{\rm R{-}X} \lesssim 0.25$ mag, much less than our detection threshold $\drx \geq 1$.
This increases our confidence that models flagged as detectable truly do differ in significant ways from the unenriched reference cases. 

An additional complication is the fact that enriched and ordinary SNe do not populate the exact same regions in the \tR-\Mr-\bej{} parameter space.
In the few instances where \rccsne{} occupy a bin containing no unenriched SNe, we prioritize comparing events of similar \tR{} and \Mr, since these are directly measured, while \bej{} must be inferred from observations.
In these cases, we define the reference color evolution for the calculation of \drx{} by selecting, from all the \rp-free SNe within the empty bin's range of \tR{} and \Mr{}, those with velocities closest to the empty bin's central velocity, and averaging over that subset.
If there are no \rp-free SNe (of any velocity) in the desired bin in (\tR, \Mr), we do not calculate \drx.

\subsubsection{Minimum observable mixing coordinate}\label{subsub:xmin}

Having established a new method for comparing enriched and unenriched SN models, we return to the \rccsne{} and determine, as a function of \tR, \Mr, and \bej, the minimum mixing coordinate that produces a detectable signal.
We define this minimum, \xmin, as the value of \xmix{} for which ${\geq}50\%$ of models in a given bin satisfy $\drx \geq 1$ mag before some threshold time $t_{\rm f}$, with $t_{\rm f}$ a parameter of the calculation.

\begin{figure*}\includegraphics[width=\textwidth]{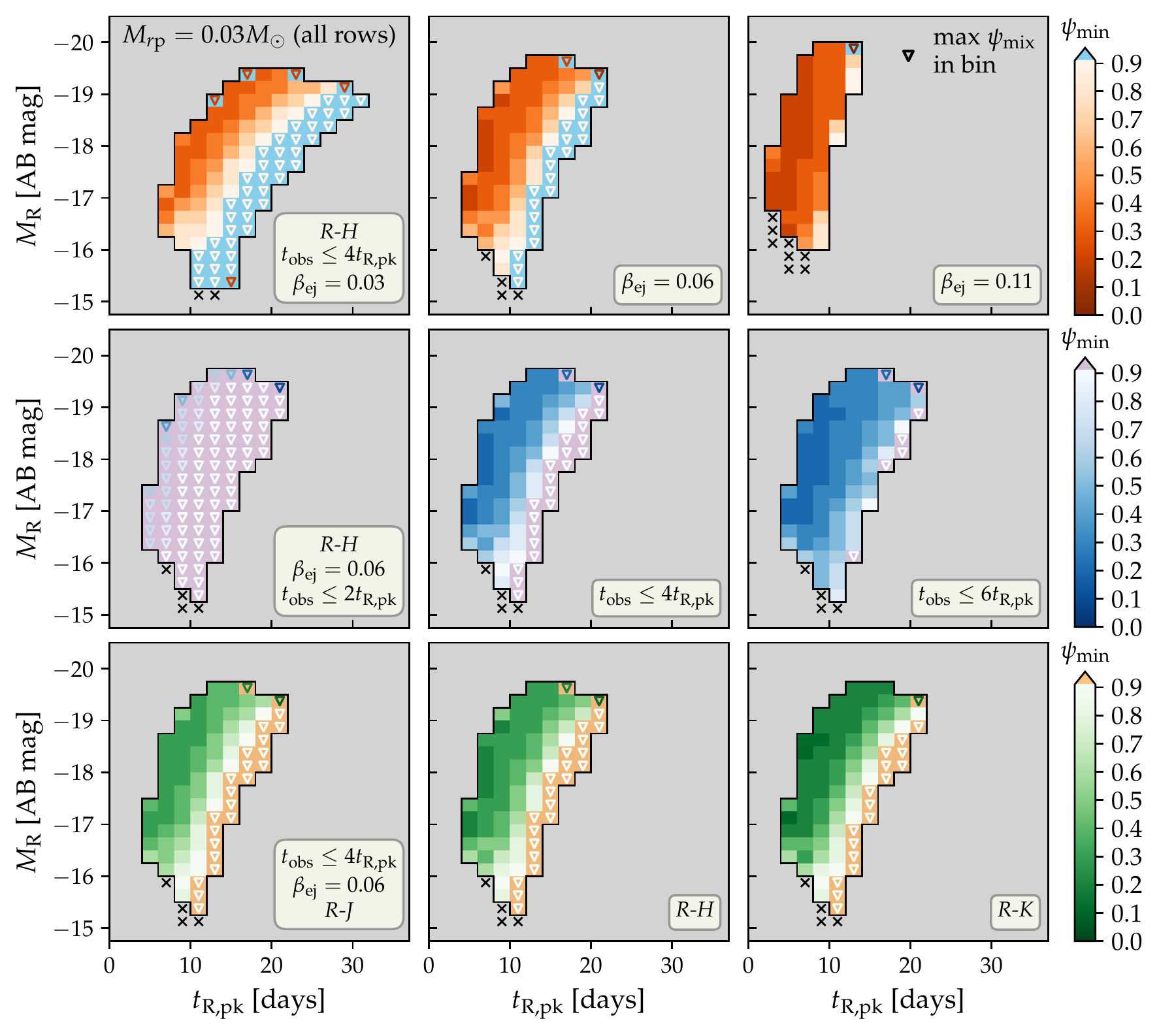}
    \caption{The minimum observable mixing coordinate, \xmin, depends on the intrinsic properties of the \rccsn{} and on the timescale and passbands of the observation.
    As elsewhere, an observable signal is one that produces an $R{-}X$ color difference of at least 1 mag compared to an average of unenriched SNe in the same bin ($\bej$, $t_{\rm R,pk}$, $M_{\rm R}$) at some point during the observation time frame.
    We define \xmin{} as the minimum \xmix{} for which ${\geq}50\%$ of models in a bin are observable.
    In bins where no \xmix{} meet this criterion, an inverted triangle indicates the maximum \xmix{} within the bin.
    Black $\times$'s indicate that a bin contained \rccsne, but no unenriched SNe for comparison.
    All enriched models have $\mrp = 0.03\msun$.
    \emph{Top panels:} the effect of \bej.
    More deeply buried \rp{} material is more easily observed for faster expanding \rccsne.
    \emph{Middle panels:} the effect of observing time frame. 
    If observations are restricted to times close to peak, the differences between enriched and unenriched SNe are difficult to discern, even for highly mixed ejecta.
    Prolonged observing campaigns provide more opportunities for detection.
    \emph{Bottom panels:} the effect of passband. 
    The signal is more easily observed when redder bands are considered in color calculations. 
    }
    \label{fig:chi_min_grid}
\end{figure*}

We find that \xmin{} depends primarily on the ejecta velocity \bej{} of the rCCSNe, the time frame over which observations are carried out, and the color considered (i.e., the choice of $X$).
The effect of each of these is illustrated in Fig.~\ref{fig:chi_min_grid}, which shows \xmin{} for each bin, or, for bins in which no \xmix{} qualified as detectable, the maximum \xmix{} represented in the bin.
For simplicity, we present in Fig.~\ref{fig:chi_min_grid} slices from the parameter space; each panel corresponds to one value of $t_{\rm f}$, one choice of $X$, and one (bin-centered) \bej{}.
We have also restricted the models of Fig.~\ref{fig:chi_min_grid} to those with $\mrp = 0.03\msun$.
However, the same trends apply to other choices of \mrp.

As expected from the discussion of \S\ref{subsec:photolytics} (e.g., Fig.~\ref{fig:xmin_mvplane}), the top row of Fig.~\ref{fig:chi_min_grid} shows that \rp{} signatures are more easily detected for \rccsne{} with higher \bej.
This means that the \rp{} collapsar hypothesis should be easiest to test for the energetic GRB-SNe most likely to form
a neutron-rich accretion disk during collapse.  
On the other hand, alternate modes of \emph{r}-production in SNe, such as lower-energy ``failed-jet'' SNe \citep{Grichener.Soker_ApJ.2019_comomen.env.jet.sn} or neutrino-driven winds from newly born magnetars \citep{Vlasov.ea_mnras.2017_weak.rproc.ms.magnetar.vwinds,Thompson.udDoula_mnras.2018_rproc.proto.ns.winds}, will be more difficult to evaluate. 

While incomplete mixing and slower expansion velocities hinder \rp{} detection, observing strategy can at least partially compensate.
The middle row of Fig.~\ref{fig:chi_min_grid} demonstrates that extending observations to later times increases the likelihood of a detection, even for models with lower \xmix.
For \rccsne{} with $\bej \approx 0.06$, we find that \rp{} signals are not reliably detected in \RH{} at times ${\leq}2\tR$, even if mixing is extensive.
(Nearly all bins in the middle row contain models with $\xmix = 0.9$.)
In contrast, for $t \leq 4\tR \: (6\tR)$, $\xmin \leq 0.3$ for 31\% (46\%) of the bins.
That said, according to our definition, a successful detection implies only that a color difference $\drx \geq 1$ mag is obtained \emph{at some point} during the observation;
it reveals nothing about the duration of that signal.
Thus, cadence is also important.

Finally, as can be seen in Fig.~\ref{fig:chi_min_grid}'s bottom row, the passbands chosen for the color comparison also matter.
We find \rp{} detection for minimally mixed ejecta is easier for \RK{} than \RH{} and \RJ{}.
However, particularly if observations are limited to photometry, multiple NIR bands may be required to rule out spurious emission features unrelated to \rp{} enrichment \citep[e.g., overtones of carbon monoxide;][]{Gerardy.ea_2002.PASJ_sn2000ew.co.nir.spec}.
In that sense, detection prospects may be limited by the performance of the weakest color considered.

\begin{figure}\includegraphics[width=\columnwidth]{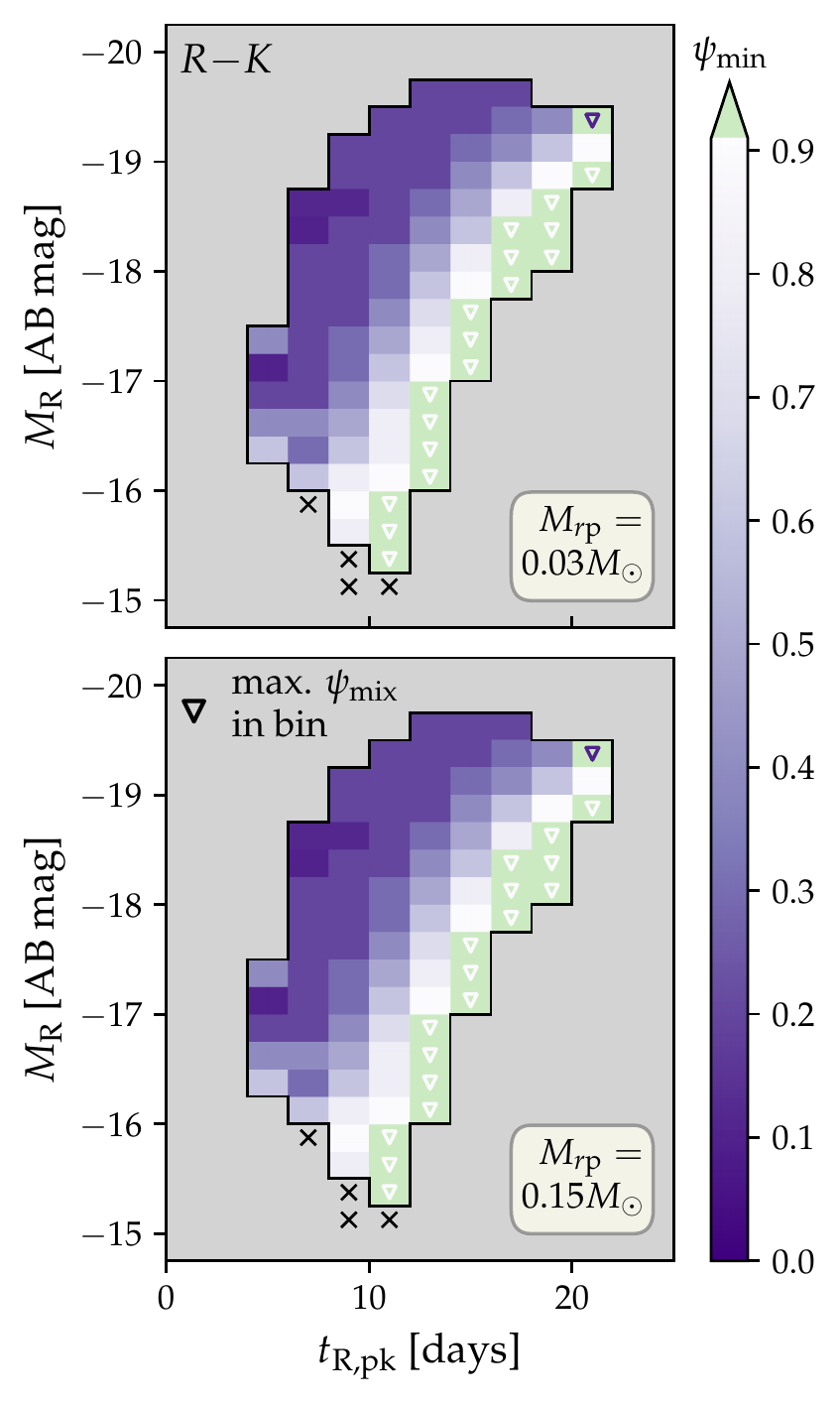}
\caption{The same as Fig.~\ref{fig:chi_min_grid}, but showing the impact of the \rp{} mass \mrp{} on \xmin, for fixed $\beta_{\rm ej} = 0.06$ and observing window $t_{\rm obs} \leq 4 \tR$. 
Detectability is determined with respect to \RK.
The minimum detectable mixing coordinate (\xmin) is largely insensitive to the quantity of \rp{} material, at least within the range of \mrp{} we consider; 
a factor of five increase in \mrp{} produces only a modest reduction in \xmin.}
 \label{fig:xmin_mrp}
\end{figure}

Not included in Fig.~\ref{fig:chi_min_grid} is the effect of \rp{} mass, which we find has a comparatively minor impact on the minimum detectable mixing coordinate.  
In Fig.~\ref{fig:xmin_mrp}, we show \xmin{} for models with bin-center $\bej = 0.06$ (the fiducial velocity of Fig.~\ref{fig:chi_min_grid}), $t_{\rm obs} \leq 4\tR$, and two \rp{} masses, $\mrp = 0.03\msun$ and $0.15\msun$.
We calculate \xmin{} based on the \RK{} color, for which the effects of \rp{} enrichment are strongest.

Despite the five-fold increase in \mrp, values of \xmin{} are fairly stable.
In part because of the velocity dependence described above, the greatest gains in detectability are for fast-evolving transients ($t_{\rm R,pk} \approx 5$--$10$ days), for which \xmin{} falls from ${\simeq}0.2$ to ${\simeq}0.1$.
For events with longer rise times, even a large \mrp{} is invisible except in cases of strong mixing.

The insensitivity of \xmin{} to \mrp{} is due to the high opacity of \rp{} material relative to the unenriched ejecta, and to the dominance of \nickel{} as an energy source in our model.
Because $\kappa_{r \rm p} \gg \kappa_{\rm sn}$,
core ejecta polluted with even trace amounts of \rp{} elements acquires an opacity much higher than that of the surrounding envelope.  Although increasing \mrp{} in the enriched core further enhances the opacity, the opacity difference between enriched and unenriched ejecta is larger than that between minimally and highly enriched ejecta.
In this way, the effects of $\kappa_{r \rm p}$ diminish with increasing \mrp{}.
Moreover, even for high \mrp{}, \nickel{} supplies most of the radioactive energy to the core, and raising \mrp{} does not meaningfully alter the energy budget of the enriched layers.
Were there reverse true, increasing \mrp{} would cause the \rp-rich layers to shine more brightly (at NIR wavelengths) and, presumably, would produce a tighter correlation between \mrp{} and \xmin.

\subsubsection{Colors and timescales}\label{subsub:cols_ts}

In order to make our analysis more concrete, we next explore the color evolution of a subset of our model suite.
As in \S\ref{subsub:xmin}, we categorize the models according to their $R$-band rise times and maxima, \tR{} and \Mr.
We focus on \rccsne{} of velocity $\bej = 0.06$ (typical for the SNe Ic-BL most likely to produce \rp{} elements in disk outflows), and a moderate level of mixing, $\xmix = 0.3$.
We use the same binning procedure as in \S\ref{subsub:xmin}, and determine for each bin a characteristic timescale and \RK{} color associated with \rp{} enrichment signals. 

\begin{figure*}\includegraphics[width=\textwidth]{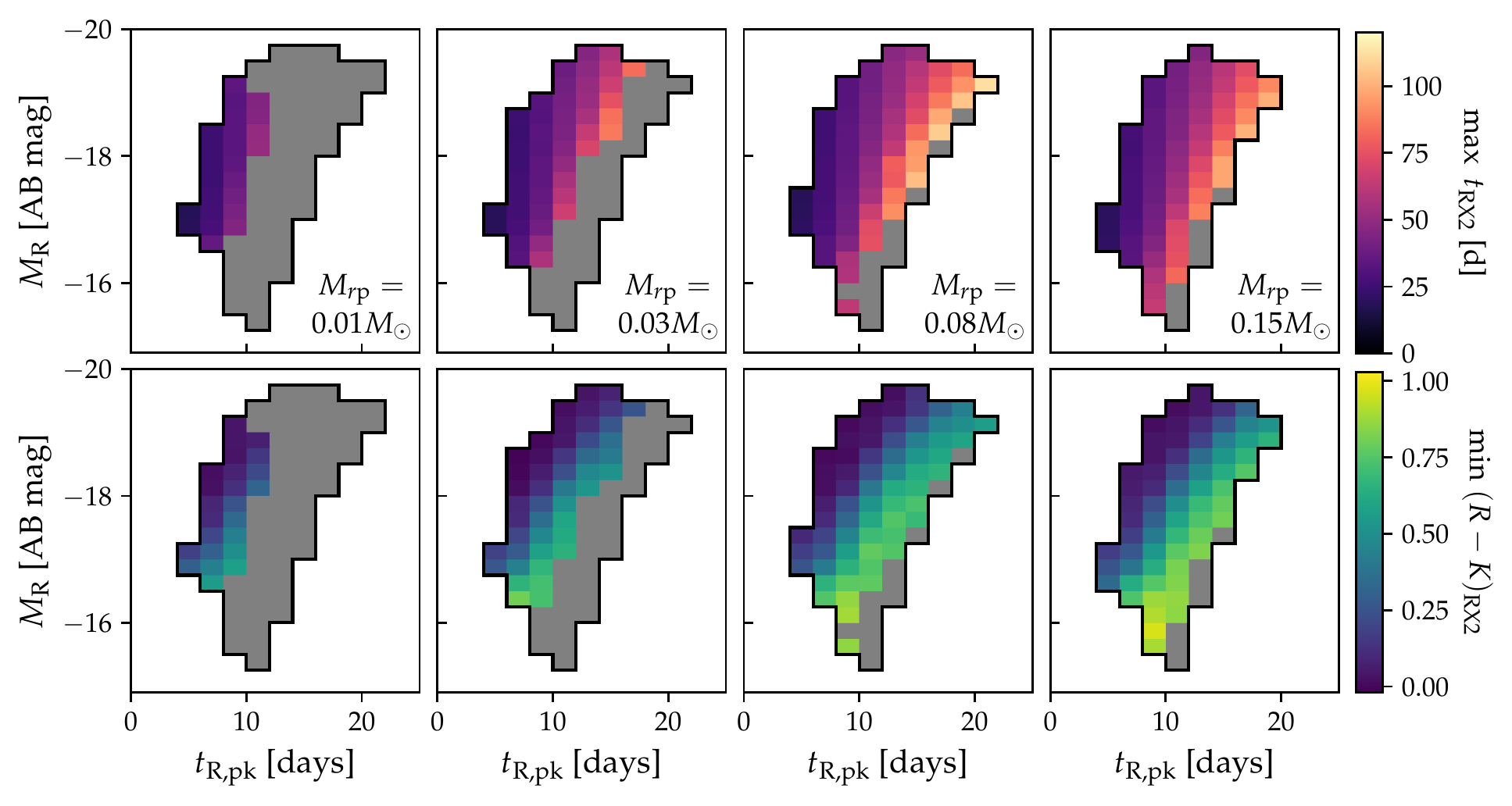}
    \caption{The range of timescales and colors associated with an \rp{} detection for \rccsne{} with $\bej = 0.06$, a moderate $\xmix = 0.3$, and different values of \mrp.
    For each model, we find the earliest time, $t_{\rm RX2}$, at which an $R{-}X$ color difference (relative to unenriched SNe) of ${\geq}1$ mag occurs for at least two $X$ in the NIR bands ($J$, $H$, and $K$). 
    We also record \RK{} at $t=t_{\rm 2RX}$.
    We display the maximum $t_{\rm RX2}$ (\emph{top panels}) and the minimum $(\RK)_{\rm RX2}$ (\emph{bottom panels}) in order to provide a conservative estimate of the strength and timing of the signal.
    Bins in which no models achieved $\drx \geq 1$ for two $X$ are colored gray.
    Evidence of the \rp{} emerges sooner and is associated with bluer \RK{}  for \rccsne{} with lower \tR{} and, to a lesser extent, brighter \Mr{}.
    The value of \mrp{} primarily influences where in $\tR$-$\Mr$ space a signal is detectable; it has only a minor effect on $t_{\rm RX2}$ and $(\RK)_{\rm RX2}$.
    }\label{fig:rk_b4_t}
\end{figure*}

For each model, we find the earliest time, $t_{\rm RX2}$, at which $\drx \geq 1$ mag for at least two $X \in \{J,H,K\}$.
(As before, the color difference is calculated with respect to an average of unenriched SNe in the same bin.)
We adopt the more stringent, two-color standard to protect against false positives due to, e.g., emission features that affect the flux in only one band.)
We also record $\RK$ at $t=t_{\rm RX2}$, which we denote $(\RK)_{\rm RX2}$. 

In Fig.~\ref{fig:rk_b4_t}, we present for different \mrp{} the latest $t_{\rm RX2}$ and the lowest value of $(\RK)_{\rm RX2}$ in each bin.
(We note that the model associated with the latest time is not necessarily also associated with the lowest color.)
In other words, the grids of Fig.~\ref{fig:rk_b4_t} can be read as saying that an \emph{r}CCSN with the specified parameters should produce a signal with \RK{} at least as red as the indicated color by no later than the indicated time.

The \rp{} signal appears first for \rccsne{} that are fast-evolving and/or bright. 
In extreme cases ($\tR \lesssim 10$ days), the signal appears within ${\sim}20$ days, or ${\sim}2\tR$, and is associated with a  moderate $\RK \approx 0.5$ mag. 
For low \mrp, these fast/bright \rccsne{} are the only ones that meet our two-band criterion. 
For higher \mrp{}, detections happen over a broader swath of the parameter space.
Detections in events with higher \tR{} and lower \Mr{} occur much later (75--100 days after explosion), but produce stronger \RK{} colors (${\approx}1$ mag).

While the details of Fig.~\ref{fig:rk_b4_t} depend on the choice of \xmix{} and \bej{}, the results nonetheless suggest that absent a high degree of mixing, minimal \rp{} enrichment ($\mrp \lesssim 0.03 \msun)$ will be difficult to detect except in the fastest-evolving transients.
In contrast, if \emph{r}-production occurs at a higher level (as \citetalias{Siegel.Barnes.Metzger_2019.Nature_rp.collapsar} argued based on the presumed more massive accretion disks formed in collapsars vis-\`{a}-vis NSMs), a signal should be visible for a much wider range of observational parameters.

Though the quantity of \rp{} material determines the breadth of the parameter space over which a detection is feasible, Fig.~\ref{fig:rk_b4_t} suggests that it may be difficult, once a detection \emph{is} made, to precisely constrain the \rp{} mass \mrp.  More specifically, it appears a detection can more easily be used to derive a lower limit for \mrp{} than an upper limit.

\begin{figure*}\includegraphics[width=\textwidth]{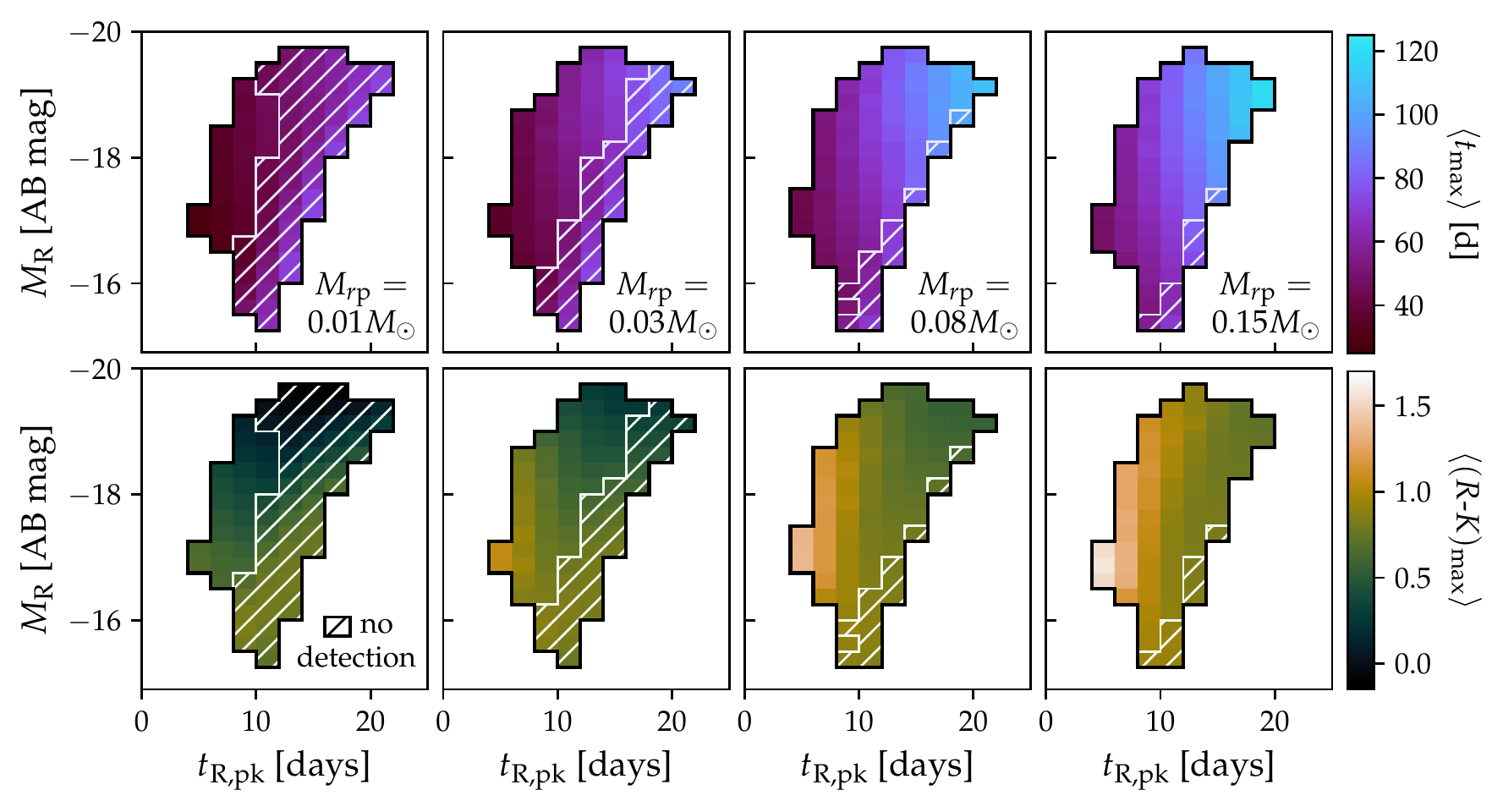}
    \caption{The maximum \RK{} color recorded for \rccsne{} with $\bej = 0.06$, $\xmix = 0.3$, and different \rp{} masses \mrp, after the \rp-free envelope has become optically thin. 
    We show the bin-averaged values of both the color maxima (\emph{top panels}), and the times at which they occur (\emph{bottom panels}).
    In all panels, hatching indicates bins for which no models have $\drx \geq 1$ for at least two bands $X \in \{J,H,K\}$ at any point before $t=200$ days.
    In bins for which detections are possible for multiple \rp{} masses, the maximum \RK{} increases with \mrp{}, as does the time at which that maximum takes place. 
    This suggests that long-term photometric follow-up can provide better constraints on \mrp{} than an initial detection alone.
    }\label{fig:rk_t_max}
\end{figure*}

However, longer-term monitoring can provide additional information, allowing a better estimate of the \rp{} mass produced in a given \rccsn. 
In Fig.~\ref{fig:rk_t_max}, we show, for the same models represented in Fig.~\ref{fig:rk_b4_t} ($\beta_{\rm ej} =0.06$, $\xmix=0.3$), the maximum values of \RK{} the \rccsne{} attain and the times at which the maxima occur, averaged within each bin.
While we do not apply any criterion for \drx{} in calculating these quantities, when determining $(R{-}X)_{\rm max}$, we consider only times after the photosphere has reached the \rp{} layers, ensuring that our color maxima are associated with the enriched ejecta.

For bins in which detections are possible for multiple values of \mrp{} 
(in each panel, light-colored hatching indicates the regions where \rp{} signatures are not detectable according to the two-color standard established for Fig.~\ref{fig:rk_b4_t}),
increasing \mrp{} results in a higher maximum \RK, occurring at a later time. 
Thus, while the \rp{} mass, at least in this mixing regime, has a small effect on the signal in its earliest stages, diligent photometric follow-up may still successfully constrain \mrp.
In fact, the relative insensitivity of the early characteristics of the signal to \mrp{} may be an advantage, as it decouples the determination of ideal observing strategies from assumptions about the level of \emph{r}-production in collapsars.

\subsection{Formulating observing strategies}\label{subsec:obsstrat}

As we have seen, even for SNe with optimal explosion properties, the odds of confidently detecting \rp{} signatures depend on the observing strategy employed.
The observing window and the bands selected are particularly important.
If we require for detection a color difference of at least 1 mag for at least two colors $R{-}X$, we find 48\% of our \rccsn{} models with SN Ic-BL-like velocities ($0.05 \leq \bej \leq 0.1$) are detectable before $t = 50$ days if the chosen colors are $R{-}J$ and $R{-}K$.
This percentage encompasses all mixing coordinates, and smooths over considerable variation with \xmix{}.
If we consider only \rccsne{} with $\xmix \geq 0.6$, 74\% are detectable, while 
for $\xmix < 0.3$, that value drops to 0.5\%.

Observing in redder bands improves the situation, as does extending the observing period. 
If $R-H$ and $R-K$ are instead used for the color comparison, 58\% of the high-velocity models are detectable by $t=50$ days, and 65\% by $t=80$ days.
If we again differentiate by \xmix, we find 14\% of the poorly-mixed models and 79\% of the well-mixed models are detectable by $t=50$, and 17\% and 85\%, respectively, by $t=80$.

Lower velocity \rccsne{} pose a greater challenge.
Of models with $\bej < 0.05$, \rp{} signatures are visible for only 10\% before day 50, even when \drx{} is calculated with respect to $R{-}H$ and $R{-}K$.
The percentage is a dismal 1\% for models with $\xmix < 0.3$, and rises to only 15\% for $\xmix \geq 0.6$.

\begin{figure}\includegraphics[width=\columnwidth]{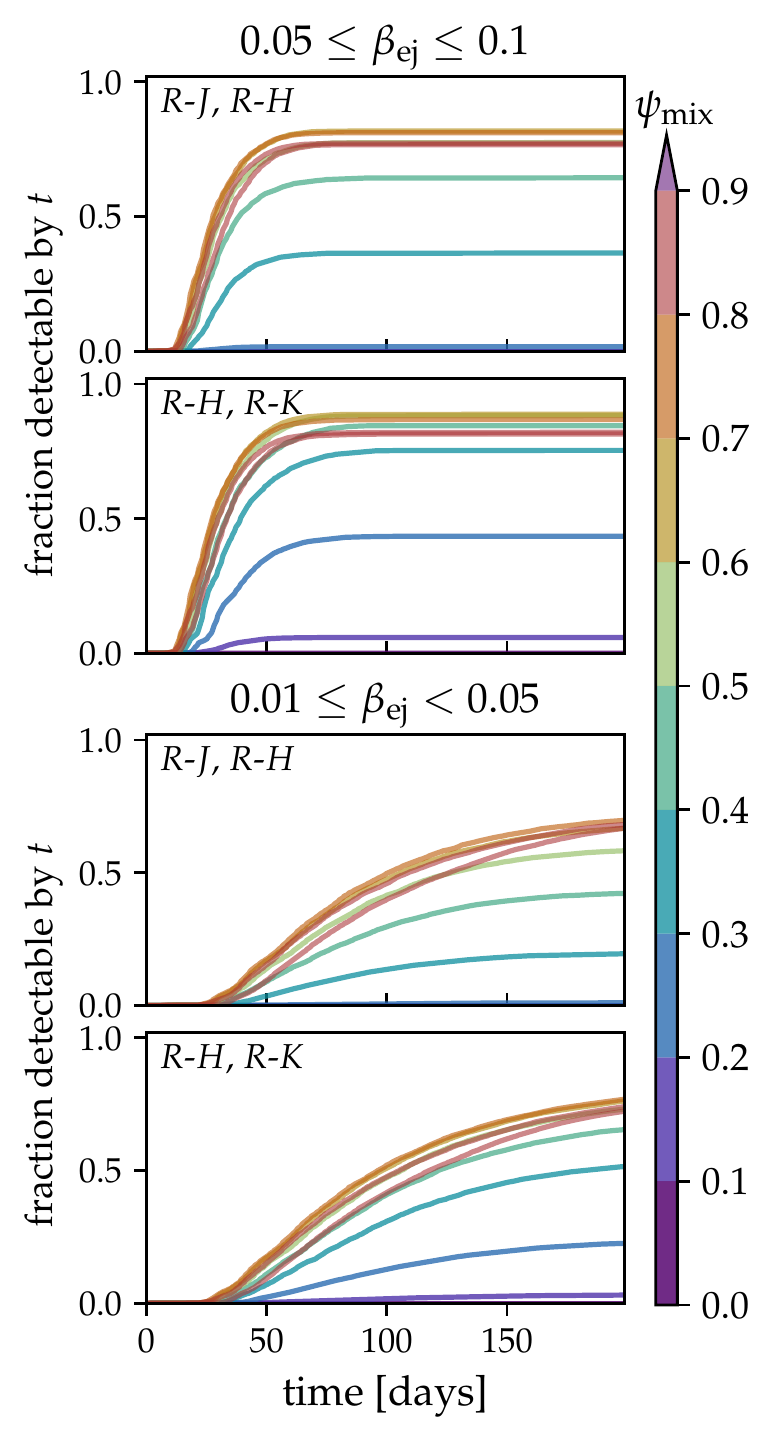}
    \caption{The time frame, relative to explosion, required to constrain \rp{} production in collapsars depends on the properties of the \rccsn{} and on the bands used in the observation.
    Above, we show the cumulative fraction of models that have been observable, according to a two-color criterion, at any point before time $t$.
    We have coarsely binned the models in \bej{} and broken them down by mixing coordinate \xmix.
    For higher-velocity \rccsne{} (\emph{top panels}), while detectability is particularly sensitive to \xmix, most of the detectable models reveal themselves within ${\sim}2$ months of explosion.
    In contrast, the odds of a detection for lower-velocity \rccsne{} (\emph{bottom panels}) rises continually out to $t=200$ days, when our simulations end.
    Regardless of velocity, color comparisons in \RH{} and \RJ{} offer better prospects for detection than \RJ{} and \RH{}, particularly for models with lower \xmix.
    }
    \label{fig:xmix_cdf}
\end{figure}

To clarify these trends, we show in Fig.~\ref{fig:xmix_cdf} the fraction of \rccsn{} models of different \xmix{} that satisfy a two-color detection criterion before a given time.
Consistent with the discussion above, we divide our models into low- ($0.01 \leq \bej < 0.05$) and high-velocity ($0.05 \leq \bej \leq 0.1$) subsets, and consider the effect of the colors used to calculate \drx.

While Fig.~\ref{fig:xmix_cdf} shows that only a limited fraction of enriched models are detectable under this framework, it also identifies the period most likely to yield a successful detection, \emph{if} one is to be forthcoming. 
For example, of models with higher \bej{} (i.e., the top two panels), very few first become detectable later than ${\sim}70$ days post-explosion. 
In contrast, the detectable fraction of lower-velocity models continues to rise steadily well past $t=100$ days. 
(We caution that while these quoted percentages reveal important relationships between observing strategies and the ability to discern \rp-enrichment signatures, their exact values depend on the distribution of explosion and enrichment parameters in our model suite, which does not necessarily reflect the distributions within the cosmological population of SNe Ic/Ic-BL.
Rather than the percentage of \rccsne{} that can be identified by a certain observation, they more accurately measure the fraction of the \emph{parameter space} an observation can probe.)

Observing \rp{} signatures in poorly-mixed explosions will be challenging regardless of other factors; still, Fig.~\ref{fig:xmix_cdf} suggests that the odds of success will be maximized if high-velocity targets are followed up in multiple bands---ideally including multiple NIR bands covering the reddest wavelengths possible---for at least two months after explosion.
By $t \approx 2.5$ months, the odds that a signal will appear for the first time decline steeply. 
If no signal has been detected by this point, observing resources would be better spent on new targets.

\begin{figure}\includegraphics[width=\columnwidth]{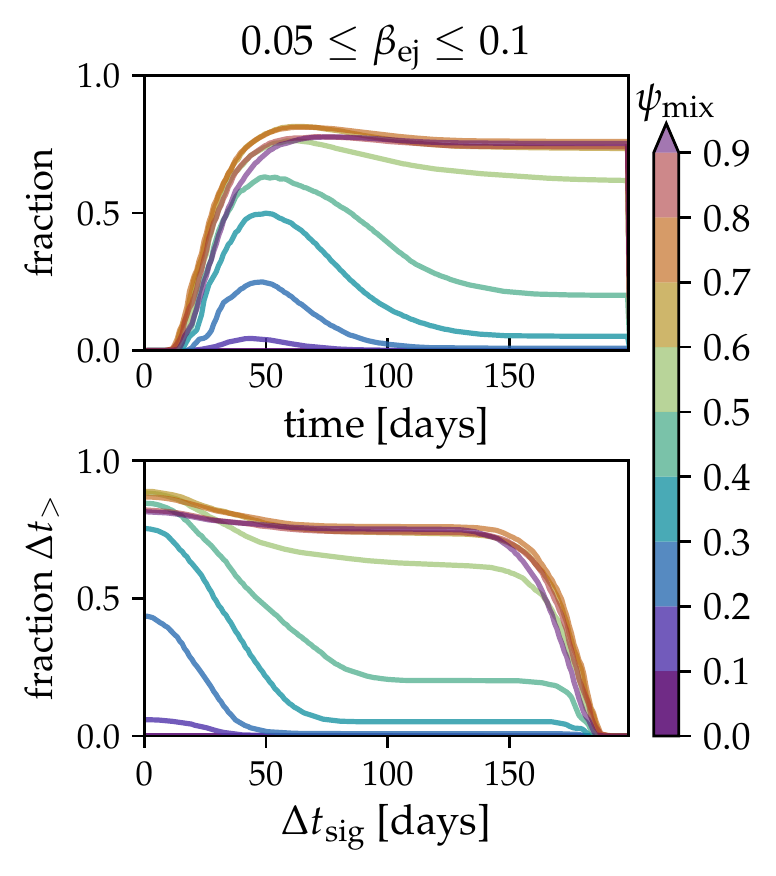}
    \caption{The ideal observing cadence depends on $\Delta t_{\rm sig}$, the lifetime of the signal, which is sensitive to the mixing coordinate \xmix{}.
    We focus here on \rccsne{} with high velocities, and define detectability as $\drx \geq 1$ for $X = H$ and $K$.
    \emph{Top panel:} The fraction of models detectable at a particular time since explosion, for different \xmix. 
    Regardless of \xmix, the fractions reach their peak around 50 days.
    (However, the values of those peaks \emph{do} depend on \xmix.)
    \emph{Bottom panel:} the fraction of models that have a signal duration $\geq\Delta t_{\rm sig}$, as a function of mixing coordinate.
    For well-mixed models, $\Delta t_{\rm sig}$ can exceed 100 days, while signals for models with lower \xmix{} are more ephemeral.
    }
    \label{fig:signal_dt}
\end{figure}

What Fig.~\ref{fig:xmix_cdf} does not elucidate is the cadence of observations needed to catch \rp{} signals, which are often transient.
To provide some sense of the signals' longevity, we calculate the fraction of models that are \emph{instantaneously} detectable as a function of time, as well as the distribution of signal lifetimes, $\Delta t_{\rm sig}$.
Here, as elsewhere, ``detectability'' refers only to the color difference relative to an \rp-free SN baseline; we do not consider telescope sensitivity, or other technicalities that would in practice constrain the acquisition of data.
We show the results in Fig.~\ref{fig:signal_dt} for \rccsne{} with $0.05 \leq \bej \leq 0.1$, for a signal constituted by $\drx \geq 1$ mag with $X = H$ and $K$.
(In other words, we selected the SNe and the $R{-}X$ colors that favor detection.)

A large fraction of well-mixed models are detectable starting at $t \sim 30$ days, and remain detectable thereafter; 
of models with $\xmix \gtrsim 0.5$,  ${\lesssim}60$\% are detectable over a period of ${\geq}100$ days.
For lower mixing coordinates, the epoch at which the largest fraction of models is detectable is similar (${\sim}50$ days after explosion), but that fraction is lower, and $\Delta t_{\rm sig}$ decreases in concert with \xmix{}.
For example, while 47\% of models with $\xmix=0.3$ have signals lasting $\Delta t_{\rm sig} \geq 30$ days, only 13\% of those with $\xmix = 0.2$ do.
Fig.~\ref{fig:signal_dt} suggests that long-lived signals enduring out into the nebular phase can be expected in a majority of cases only if $\xmix \gtrsim 0.5$.

Taken together, Figs.~\ref{fig:xmix_cdf} and \ref{fig:signal_dt} indicate that observations targeting high-velocity SNe in the first few months after explosion and at reasonably high cadence will maximize the chances of identifying \rccsne{}.
The most poorly-mixed \rccsne{} that offer any hope of detection (those with $\xmix = 0.1$) have signal durations sharply concentrated at $\Delta t_{\rm sig} \lesssim 25$ days.
Ideally, observing campaigns would return to such a target multiple times during this window to maximize confidence in a detection, suggesting a delay of no greater than ${\sim}1$ week (and optimally ${\sim}3$--4 days) between consecutive visits.

\section{Conclusions}\label{sec:conclusions}

Our analysis suggests it may be possible to directly detect signs of the \rp{} in photometric data from \rccsne.  Significantly, observations can constrain \rp{} enrichment at a level that would make collapsars competitive with NSMs as \rp{} sources.
(While the exact mass-per-event required for collapsars to overtake mergers depends on the uncertain rates of NSMs, NS-black hole mergers, and GRB-SNe, we estimate \citep[e.g.][]{Siegel.Barnes.Metzger_2019.Nature_rp.collapsar, Brauer.ea_2021.ApJ_proc.EuFe.scatter} that $\langle \mrp \rangle \approx $ 0.01--0.1$M_{\odot}$ would put collapsars in contention.)
In contrast, observations of this kind are not suitable for probing \emph{r}-production in garden-variety core-collapse SNe, which, if it occurs, is predicted to burn only $10^{-4}$--$10^{-5}\msun$ of \rp{} elements \citep{Qian.Woosley_1996.ApJ_neutrino.winds.ccsne.nucleosyn}.

Though previous studies \citepalias{Siegel.Barnes.Metzger_2019.Nature_rp.collapsar} advocated testing the \rp{} collapsar hypothesis with observations during the nebular phase, we find that, in many cases, the \rp{} signal manifests much earlier in the \rccsn's evolution.
Thus, in addition to being an important tool for the discovery of \rccsne{} in its own right, photospheric-phase observation can also identify the events most worthy of follow-up in the nebular phase.
Such triage is particularly important given that nebular-phase observations are resource intensive, and the tool best-suited to take them---the Mid-Infrared Instrument (MIRI) aboard \emph{JWST}---will have limited availability for rapid-response, target-of-opportunity campaigns.
Early identification will thus maximize the science returns from what will likely be a finite number of visits to candidate \rccsn{} nebulae.

Because our models are not fully mixed, their photospheric phase consists of an early optical stage bright and long-lived enough that, in most cases, its (optical) peak magnitude and rise time are consistent with those of observed SNe (e.g. Fig.~\ref{fig:chi_min_grid}).
This means that \rccsne{} are---except in extreme cases---likely to be found in blind searches and confidently identified as SNe Ic/Ic-BL.
After this initial period, which is powered by the \rp-free outer envelope, \rccsn{} emission becomes dominated by radiation from the \rp-enriched core.
Due to the high opacity of \rp{} elements, the signal at this point appears as a NIR excess, and is most clearly distinguishable by its optical-NIR colors.

Multi-band photometry extending out to late times and long wavelengths (ideally including $K$-band) is therefore critical in the hunt for \rccsne.
The most promising targets are highly kinetic GRB-SNe/SNe Ic-BL.
Not only are these the SNe most closely tied, theoretically, to the collapsar model of the \rp, they are also characterized by rapid expansion that causes any \rp-free outer layers to quickly become transparent, revealing the inner, enriched core on a fairly short timescale. 
(Crucially, however, the signal is delayed enough that contamination by an afterglow---in the case of SNe Ic-BL discovered through a long GRB trigger---is not a concern.)

As we argued in \S\ref{subsec:obsstrat}, frequent monitoring of these high-velocity SNe in optical and at least $H$ and $K$ bands for the first ${\sim}75$ days after explosion is an efficient strategy for \rccsn{} searches. 
While such an observation will not catch all \rccsne{}, its misses will primarily be due to the intrinsic difficulty of identifying \emph{r}-production in cases of minimal \xmix, and not to an insufficiently long observing window, or reliance on non-optimal colors to differentiate enriched from unenriched SNe.

The potential of such an observing strategy was recently demonstrated by an effort---the first of its kind---to systematically follow up SNe Ic-BL in the NIR in pursuit of photometric \rp-enrichment signatures.
The observations, to be presented in a companion paper \citep{Anand.ea_in.press_Sne.IcBL.nir.survey}, show evidence for diverse \rp-enrichment outcomes in these energetic SNe.
Upcoming surveys by facilities with extensive infrared capabilities, e.g. WINTER \citep{Lourie.ea_2020.SPIE_WINTER.goals.design} and the \emph{Roman Space Telescope} \citep{Mutchler.ea_2021.AAS_Roman.sci.ops}, will provide new opportunities to search for \rccsne.

Our ability to identify \rccsne{} at even low levels of mixing will improve as our understanding of late-time SN emission solidifies.
Regardless of the parameters of an observation and the properties of its target, a detection requires a clear picture of NIR emission from unenriched SNe in the nebular phase, against which new observations can be compared.
We constructed our models of ordinary SNe using the limited data available.
However, further observations of SNe Ib/c over the course of their evolution,  with a focus on lower-energy events less likely to undergo an \rp, are necessary to more firmly establish baseline expectations of colors in the \rp-free case.

Nebular-phase \rp{} emission is another area where additional data would increase confidence in the models and, perhaps, inform observing strategies.
If, for example, \rp{} nebulae shine predominantly at longer wavelengths than predicted by \citet[][the study on which our models are based]{Hotokezaka.ea_2021.MNRAS_kn.nebular.model}, emission even in $J$, $H$, and $K$ bands could be negligible (Fig.~\ref{fig:rp-neb-mod}), and facilities with mid-infrared capabilities may be required for smoking-gun detections in the nebular phase.

Despite these uncertainties, taken as a whole, our findings suggest that signs of enrichment in \rccsne{} may be visible across a significant fraction of the parameter space, which makes SN observation an important tool for assessing collapsars as sites of \rp{} nucleosynthesis.
Observations of additional SNe Ic/Ic-BL and---one hopes---additional kilonovae will further refine this tool, providing a new method to uncover the means by which the Universe becomes seeded with the heaviest elements.

\section{Acknowledgments}

The authors thank S. Anand and M. Kasliwal for helpful discussions.
J.B. gratefully acknowledges support from the Gordon and Betty Moore Foundation through Grant GBMF5076, and from the NASA Einstein Fellowship Program through Grant PF7-180162.
B.D.M. is supported in part by the National Science Foundation (Grants AST-2009255, AST-2002577).

\bibliographystyle{apj} 
\bibliography{refs}

\newcommand{\noop}[1]{}
\begin{thebibliography}{}
\expandafter\ifx\csname natexlab\endcsname\relax\def\natexlab#1{#1}\fi

\bibitem[{{Abbott} {et~al.}(2017{\natexlab{a}}){Abbott}, {Abbott}, {Abbott}, \&
  {Acernese}}]{Abbott.ea_2017ApJL_gw.170817.multimess}
{Abbott}, B.~P., {Abbott}, R., {Abbott}, T.~D., \& {Acernese}, F.
  2017{\natexlab{a}}, \apjl, 848, L12

\bibitem[{{Abbott} {et~al.}(2017{\natexlab{b}}){Abbott}, {Abbott}, {Abbott},
  {Acernese}, {Ackley}, {Adams}, {Adams}, {Addesso}, {Adhikari}, {Adya}, \&
  et~al.}]{Abbott.ea_2017ApJ_gw170817.grb}
{Abbott}, B.~P., {Abbott}, R., {Abbott}, T.~D., {et~al.} 2017{\natexlab{b}},
  \apjl, 848, L13

\bibitem[{{Abbott} {et~al.}(2017{\natexlab{c}}){Abbott}, {Abbott}, {Abbott},
  {Acernese}, {Ackley}, {Adams}, {Adams}, {Addesso}, {Adhikari}, {Adya}, \&
  et~al.}]{Abbott.ea_2017.PRL_gw170817.disc}
---. 2017{\natexlab{c}}, Physical Review Letters, 119, 161101

\bibitem[{{Aloy} {et~al.}(2000){Aloy}, {M{\"u}ller}, {Ib{\'a}{\~n}ez},
  {Mart{\'\i}}, \& {MacFadyen}}]{Aloy.ea_2000.ApJL_grb.jets.collapsars}
{Aloy}, M.~A., {M{\"u}ller}, E., {Ib{\'a}{\~n}ez}, J.~M., {Mart{\'\i}}, J.~M.,
  \& {MacFadyen}, A. 2000, \apjl, 531, L119

\bibitem[{{Anand} {et~al.}(\noop{3001}in prep.){Anand}, {Kasliwal}, \& {the ZTF
  Team}}]{Anand.ea_in.press_Sne.IcBL.nir.survey}
{Anand}, S., {Kasliwal}, M., \& {the ZTF Team}. \noop{3001}in prep.

\bibitem[{{Arcavi} {et~al.}(2017){Arcavi}, {Hosseinzadeh}, {Howell}, {McCully},
  {Poznanski}, {Kasen}, {Barnes}, {Zaltzman}, {Vasylyev}, {Maoz}, \&
  {Valenti}}]{Aracavi.ea_2017Natur_gw170817.lco.emcp.disc}
{Arcavi}, I., {Hosseinzadeh}, G., {Howell}, D.~A., {et~al.} 2017, \nat, 551, 64

\bibitem[{{Arcones} \&
  {Montes}(2011)}]{Arcones.Montes_2011.ApJ_light.elem.prod.neutrino.winds}
{Arcones}, A., \& {Montes}, F. 2011, \apj, 731, 5

\bibitem[{{Arnett}(1980)}]{Arnett_1980}
{Arnett}, W.~D. 1980, \apj, 237, 541

\bibitem[{{Arnett}(1982)}]{Arnett_1982_Sne}
---. 1982, \apj, 253, 785

\bibitem[{{Barbarino} {et~al.}(2021){Barbarino}, {Sollerman}, {Taddia},
  {Fremling}, {Karamehmetoglu}, {Arcavi}, {Gal-Yam}, {Laher}, {Schulze},
  {Wozniak}, \& {Yan}}]{Barbarino.ea_AandA2021_reg.Ic.iPTF.survey}
{Barbarino}, C., {Sollerman}, J., {Taddia}, F., {et~al.} 2021, \aap, 651, A81

\bibitem[{{Barnes}(2020)}]{Barnes_2020.FrontPhys_knova.review}
{Barnes}, J. 2020, Frontiers in Physics, 8, 355

\bibitem[{{Barnes} {et~al.}(2018){Barnes}, {Duffell}, {Liu}, {Modjaz},
  {Bianco}, {Kasen}, \&
  {MacFadyen}}]{Barnes.Duffell.ea_2018.ApJ_grb.icbl.engine}
{Barnes}, J., {Duffell}, P.~C., {Liu}, Y., {et~al.} 2018, \apj, 860, 38

\bibitem[{{Barnes} {et~al.}(2016){Barnes}, {Kasen}, {Wu}, \&
  {Mart{\'{\i}}nez-Pinedo}}]{Barnes_etal_2016}
{Barnes}, J., {Kasen}, D., {Wu}, M.-R., \& {Mart{\'{\i}}nez-Pinedo}, G. 2016,
  \apj, 829, 110

\bibitem[{{Barnes} {et~al.}(2021){Barnes}, {Zhu}, {Lund}, {Sprouse}, {Vassh},
  {McLaughlin}, {Mumpower}, \&
  {Surman}}]{Barnes.ea_2021.ApJ_knova.nuc.landscape}
{Barnes}, J., {Zhu}, Y.~L., {Lund}, K.~A., {et~al.} 2021, \apj, 918, 44

\bibitem[{{Bartos} \&
  {M{\'a}rka}(2019)}]{Bartos.Marka_2019.ApJL_rproc.solar.rccsn.constraints}
{Bartos}, I., \& {M{\'a}rka}, S. 2019, \apjl, 881, L4

\bibitem[{{Bauswein} {et~al.}(2013){Bauswein}, {Goriely}, \&
  {Janka}}]{Bauswein_2013}
{Bauswein}, A., {Goriely}, S., \& {Janka}, H.-T. 2013, \apj, 773, 78

\bibitem[{{Beloborodov}(2003)}]{Beloborodov_2003.ApJ_grb.nuclear.comp}
{Beloborodov}, A.~M. 2003, \apj, 588, 931

\bibitem[{{Beniamini} {et~al.}(2018){Beniamini}, {Dvorkin}, \&
  {Silk}}]{Beniamini.ea_2018.mnras_rproc.retention.dwarf.gal}
{Beniamini}, P., {Dvorkin}, I., \& {Silk}, J. 2018, \mnras, 478, 1994

\bibitem[{{Berger}(2014)}]{Berger_2014.aara_sgrbs}
{Berger}, E. 2014, \araa, 52, 43

\bibitem[{{Bianco} {et~al.}(2014){Bianco}, {Modjaz}, {Hicken}, {Friedman},
  {Kirshner}, {Bloom}, {Challis}, {Marion}, {Wood-Vasey}, \&
  {Rest}}]{Bianco.Modjaz.ea_2014.ApJS_sesn.ccsne.lcs}
{Bianco}, F.~B., {Modjaz}, M., {Hicken}, M., {et~al.} 2014, \apjs, 213, 19

\bibitem[{{Bloom} {et~al.}(2002){Bloom}, {Kulkarni}, {Price}, {Reichart},
  {Galama}, {Schmidt}, {Frail}, {Berger}, {McCarthy}, {Chevalier}, {Wheeler},
  {Halpern}, {Fox}, {Djorgovski}, {Harrison}, {Sari}, {Axelrod}, {Kimble},
  {Holtzman}, {Hurley}, {Frontera}, {Piro}, \&
  {Costa}}]{Bloom.ea_2002.ApJL_sn.grb.assoc.011121}
{Bloom}, J.~S., {Kulkarni}, S.~R., {Price}, P.~A., {et~al.} 2002, \apjl, 572,
  L45

\bibitem[{{Bovard} {et~al.}(2017){Bovard}, {Martin}, {Guercilena}, {Arcones},
  {Rezzolla}, \& {Korobkin}}]{Bovard.ea_2017.PhRvD_rproc.nsm.mass.ej}
{Bovard}, L., {Martin}, D., {Guercilena}, F., {et~al.} 2017, \prd, 96, 124005

\bibitem[{{Brauer} {et~al.}(2021){Brauer}, {Ji}, {Drout}, \&
  {Frebel}}]{Brauer.ea_2021.ApJ_proc.EuFe.scatter}
{Brauer}, K., {Ji}, A.~P., {Drout}, M.~R., \& {Frebel}, A. 2021, \apj, 915, 81

\bibitem[{{Bromberg} \&
  {Tchekhovskoy}(2016)}]{Bromberg.Tchekhovskoy_2016.mnras_rel.mhd.ccsn.grb}
{Bromberg}, O., \& {Tchekhovskoy}, A. 2016, \mnras, 456, 1739

\bibitem[{{Bruenn} {et~al.}(2016){Bruenn}, {Lentz}, {Hix}, {Mezzacappa},
  {Harris}, {Messer}, {Endeve}, {Blondin}, {Chertkow}, {Lingerfelt},
  {Marronetti}, \& {Yakunin}}]{Bruenn.ea_2016.ApJ_ccsn.sims}
{Bruenn}, S.~W., {Lentz}, E.~J., {Hix}, W.~R., {et~al.} 2016, \apj, 818, 123

\bibitem[{{Burbidge} {et~al.}(1957){Burbidge}, {Burbidge}, {Fowler}, \&
  {Hoyle}}]{Burbidge2.Howler.Foyle_1957.RMP_rprocess}
{Burbidge}, E.~M., {Burbidge}, G.~R., {Fowler}, W.~A., \& {Hoyle}, F. 1957,
  Reviews of Modern Physics, 29, 547

\bibitem[{{Cameron}(1957)}]{Cameron_1957.AJ_rprocess}
{Cameron}, A.~G.~W. 1957, \aj, 62, 9

\bibitem[{{Chatzopoulos} {et~al.}(2012){Chatzopoulos}, {Wheeler}, \&
  {Vinko}}]{Chatzopoulos.ea_2012.ApJ_sn.lc.analytic}
{Chatzopoulos}, E., {Wheeler}, J.~C., \& {Vinko}, J. 2012, \apj, 746, 121

\bibitem[{{Chornock} {et~al.}(2017){Chornock}, {Berger}, {Kasen},
  {Cowperthwaite}, {Nicholl}, {Villar}, {Alexand er}, {Blanchard}, {Eftekhari},
  {Fong}, {Margutti}, {Williams}, {Annis}, {Brout}, {Brown}, {Chen}, {Drout},
  {Farr}, {Foley}, {Frieman}, {Fryer}, {Herner}, {Holz}, {Kessler}, {Matheson},
  {Metzger}, {Quataert}, {Rest}, {Sako}, {Scolnic}, {Smith}, \&
  {Soares-Santos}}]{Chornock.ea_2017ApJ_gw.170817.em.red.spec}
{Chornock}, R., {Berger}, E., {Kasen}, D., {et~al.} 2017, \apjl, 848, L19

\bibitem[{{Colgate} {et~al.}(1980){Colgate}, {Petschek}, \&
  {Kriese}}]{Colgate.ea_1980.ApJ_sn.grays.deposition.lum}
{Colgate}, S.~A., {Petschek}, A.~G., \& {Kriese}, J.~T. 1980, \apjl, 237, L81

\bibitem[{{C{\^o}t{\'e}} {et~al.}(2019){C{\^o}t{\'e}}, {Eichler}, {Arcones},
  {Hansen}, {Simonetti}, {Frebel}, {Fryer}, {Pignatari}, {Reichert},
  {Belczynski}, \& {Matteucci}}]{Cote.ea_2019.ApJ_other.rproc.sources}
{C{\^o}t{\'e}}, B., {Eichler}, M., {Arcones}, A., {et~al.} 2019, \apj, 875, 106

\bibitem[{{Coulter} {et~al.}(2017){Coulter}, {Foley}, {Kilpatrick}, {Drout},
  {Piro}, {Shappee}, {Siebert}, {Simon}, {Ulloa}, {Kasen}, {Madore},
  {Murguia-Berthier}, {Pan}, {Prochaska}, {Ramirez-Ruiz}, {Rest}, \&
  {Rojas-Bravo}}]{Coulter.ea_2017Sci_gw.170817.emcp.disc}
{Coulter}, D.~A., {Foley}, R.~J., {Kilpatrick}, C.~D., {et~al.} 2017, Science,
  358, 1556

\bibitem[{{Cowan} {et~al.}(2021){Cowan}, {Sneden}, {Lawler}, {Aprahamian},
  {Wiescher}, {Langanke}, {Mart{\'\i}nez-Pinedo}, \&
  {Thielemann}}]{Cowan.ea_2021.RMP_rprocess.origins}
{Cowan}, J.~J., {Sneden}, C., {Lawler}, J.~E., {et~al.} 2021, Reviews of Modern
  Physics, 93, 015002

\bibitem[{{Cowperthwaite} {et~al.}(2017){Cowperthwaite}, {Berger}, {Villar},
  {et~al.}}]{Cowperthwaite+17}
{Cowperthwaite}, P.~S., {Berger}, E., {Villar}, V.~A., {et~al.} 2017, \apjl,
  848, L17

\bibitem[{{de los Reyes} {et~al.}(2022){de los Reyes}, {Kirby}, {Ji}, \&
  {Nu{\~n}ez}}]{delosReyes.ea_2022.apj_sfh.nucleosyn.sculptor}
{de los Reyes}, M. A.~C., {Kirby}, E.~N., {Ji}, A.~P., \& {Nu{\~n}ez}, E.~H.
  2022, \apj, 925, 66

\bibitem[{{Deng} {et~al.}(2003){Deng}, {Mazzali}, {Maeda}, \&
  {Nomoto}}]{Deng.ea_2003.NuPhA_2002ap.mods}
{Deng}, J., {Mazzali}, P.~A., {Maeda}, K., \& {Nomoto}, K. 2003, \nphysa, 718,
  569

\bibitem[{{Desai} {et~al.}(2022){Desai}, {Siegel}, \&
  {Metzger}}]{Desai.ea_2022.arXiv_neutrino.winds.proto.ns}
{Desai}, D.~K., {Siegel}, D.~M., \& {Metzger}, B.~D. 2022, arXiv e-prints,
  arXiv:2203.16560

\bibitem[{{Drout} {et~al.}(2011){Drout}, {Soderberg}, {Gal-Yam}, {Cenko},
  {Fox}, {Leonard}, {Sand}, {Moon}, {Arcavi}, \&
  {Green}}]{Drout.ea_2011.ApJ_sn.type.ibc.stats}
{Drout}, M.~R., {Soderberg}, A.~M., {Gal-Yam}, A., {et~al.} 2011, \apj, 741, 97

\bibitem[{{Drout} {et~al.}(2017){Drout}, {Piro}, {Shappee}, {Kilpatrick},
  {Simon}, {Contreras}, {Coulter}, {Foley}, {Siebert}, {Morrell}, {Boutsia},
  {Di Mille}, {Holoien}, {Kasen}, {Kollmeier}, {Madore}, {Monson},
  {Murguia-Berthier}, {Pan}, {Prochaska}, {Ramirez-Ruiz}, {Rest}, {Adams},
  {Alatalo}, {Ba{\~n}ados}, {Baughman}, {Beers}, {Bernstein}, {Bitsakis},
  {Campillay}, {Hansen}, {Higgs}, {Ji}, {Maravelias}, {Marshall}, {Moni Bidin},
  {Prieto}, {Rasmussen}, {Rojas-Bravo}, {Strom}, {Ulloa},
  {Vargas-Gonz{\'a}lez}, {Wan}, \&
  {Whitten}}]{Drout.ea_2017Sci_gw.170817.emcp.disc}
{Drout}, M.~R., {Piro}, A.~L., {Shappee}, B.~J., {et~al.} 2017, Science, 358,
  1570

\bibitem[{{Duggan} {et~al.}(2018){Duggan}, {Kirby}, {Andrievsky}, \&
  {Korotin}}]{Duggan.ea_2018.ApJ_nsm.main.rproc.src.dwarf}
{Duggan}, G.~E., {Kirby}, E.~N., {Andrievsky}, S.~M., \& {Korotin}, S.~A. 2018,
  \apj, 869, 50

\bibitem[{{Eichler} {et~al.}(1989){Eichler}, {Livio}, {Piran}, \&
  {Schramm}}]{Eichler_1989.nat_sgrb.nsm.rproc}
{Eichler}, D., {Livio}, M., {Piran}, T., \& {Schramm}, D.~N. 1989, \nat, 340,
  126

\bibitem[{{Evans} {et~al.}(2017){Evans}, {Cenko}, {Kennea}, {Emery}, {Kuin},
  {Korobkin}, {Wollaeger}, {Fryer}, {Madsen}, {Harrison}, {Xu}, {Nakar},
  {Hotokezaka}, {Lien}, {Campana}, {Oates}, {Troja}, {Breeveld}, {Marshall},
  {Barthelmy}, {Beardmore}, {Burrows}, {Cusumano}, {D'A{\`\i}}, {D'Avanzo},
  {D'Elia}, {de Pasquale}, {Even}, {Fontes}, {Forster}, {Garcia}, {Giommi},
  {Grefenstette}, {Gronwall}, {Hartmann}, {Heida}, {Hungerford}, {Kasliwal},
  {Krimm}, {Levan}, {Malesani}, {Melandri}, {Miyasaka}, {Nousek}, {O'Brien},
  {Osborne}, {Pagani}, {Page}, {Palmer}, {Perri}, {Pike}, {Racusin}, {Rosswog},
  {Siegel}, {Sakamoto}, {Sbarufatti}, {Tagliaferri}, {Tanvir}, \&
  {Tohuvavohu}}]{Evans.ea_2017Sci_gw.170817.em.blue.spec}
{Evans}, P.~A., {Cenko}, S.~B., {Kennea}, J.~A., {et~al.} 2017, Science, 358,
  1565

\bibitem[{{Fern{\'a}ndez} \&
  {Metzger}(2013)}]{Fernandez.Metzger_2013.mnras_disk.outlfows.nsm}
{Fern{\'a}ndez}, R., \& {Metzger}, B.~D. 2013, \mnras, 435, 502

\bibitem[{{Fern{\'a}ndez} {et~al.}(2019){Fern{\'a}ndez}, {Tchekhovskoy},
  {Quataert}, {Foucart}, \&
  {Kasen}}]{Fernandez.ea_2019.mnras_grmhd.sims.nsm.disks}
{Fern{\'a}ndez}, R., {Tchekhovskoy}, A., {Quataert}, E., {Foucart}, F., \&
  {Kasen}, D. 2019, \mnras, 482, 3373

\bibitem[{{Filippenko} \&
  {Chornock}(2002)}]{Filippenko.Chornock_2002.IAUC_sn2002ap.disc.icbl.id}
{Filippenko}, A.~V., \& {Chornock}, R. 2002, \iaucirc, 7825, 1

\bibitem[{{Fontes} {et~al.}(2020){Fontes}, {Fryer}, {Hungerford}, {Wollaeger},
  \& {Korobkin}}]{Fontes.ea_2020.MNRAS_knova.opacs.lanl}
{Fontes}, C.~J., {Fryer}, C.~L., {Hungerford}, A.~L., {Wollaeger}, R.~T., \&
  {Korobkin}, O. 2020, \mnras, 493, 4143

\bibitem[{{Fraser} \&
  {Sch{\"o}nrich}(2022)}]{Fraser.Schonrich_2022.mnras_no.rccsne.MW.metallicity}
{Fraser}, J., \& {Sch{\"o}nrich}, R. 2022, \mnras, 509, 6008

\bibitem[{{Freiburghaus} {et~al.}(1999){Freiburghaus}, {Rosswog}, \&
  {Thielemann}}]{Freiburghaus.ea_1999.ApJL_rproc.nsm}
{Freiburghaus}, C., {Rosswog}, S., \& {Thielemann}, F.-K. 1999, \apjl, 525,
  L121

\bibitem[{{Fr{\"o}hlich} {et~al.}(2006){Fr{\"o}hlich}, {Hauser},
  {Liebend{\"o}rfer}, {Mart{\'{\i}}nez-Pinedo}, {Thielemann}, {Bravo},
  {Zinner}, {Hix}, {Langanke}, {Mezzacappa}, \&
  {Nomoto}}]{Frohlich.ea_2006.ApJ_ccsne.inner.ejecta.comp}
{Fr{\"o}hlich}, C., {Hauser}, P., {Liebend{\"o}rfer}, M., {et~al.} 2006, \apj,
  637, 415

\bibitem[{{Fujibayashi} {et~al.}(2018){Fujibayashi}, {Kiuchi}, {Nishimura},
  {Sekiguchi}, \&
  {Shibata}}]{Fujibayashi.ea_2018.ApJ_mass.ejection.nsm.remnant.disks}
{Fujibayashi}, S., {Kiuchi}, K., {Nishimura}, N., {Sekiguchi}, Y., \&
  {Shibata}, M. 2018, \apj, 860, 64

\bibitem[{{Gal-Yam} {et~al.}(2002){Gal-Yam}, {Ofek}, \&
  {Shemmer}}]{GalYam.ea._2002_02ap.data}
{Gal-Yam}, A., {Ofek}, E.~O., \& {Shemmer}, O. 2002, \mnras, 332, L73

\bibitem[{{Galama} {et~al.}(1998){Galama}, {Vreeswijk}, {van Paradijs},
  {Kouveliotou}, {Augusteijn}, {B{\"o}hnhardt}, {Brewer}, {Doublier},
  {Gonzalez}, {Leibundgut}, {Lidman}, {Hainaut}, {Patat}, {Heise}, {in't Zand},
  {Hurley}, {Groot}, {Strom}, {Mazzali}, {Iwamoto}, {Nomoto}, {Umeda},
  {Nakamura}, {Young}, {Suzuki}, {Shigeyama}, {Koshut}, {Kippen}, {Robinson},
  {de Wildt}, {Wijers}, {Tanvir}, {Greiner}, {Pian}, {Palazzi}, {Frontera},
  {Masetti}, {Nicastro}, {Feroci}, {Costa}, {Piro}, {Peterson}, {Tinney},
  {Boyle}, {Cannon}, {Stathakis}, {Sadler}, {Begam}, \&
  {Ianna}}]{Galama98_bwDisc}
{Galama}, T.~J., {Vreeswijk}, P.~M., {van Paradijs}, J., {et~al.} 1998, \nat,
  395, 670

\bibitem[{{Gerardy} {et~al.}(2002){Gerardy}, {Fesen}, {Nomoto}, {Maeda},
  {Hoflich}, \& {Wheeler}}]{Gerardy.ea_2002.PASJ_sn2000ew.co.nir.spec}
{Gerardy}, C.~L., {Fesen}, R.~A., {Nomoto}, K., {et~al.} 2002, \pasj, 54, 905

\bibitem[{{Goldstein} {et~al.}(2017){Goldstein}, {Veres}, {Burns}, {Briggs},
  {Hamburg}, {Kocevski}, {Wilson-Hodge}, {Preece}, {Poolakkil}, {Roberts},
  {Hui}, {Connaughton}, {Racusin}, {von Kienlin}, {Dal Canton}, {Christensen},
  {Littenberg}, {Siellez}, {Blackburn}, {Broida}, {Bissaldi}, {Cleveland},
  {Gibby}, {Giles}, {Kippen}, {McBreen}, {McEnery}, {Meegan}, {Paciesas}, \&
  {Stanbro}}]{Goldstein.ea_2017.ApJ_grb.gw170817}
{Goldstein}, A., {Veres}, P., {Burns}, E., {et~al.} 2017, \apjl, 848, L14

\bibitem[{{G{\'o}mez} \&
  {L{\'o}pez}(2002)}]{Gomez.Lopez_2002.AJ_neb.spec.sne.ic}
{G{\'o}mez}, G., \& {L{\'o}pez}, R. 2002, \aj, 123, 328

\bibitem[{{Gottlieb} {et~al.}(2022){Gottlieb}, {Lalakos}, {Bromberg}, {Liska},
  \& {Tchekhovskoy}}]{Gottlieb.ea_2022.mnras_bh_3d.grmhd.ccsn.jets}
{Gottlieb}, O., {Lalakos}, A., {Bromberg}, O., {Liska}, M., \& {Tchekhovskoy},
  A. 2022, \mnras, 510, 4962

\bibitem[{{Grichener} \&
  {Soker}(2019)}]{Grichener.Soker_ApJ.2019_comomen.env.jet.sn}
{Grichener}, A., \& {Soker}, N. 2019, \apj, 878, 24

\bibitem[{{Grossman} {et~al.}(2014){Grossman}, {Korobkin}, {Rosswog}, \&
  {Piran}}]{Grossman_2014_kNe}
{Grossman}, D., {Korobkin}, O., {Rosswog}, S., \& {Piran}, T. 2014, \mnras,
  439, 757

\bibitem[{{Guillochon} {et~al.}(2017){Guillochon}, {Parrent}, {Kelley}, \&
  {Margutti}}]{Guillochon.ea_2017.ApJ_open.sn.cat}
{Guillochon}, J., {Parrent}, J., {Kelley}, L.~Z., \& {Margutti}, R. 2017, \apj,
  835, 64

\bibitem[{{Halevi} \&
  {M{\"o}sta}(2018)}]{Halevi.Mosta_2018.mnras_jet.sn.rproc.3d}
{Halevi}, G., \& {M{\"o}sta}, P. 2018, \mnras, 477, 2366

\bibitem[{{Hjorth} {et~al.}(2003){Hjorth}, {Sollerman}, {M{\o}ller}, {Fynbo},
  {Woosley}, {Kouveliotou}, {Tanvir}, {Greiner}, {Andersen}, {Castro-Tirado},
  {Castro Cer{\'o}n}, {Fruchter}, {Gorosabel}, {Jakobsson}, {Kaper}, {Klose},
  {Masetti}, {Pedersen}, {Pedersen}, {Pian}, {Palazzi}, {Rhoads}, {Rol}, {van
  den Heuvel}, {Vreeswijk}, {Watson}, \&
  {Wijers}}]{Hjorth.ea_2003_2003dh.disc.nature}
{Hjorth}, J., {Sollerman}, J., {M{\o}ller}, P., {et~al.} 2003, \nat, 423, 847

\bibitem[{{Hoffman} {et~al.}(1997){Hoffman}, {Woosley}, \&
  {Qian}}]{Hoffman.eq_1997.ApJ_neutrino.winds.nucleosyn}
{Hoffman}, R.~D., {Woosley}, S.~E., \& {Qian}, Y.-Z. 1997, \apj, 482, 951

\bibitem[{{Horowitz} {et~al.}(2019){Horowitz}, {Arcones}, {C{\^o}t{\'e}},
  {Dillmann}, {Nazarewicz}, {Roederer}, {Schatz}, {Aprahamian}, {Atanasov},
  {Bauswein}, {Beers}, {Bliss}, {Brodeur}, {Clark}, {Frebel}, {Foucart},
  {Hansen}, {Just}, {Kankainen}, {McLaughlin}, {Kelly}, {Liddick}, {Lee},
  {Lippuner}, {Martin}, {Mendoza-Temis}, {Metzger}, {Mumpower}, {Perdikakis},
  {Pereira}, {O'Shea}, {Reifarth}, {Rogers}, {Siegel}, {Spyrou}, {Surman},
  {Tang}, {Uesaka}, \& {Wang}}]{Horowitz.ea_2019.JPG_rproc.frib.connect}
{Horowitz}, C.~J., {Arcones}, A., {C{\^o}t{\'e}}, B., {et~al.} 2019, Journal of
  Physics G Nuclear Physics, 46, 083001

\bibitem[{{Hotokezaka} {et~al.}(2013){Hotokezaka}, {Kiuchi}, {Kyutoku},
  {Okawa}, {Sekiguchi}, {Shibata}, \& {Taniguchi}}]{Hotokezaka_2013_massEj}
{Hotokezaka}, K., {Kiuchi}, K., {Kyutoku}, K., {et~al.} 2013, \prd, 87, 024001

\bibitem[{{Hotokezaka} {et~al.}(2021){Hotokezaka}, {Tanaka}, {Kato}, \&
  {Gaigalas}}]{Hotokezaka.ea_2021.MNRAS_kn.nebular.model}
{Hotokezaka}, K., {Tanaka}, M., {Kato}, D., \& {Gaigalas}, G. 2021, \mnras,
  506, 5863

\bibitem[{{Hunter} {et~al.}(2009){Hunter}, {Valenti}, {Kotak}, {Meikle},
  {Taubenberger}, {Pastorello}, {Benetti}, {Stanishev}, {Smartt}, {Trundle},
  {Arkharov}, {Bufano}, {Cappellaro}, {Di Carlo}, {Dolci}, {Elias-Rosa},
  {Frandsen}, {Fynbo}, {Hopp}, {Larionov}, {Laursen}, {Mazzali}, {Navasardyan},
  {Ries}, {Riffeser}, {Rizzi}, {Tsvetkov}, {Turatto}, \&
  {Wilke}}]{Hunter.ea_2009.AandA_sn2007gr.phot}
{Hunter}, D.~J., {Valenti}, S., {Kotak}, R., {et~al.} 2009, \aap, 508, 371

\bibitem[{{Janka} {et~al.}(2016){Janka}, {Melson}, \&
  {Summa}}]{Janka.ea_2016.arnps_ccsn.physics.3d}
{Janka}, H.-T., {Melson}, T., \& {Summa}, A. 2016, Annual Review of Nuclear and
  Particle Science, 66, 341

\bibitem[{{Jeon} {et~al.}(2021){Jeon}, {Besla}, \&
  {Bromm}}]{Jeon.ea_2021.mnras_rproc.stars.ufdwarf.gal}
{Jeon}, M., {Besla}, G., \& {Bromm}, V. 2021, \mnras, 506, 1850

\bibitem[{{Jerkstrand}(2017)}]{Jerkstrand.A_2017.hsn.book_neb.phase.sne}
{Jerkstrand}, A. 2017, in Handbook of Supernovae, ed. A.~W. {Alsabti} \&
  P.~{Murdin}, 795

\bibitem[{{Ji} {et~al.}(2016){Ji}, {Frebel}, {Chiti}, \&
  {Simon}}]{Ji.ea_2016.Nature_rproc.single.event.retII}
{Ji}, A.~P., {Frebel}, A., {Chiti}, A., \& {Simon}, J.~D. 2016, \nat, 531, 610

\bibitem[{{Just} {et~al.}(2015){Just}, {Bauswein}, {Ardevol Pulpillo},
  {Goriely}, \& {Janka}}]{Just.ea_2015.arxiv_nsm.torus.nucleosyn}
{Just}, O., {Bauswein}, A., {Ardevol Pulpillo}, R., {Goriely}, S., \& {Janka},
  H.-T. 2015, ArXiv e-prints, arXiv:1504.05448

\bibitem[{{Just} {et~al.}(2022){Just}, {Goriely}, {Janka}, {Nagataki}, \&
  {Bauswein}}]{Just.ea_2022.mnras_nuetrino.cooled.bh.acc.disks}
{Just}, O., {Goriely}, S., {Janka}, H.~T., {Nagataki}, S., \& {Bauswein}, A.
  2022, \mnras, 509, 1377

\bibitem[{{Kasen} {et~al.}(2013){Kasen}, {Badnell}, \&
  {Barnes}}]{Kasen_2013_AS}
{Kasen}, D., {Badnell}, N.~R., \& {Barnes}, J. 2013, \apj, 774, 25

\bibitem[{{Kasen} {et~al.}(2017){Kasen}, {Metzger}, {Barnes}, {Quataert}, \&
  {Ramirez-Ruiz}}]{Kasen.ea_2017Natur_gw.170817.knova.theory}
{Kasen}, D., {Metzger}, B., {Barnes}, J., {Quataert}, E., \& {Ramirez-Ruiz}, E.
  2017, \nat, 551, 80

\bibitem[{{Kasliwal} {et~al.}(2017){Kasliwal}, {Nakar}, {Singer}, {Kaplan},
  {Cook}, {Van Sistine}, {Lau}, {Fremling}, {Gottlieb}, {Jencson}, {Adams},
  {Feindt}, {Hotokezaka}, {Ghosh}, {Perley}, {Yu}, {Piran}, {Allison},
  {Anupama}, {Balasubramanian}, {Bannister}, {Bally}, {Barnes}, {Barway},
  {Bellm}, {Bhalerao}, {Bhattacharya}, {Blagorodnova}, {Bloom}, {Brady},
  {Cannella}, {Chatterjee}, {Cenko}, {Cobb}, {Copperwheat}, {Corsi}, {De},
  {Dobie}, {Emery}, {Evans}, {Fox}, {Frail}, {Frohmaier}, {Goobar}, {Hallinan},
  {Harrison}, {Helou}, {Hinderer}, {Ho}, {Horesh}, {Ip}, {Itoh}, {Kasen},
  {Kim}, {Kuin}, {Kupfer}, {Lynch}, {Madsen}, {Mazzali}, {Miller}, {Mooley},
  {Murphy}, {Ngeow}, {Nichols}, {Nissanke}, {Nugent}, {Ofek}, {Qi}, {Quimby},
  {Rosswog}, {Rusu}, {Sadler}, {Schmidt}, {Sollerman}, {Steele}, {Williamson},
  {Xu}, {Yan}, {Yatsu}, {Zhang}, \&
  {Zhao}}]{Kasliwal.ea_2017Sci_gw.170817.em.interp}
{Kasliwal}, M.~M., {Nakar}, E., {Singer}, L.~P., {et~al.} 2017, Science, 358,
  1559

\bibitem[{{Kasliwal} {et~al.}(2022){Kasliwal}, {Kasen}, {Lau}, {Perley},
  {Rosswog}, {Ofek}, {Hotokezaka}, {Chary}, {Sollerman}, {Goobar}, \&
  {Kaplan}}]{Kasliwal.ea_2022.MNRAS_spitzer.late.obs.kn170817}
{Kasliwal}, M.~M., {Kasen}, D., {Lau}, R.~M., {et~al.} 2022, \mnras, 510, L7

\bibitem[{{Kilpatrick} {et~al.}(2017){Kilpatrick}, {Foley}, {Kasen},
  {Murguia-Berthier}, {Ramirez-Ruiz}, {Coulter}, {Drout}, {Piro}, {Shappee},
  {Boutsia}, {Contreras}, {Di Mille}, {Madore}, {Morrell}, {Pan}, {Prochaska},
  {Rest}, {Rojas-Bravo}, {Siebert}, {Simon}, \&
  {Ulloa}}]{Kilpatrick.ea_2017.Sci_gw.170817.spectrum.opt.nir}
{Kilpatrick}, C.~D., {Foley}, R.~J., {Kasen}, D., {et~al.} 2017, Science, 358,
  1583

\bibitem[{{Korobkin} {et~al.}(2012){Korobkin}, {Rosswog}, {Arcones}, \&
  {Winteler}}]{Korobkin_NSM_rp}
{Korobkin}, O., {Rosswog}, S., {Arcones}, A., \& {Winteler}, C. 2012, \mnras,
  426, 1940

\bibitem[{{Kuroda} {et~al.}(2020){Kuroda}, {Arcones}, {Takiwaki}, \&
  {Kotake}}]{Kuroda.ea_2020.ApJ_magnetorot.sn.neutrino.gr3d}
{Kuroda}, T., {Arcones}, A., {Takiwaki}, T., \& {Kotake}, K. 2020, \apj, 896,
  102

\bibitem[{{Kyutoku} {et~al.}(2015){Kyutoku}, {Ioka}, {Okawa}, {Shibata}, \&
  {Taniguchi}}]{Kyutoku_2015_massEj}
{Kyutoku}, K., {Ioka}, K., {Okawa}, H., {Shibata}, M., \& {Taniguchi}, K. 2015,
  \prd, 92, 044028

\bibitem[{{Lattimer} \&
  {Schramm}(1974)}]{Lattimer.Schramm_1974.ApJL_rproc.nsbh.merger}
{Lattimer}, J.~M., \& {Schramm}, D.~N. 1974, \apjl, 192, L145

\bibitem[{{Lourie} {et~al.}(2020){Lourie}, {Baker}, {Burruss}, {Egan},
  {F{\.z}r{\'e}sz}, {Frostig}, {Garcia-Zych}, {Ganciu}, {Haworth},
  {Hinrichsen}, {Kasliwal}, {Karambelkar}, {Malonis}, {Simcoe}, \&
  {Zolkower}}]{Lourie.ea_2020.SPIE_WINTER.goals.design}
{Lourie}, N.~P., {Baker}, J.~W., {Burruss}, R.~S., {et~al.} 2020, in Society of
  Photo-Optical Instrumentation Engineers (SPIE) Conference Series, Vol. 11447,
  Society of Photo-Optical Instrumentation Engineers (SPIE) Conference Series,
  114479K

\bibitem[{{MacFadyen} \& {Woosley}(1999)}]{MacFadyenWoosley99_collapsar}
{MacFadyen}, A.~I., \& {Woosley}, S.~E. 1999, \apj, 524, 262

\bibitem[{{Macias} \&
  {Ramirez-Ruiz}(2019)}]{Macias.RamirezRuiz_2019.ApJL_rccsn.constraints.stellar.abund}
{Macias}, P., \& {Ramirez-Ruiz}, E. 2019, \apjl, 877, L24

\bibitem[{{Madison} \& {Li}(2007)}]{Madison.Li_2007.CBET_sn.2007gr.disc}
{Madison}, D., \& {Li}, W. 2007, Central Bureau Electronic Telegrams, 1034, 1

\bibitem[{{Maeda} {et~al.}(2003){Maeda}, {Mazzali}, {Deng}, {Nomoto}, {Yoshii},
  {Tomita}, \& {Kobayashi}}]{Maeda03ApJ_twoCompNi_blic}
{Maeda}, K., {Mazzali}, P.~A., {Deng}, J., {et~al.} 2003, \apj, 593, 931

\bibitem[{{Maeda} {et~al.}(2002){Maeda}, {Nakamura}, {Nomoto}, {Mazzali},
  {Patat}, \& {Hachisu}}]{Maeda.ea_2002.ApJ_nucleosyn.hypernovae.1998bw.spec}
{Maeda}, K., {Nakamura}, T., {Nomoto}, K., {et~al.} 2002, \apj, 565, 405

\bibitem[{{Mazzali} {et~al.}(2002){Mazzali}, {Deng}, {Maeda}, {Nomoto},
  {Umeda}, {Hatano}, {Iwamoto}, {Yoshii}, {Kobayashi}, {Minezaki}, {Doi},
  {Enya}, {Tomita}, {Smartt}, {Kinugasa}, {Kawakita}, {Ayani}, {Kawabata},
  {Yamaoka}, {Qiu}, {Motohara}, {Gerardy}, {Fesen}, {Kawabata}, {Iye},
  {Kashikawa}, {Kosugi}, {Ohyama}, {Takada-Hidai}, {Zhao}, {Chornock},
  {Filippenko}, {Benetti}, \& {Turatto}}]{Mazzali.ea_2002ApJL_snic.2002ap}
{Mazzali}, P.~A., {Deng}, J., {Maeda}, K., {et~al.} 2002, \apjl, 572, L61

\bibitem[{{Mazzali} {et~al.}(2003){Mazzali}, {Deng}, {Tominaga}, {Maeda},
  {Nomoto}, {Matheson}, {Kawabata}, {Stanek}, \&
  {Garnavich}}]{Mazzali.ea_2003.ApjL_icbl.sn.2003dh}
{Mazzali}, P.~A., {Deng}, J., {Tominaga}, N., {et~al.} 2003, \apjl, 599, L95

\bibitem[{{McCully} {et~al.}(2017){McCully}, {Hiramatsu}, {Howell},
  {Hosseinzadeh}, {Arcavi}, {Kasen}, {Barnes}, {Shara}, {Williams},
  {V{\"a}is{\"a}nen}, {Potter}, {Romero-Colmenero}, {Crawford}, {Buckley},
  {Cooke}, {Andreoni}, {Pritchard}, {Mao}, {Gromadzki}, \&
  {Burke}}]{McCully.ea_2017ApJ_gw.170817.blue.spec}
{McCully}, C., {Hiramatsu}, D., {Howell}, D.~A., {et~al.} 2017, \apjl, 848, L32

\bibitem[{{Meikle} {et~al.}(2002){Meikle}, {Lucy}, {Smartt}, {Leibundgut},
  {Lundqvist}, \& {Ostensen}}]{Meikle.ea_2002.IAUC_sn2002ap.disc.icbl.id}
{Meikle}, P., {Lucy}, L., {Smartt}, S., {et~al.} 2002, \iaucirc, 7811, 2

\bibitem[{{Metzger} {et~al.}(2008){Metzger}, {Piro}, \&
  {Quataert}}]{Metzger.ea_2008.mnras_accretion.disk.co.merg.tdep}
{Metzger}, B.~D., {Piro}, A.~L., \& {Quataert}, E. 2008, \mnras, 390, 781

\bibitem[{{Metzger} {et~al.}(2007){Metzger}, {Thompson}, \&
  {Quataert}}]{Metzger.ea_2007.ApJ_proto.ns.winds.bfields.rot}
{Metzger}, B.~D., {Thompson}, T.~A., \& {Quataert}, E. 2007, \apj, 659, 561

\bibitem[{{Metzger} {et~al.}(2010){Metzger}, {Mart{\'{\i}}nez-Pinedo},
  {Darbha}, {Quataert}, {Arcones}, {Kasen}, {Thomas}, {Nugent}, {Panov}, \&
  {Zinner}}]{Metzger_2010}
{Metzger}, B.~D., {Mart{\'{\i}}nez-Pinedo}, G., {Darbha}, S., {et~al.} 2010,
  \mnras, 406, 2650

\bibitem[{{Meyer} \& {Brown}(1997)}]{Meyer.Brown_1997.ApJS_rproc.mods.survey}
{Meyer}, B.~S., \& {Brown}, J.~S. 1997, \apjs, 112, 199

\bibitem[{{Miller} {et~al.}(2020){Miller}, {Sprouse}, {Fryer}, {Ryan},
  {Dolence}, {Mumpower}, \& {Surman}}]{Miller.ea_2020.ApJ_rproc.collapsar.blue}
{Miller}, J.~M., {Sprouse}, T.~M., {Fryer}, C.~L., {et~al.} 2020, \apj, 902, 66

\bibitem[{{Molero} {et~al.}(2021){Molero}, {Romano}, {Reichert}, {Matteucci},
  {Arcones}, {Cescutti}, {Simonetti}, {Hansen}, \&
  {Lanfranchi}}]{Molero.ea_2021.mnras_evol.rproc.dwarf.gal}
{Molero}, M., {Romano}, D., {Reichert}, M., {et~al.} 2021, \mnras, 505, 2913

\bibitem[{{M{\"o}sta} {et~al.}(2018){M{\"o}sta}, {Roberts}, {Halevi}, {Ott},
  {Lippuner}, {Haas}, \&
  {Schnetter}}]{Mosta.ea_2018.ApJ_rproc.3d.magnetorot.sne}
{M{\"o}sta}, P., {Roberts}, L.~F., {Halevi}, G., {et~al.} 2018, \apj, 864, 171

\bibitem[{{M{\"o}sta} {et~al.}(2014){M{\"o}sta}, {Richers}, {Ott}, {Haas},
  {Piro}, {Boydstun}, {Abdikamalov}, {Reisswig}, \&
  {Schnetter}}]{Mosta.ea_2014.ApJL_magnetorot.sne.3d}
{M{\"o}sta}, P., {Richers}, S., {Ott}, C.~D., {et~al.} 2014, Astrophys. J.
  Lett., 785, L29

\bibitem[{{M{\"u}ller} {et~al.}(2017){M{\"u}ller}, {Melson}, {Heger}, \&
  {Janka}}]{Muller.ea_2017.mnras_sn.sim.3d.prog}
{M{\"u}ller}, B., {Melson}, T., {Heger}, A., \& {Janka}, H.-T. 2017, \mnras,
  472, 491

\bibitem[{{Mutchler} {et~al.}(2021){Mutchler}, {Bellini}, {Casertano},
  {Christian}, {De Rosa}, {Desjardins}, {Deustua}, {Geda}, {Hargis},
  {Koekemoer}, {Lajoie}, {Nelan}, {Peeples}, {Petric}, {Ryan}, \&
  {York}}]{Mutchler.ea_2021.AAS_Roman.sci.ops}
{Mutchler}, M., {Bellini}, A., {Casertano}, S., {et~al.} 2021, in American
  Astronomical Society Meeting Abstracts, Vol.~53, American Astronomical
  Society Meeting Abstracts, 216.01

\bibitem[{{Naidu} {et~al.}(2022){Naidu}, {Ji}, {Conroy}, {Bonaca}, {Ting},
  {Zaritsky}, {van Son}, {Broekgaarden}, {Tacchella}, {Chandra}, {Caldwell},
  {Cargile}, \& {Speagle}}]{Naidu.ea_2022.ApJ_disrupted.halos.rproc.sources}
{Naidu}, R.~P., {Ji}, A.~P., {Conroy}, C., {et~al.} 2022, \apjl, 926, L36

\bibitem[{{Nakar}(2007)}]{Nakar_2007.PRep_sgrbs}
{Nakar}, E. 2007, \physrep, 442, 166

\bibitem[{{Nicholl} {et~al.}(2017){Nicholl}, {Berger}, {Kasen}, {Metzger},
  {Elias}, {Brice{\~n}o}, {Alexander}, {Blanchard}, {Chornock},
  {Cowperthwaite}, {Eftekhari}, {Fong}, {Margutti}, {Villar}, {Williams},
  {Brown}, {Annis}, {Bahramian}, {Brout}, {Brown}, {Chen}, {Clemens},
  {Dennihy}, {Dunlap}, {Holz}, {Marchesini}, {Massaro}, {Moskowitz},
  {Pelisoli}, {Rest}, {Ricci}, {Sako}, {Soares-Santos}, \&
  {Strader}}]{Nicholl.ea_2017ApJ_gw.170817.blue.spec}
{Nicholl}, M., {Berger}, E., {Kasen}, D., {et~al.} 2017, \apjl, 848, L18

\bibitem[{{Perego} {et~al.}(2014){Perego}, {Rosswog}, {Cabez{\'o}n},
  {Korobkin}, {K{\"a}ppeli}, {Arcones}, \&
  {Liebend{\"o}rfer}}]{Perego.ea_2014.mnras_neutrino.winds.nsm}
{Perego}, A., {Rosswog}, S., {Cabez{\'o}n}, R.~M., {et~al.} 2014, \mnras, 443,
  3134

\bibitem[{{Perley} {et~al.}(2020){Perley}, {Fremling}, {Sollerman}, {Miller},
  {Dahiwale}, {Sharma}, {Bellm}, {Biswas}, {Brink}, {Bruch}, {De}, {Dekany},
  {Drake}, {Duev}, {Filippenko}, {Gal-Yam}, {Goobar}, {Graham}, {Graham}, {Ho},
  {Irani}, {Kasliwal}, {Kim}, {Kulkarni}, {Mahabal}, {Masci}, {Modak}, {Neill},
  {Nordin}, {Riddle}, {Soumagnac}, {Strotjohann}, {Schulze}, {Taggart},
  {Tzanidakis}, {Walters}, \& {Yan}}]{Perley.ea_2020.ApJ_ZTF.sn.sample}
{Perley}, D.~A., {Fremling}, C., {Sollerman}, J., {et~al.} 2020, \apj, 904, 35

\bibitem[{{Prentice} {et~al.}(2019){Prentice}, {Ashall}, {James}, {Short},
  {Mazzali}, {Bersier}, {Crowther}, {Barbarino}, {Chen}, {Copperwheat},
  {Darnley}, {Denneau}, {Elias-Rosa}, {Fraser}, {Galbany}, {Gal-Yam},
  {Harmanen}, {Howell}, {Hosseinzadeh}, {Inserra}, {Kankare}, {Karamehmetoglu},
  {Lamb}, {Limongi}, {Maguire}, {McCully}, {Olivares E}, {Piascik}, {Pignata},
  {Reichart}, {Rest}, {Reynolds}, {Rodr{\'\i}guez}, {Saario}, {Schulze},
  {Smartt}, {Smith}, {Sollerman}, {Stalder}, {Sullivan}, {Taddia}, {Valenti},
  {Vergani}, {Williams}, \&
  {Young}}]{Prentice.ea_2019.MNRAS_sesne.properties.progens}
{Prentice}, S.~J., {Ashall}, C., {James}, P.~A., {et~al.} 2019, \mnras, 485,
  1559

\bibitem[{{Qian} \&
  {Woosley}(1996)}]{Qian.Woosley_1996.ApJ_neutrino.winds.ccsne.nucleosyn}
{Qian}, Y., \& {Woosley}, S.~E. 1996, \apj, 471, 331

\bibitem[{{Radice} {et~al.}(2018){Radice}, {Perego}, {Hotokezaka}, {Fromm},
  {Bernuzzi}, \& {Roberts}}]{Radice.ea_2018.ApJ_nsm.mass.eject}
{Radice}, D., {Perego}, A., {Hotokezaka}, K., {et~al.} 2018, \apj, 869, 130

\bibitem[{{Scheck} {et~al.}(2006){Scheck}, {Kifonidis}, {Janka}, \&
  {M{\"u}ller}}]{Scheck.ea_2006/AAP_multiD.sn.sim.neutrino.transport}
{Scheck}, L., {Kifonidis}, K., {Janka}, H.~T., \& {M{\"u}ller}, E. 2006, \aap,
  457, 963

\bibitem[{{Sch{\"o}nrich} \&
  {Weinberg}(2019)}]{Schonrich.Weinberg_2019.mnras_rproc.nsm.ism}
{Sch{\"o}nrich}, R.~A., \& {Weinberg}, D.~H. 2019, \mnras, 487, 580

\bibitem[{{Shappee} {et~al.}(2017){Shappee}, {Simon}, {Drout}, {Piro},
  {Morrell}, {Prieto}, {Kasen}, {Holoien}, {Kollmeier}, {Kelson}, {Coulter},
  {Foley}, {Kilpatrick}, {Siebert}, {Madore}, {Murguia-Berthier}, {Pan},
  {Prochaska}, {Ramirez-Ruiz}, {Rest}, {Adams}, {Alatalo}, {Ba{\~n}ados},
  {Baughman}, {Bernstein}, {Bitsakis}, {Boutsia}, {Bravo}, {Di Mille}, {Higgs},
  {Ji}, {Maravelias}, {Marshall}, {Placco}, {Prieto}, \&
  {Wan}}]{Shappee.ea_2017.Science_kn170817}
{Shappee}, B.~J., {Simon}, J.~D., {Drout}, M.~R., {et~al.} 2017, Science, 358,
  1574

\bibitem[{{Siegel}(2019)}]{Siegel_2019.EPJA_knova.review}
{Siegel}, D.~M. 2019, European Physical Journal A, 55, 203

\bibitem[{{Siegel} {et~al.}(2021){Siegel}, {Agarwal}, {Barnes}, {Metzger},
  {Renzo}, \& {Villar}}]{Siegel.ea_2021.arXiv_super.kilonovae}
{Siegel}, D.~M., {Agarwal}, A., {Barnes}, J., {et~al.} 2021, arXiv e-prints,
  arXiv:2111.03094

\bibitem[{{Siegel} {et~al.}(2019){Siegel}, {Barnes}, \&
  {Metzger}}]{Siegel.Barnes.Metzger_2019.Nature_rp.collapsar}
{Siegel}, D.~M., {Barnes}, J., \& {Metzger}, B.~D. 2019, \nat, 569, 241

\bibitem[{{Siegel} \&
  {Metzger}(2017)}]{Siegel.Metzger_2017.PRL_nsm.accretion.disks.rproc}
{Siegel}, D.~M., \& {Metzger}, B.~D. 2017, \prl, 119, 231102

\bibitem[{{Smartt} {et~al.}(2017){Smartt}, {Chen}, {Jerkstrand}, {Coughlin},
  {Kankare}, {Sim}, {Fraser}, {Inserra}, {Maguire}, {Chambers}, {Huber},
  {Kr{\"u}hler}, {Leloudas}, {Magee}, {Shingles}, {Smith}, {Young}, {Tonry},
  {Kotak}, {Gal-Yam}, {Lyman}, {Homan}, {Agliozzo}, {Anderson}, {Angus},
  {Ashall}, {Barbarino}, {Bauer}, {Berton}, {Botticella}, {Bulla}, {Bulger},
  {Cannizzaro}, {Cano}, {Cartier}, {Cikota}, {Clark}, {De Cia}, {Della Valle},
  {Denneau}, {Dennefeld}, {Dessart}, {Dimitriadis}, {Elias-Rosa}, {Firth},
  {Flewelling}, {Fl{\"o}rs}, {Franckowiak}, {Frohmaier}, {Galbany},
  {Gonz{\'a}lez-Gait{\'a}n}, {Greiner}, {Gromadzki}, {Guelbenzu},
  {Guti{\'e}rrez}, {Hamanowicz}, {Hanlon}, {Harmanen}, {Heintz}, {Heinze},
  {Hernandez}, {Hodgkin}, {Hook}, {Izzo}, {James}, {Jonker}, {Kerzendorf},
  {Klose}, {Kostrzewa-Rutkowska}, {Kowalski}, {Kromer}, {Kuncarayakti},
  {Lawrence}, {Lowe}, {Magnier}, {Manulis}, {Martin-Carrillo}, {Mattila},
  {McBrien}, {M{\"u}ller}, {Nordin}, {O'Neill}, {Onori}, {Palmerio},
  {Pastorello}, {Patat}, {Pignata}, {Podsiadlowski}, {Pumo}, {Prentice}, {Rau},
  {Razza}, {Rest}, {Reynolds}, {Roy}, {Ruiter}, {Rybicki}, {Salmon}, {Schady},
  {Schultz}, {Schweyer}, {Seitenzahl}, {Smith}, {Sollerman}, {Stalder},
  {Stubbs}, {Sullivan}, {Szegedi}, {Taddia}, {Taubenberger}, {Terreran}, {van
  Soelen}, {Vos}, {Wainscoat}, {Walton}, {Waters}, {Weiland}, {Willman},
  {Wiseman}, {Wright}, {Wyrzykowski}, \&
  {Yaron}}]{Smartt.ea_2017Natur_gw170817.empc.disc}
{Smartt}, S.~J., {Chen}, T.~W., {Jerkstrand}, A., {et~al.} 2017, \nat, 551, 75

\bibitem[{{Soares-Santos} {et~al.}(2017){Soares-Santos}, {Holz}, {Annis}, {Dark
  Energy Survey}, \& {Dark Energy Camera GW-EM
  Collaboration}}]{SoaresSantos.ea_2017ApJ_gw.170817.empc.decam.disc}
{Soares-Santos}, M., {Holz}, D.~E., {Annis}, J., {Dark Energy Survey}, \& {Dark
  Energy Camera GW-EM Collaboration}. 2017, \apjl, 848, L16

\bibitem[{{Stanek} {et~al.}(2003){Stanek}, {Matheson}, {Garnavich}, {Martini},
  {Berlind}, {Caldwell}, {Challis}, {Brown}, {Schild}, {Krisciunas}, {Calkins},
  {Lee}, {Hathi}, {Jansen}, {Windhorst}, {Echevarria}, {Eisenstein}, {Pindor},
  {Olszewski}, {Harding}, {Holland}, \&
  {Bersier}}]{Stanek.ea_2003.ApJL_sn2003dh.data}
{Stanek}, K.~Z., {Matheson}, T., {Garnavich}, P.~M., {et~al.} 2003, \apjl, 591,
  L17

\bibitem[{{Symbalisty} \&
  {Schramm}(1982)}]{Symbalisty.Schramm_1982.ApL_rproc.ns.merger}
{Symbalisty}, E., \& {Schramm}, D.~N. 1982, \aplett, 22, 143

\bibitem[{{Taddia} {et~al.}(2019){Taddia}, {Sollerman}, {Fremling},
  {Barbarino}, {Karamehmetoglu}, {Arcavi}, {Cenko}, {Filippenko}, {Gal-Yam},
  {Hiramatsu}, {Hosseinzadeh}, {Howell}, {Kulkarni}, {Laher}, {Lunnan},
  {Masci}, {Nugent}, {Nyholm}, {Perley}, {Quimby}, \&
  {Silverman}}]{Taddia.ea_2019.AandA_IcBL.iPTF.survey}
{Taddia}, F., {Sollerman}, J., {Fremling}, C., {et~al.} 2019, \aap, 621, A71

\bibitem[{{Tanaka} \& {Hotokezaka}(2013)}]{Tanaka_Hotok_rpOps}
{Tanaka}, M., \& {Hotokezaka}, K. 2013, \apj, 775, 113

\bibitem[{{Tanaka} {et~al.}(2020){Tanaka}, {Kato}, {Gaigalas}, \&
  {Kawaguchi}}]{Tanaka.ea_2020.MNRAS_knova.kappas}
{Tanaka}, M., {Kato}, D., {Gaigalas}, G., \& {Kawaguchi}, K. 2020, \mnras, 496,
  1369

\bibitem[{{Tanaka} {et~al.}(2008){Tanaka}, {Kawabata}, {Maeda}, {Hattori}, \&
  {Nomoto}}]{Tanaka.ea_2008.ApJ_sn.2002ap.asphericity.rt}
{Tanaka}, M., {Kawabata}, K.~S., {Maeda}, K., {Hattori}, T., \& {Nomoto}, K.
  2008, \apj, 689, 1191

\bibitem[{{Tanaka} {et~al.}(2017){Tanaka}, {Utsumi}, {Mazzali}, {Tominaga},
  {Yoshida}, {Sekiguchi}, {Morokuma}, {Motohara}, {Ohta}, {Kawabata}, {Abe},
  {Aoki}, {Asakura}, {Baar}, {Barway}, {Bond}, {Doi}, {Fujiyoshi}, {Furusawa},
  {Honda}, {Itoh}, {Kawabata}, {Kawai}, {Kim}, {Lee}, {Miyazaki}, {Morihana},
  {Nagashima}, {Nagayama}, {Nakaoka}, {Nakata}, {Ohsawa}, {Ohshima}, {Okita},
  {Saito}, {Sumi}, {Tajitsu}, {Takahashi}, {Takayama}, {Tamura}, {Tanaka},
  {Terai}, {Tristram}, {Yasuda}, \&
  {Zenko}}]{Tanaka.ea_2017PASJ_gw.170817.knova.interp}
{Tanaka}, M., {Utsumi}, Y., {Mazzali}, P.~A., {et~al.} 2017, \pasj, 69, 102

\bibitem[{{Tanvir} {et~al.}(2017){Tanvir}, {Levan},
  {Gonz{\'a}lez-Fern{\'a}ndez}, {Korobkin}, {Mandel}, {Rosswog}, {Hjorth},
  {D'Avanzo}, {Fruchter}, {Fryer}, {Kangas}, {Milvang-Jensen}, {Rosetti},
  {Steeghs}, {Wollaeger}, {Cano}, {Copperwheat}, {Covino}, {D'Elia}, {de Ugarte
  Postigo}, {Evans}, {Even}, {Fairhurst}, {Figuera Jaimes}, {Fontes}, {Fujii},
  {Fynbo}, {Gompertz}, {Greiner}, {Hodosan}, {Irwin}, {Jakobsson},
  {J{\o}rgensen}, {Kann}, {Lyman}, {Malesani}, {McMahon}, {Melandri},
  {O'Brien}, {Osborne}, {Palazzi}, {Perley}, {Pian}, {Piranomonte}, {Rabus},
  {Rol}, {Rowlinson}, {Schulze}, {Sutton}, {Th{\"o}ne}, {Ulaczyk}, {Watson},
  {Wiersema}, \& {Wijers}}]{Tanvir.ea+2017.ApJL_gw170817.emcp.disc}
{Tanvir}, N.~R., {Levan}, A.~J., {Gonz{\'a}lez-Fern{\'a}ndez}, C., {et~al.}
  2017, \apjl, 848, L27

\bibitem[{{Taubenberger} {et~al.}(2009){Taubenberger}, {Valenti}, {Benetti},
  {Cappellaro}, {Della Valle}, {Elias-Rosa}, {Hachinger}, {Hillebrandt},
  {Maeda}, {Mazzali}, {Pastorello}, {Patat}, {Sim}, \&
  {Turatto}}]{Taubenberger.ea_2009.MNRAS_sne.ibc.emiss.lines.aspher}
{Taubenberger}, S., {Valenti}, S., {Benetti}, S., {et~al.} 2009, \mnras, 397,
  677

\bibitem[{{Thielemann} {et~al.}(2020){Thielemann}, {Wehmeyer}, \&
  {Wu}}]{Thielemann.ea_2020.JPCS_rproc.sites.review}
{Thielemann}, F.-K., {Wehmeyer}, B., \& {Wu}, M.-R. 2020, in Journal of Physics
  Conference Series, Vol. 1668, Journal of Physics Conference Series, 012044

\bibitem[{{Thompson} {et~al.}(2001){Thompson}, {Burrows}, \&
  {Meyer}}]{Thompson.ea_2001.ApJ_proto.ns.winds.rproc}
{Thompson}, T.~A., {Burrows}, A., \& {Meyer}, B.~S. 2001, \apj, 562, 887

\bibitem[{{Thompson} {et~al.}(2004){Thompson}, {Chang}, \&
  {Quataert}}]{Thompson.ea_2004.ApJ_magnetar.sne.grb}
{Thompson}, T.~A., {Chang}, P., \& {Quataert}, E. 2004, \apj, 611, 380

\bibitem[{{Thompson} \&
  {ud-Doula}(2018)}]{Thompson.udDoula_mnras.2018_rproc.proto.ns.winds}
{Thompson}, T.~A., \& {ud-Doula}, A. 2018, \mnras, 476, 5502

\bibitem[{{Tomita} {et~al.}(2006){Tomita}, {Deng}, {Maeda}, {Yoshii}, {Nomoto},
  {Mazzali}, {Suzuki}, {Kobayashi}, {Minezaki}, {Aoki}, {Enya}, \&
  {Suganuma}}]{Tomita.ea_2006.ApJ_2002ap.opt.nir.500.days}
{Tomita}, H., {Deng}, J., {Maeda}, K., {et~al.} 2006, \apj, 644, 400

\bibitem[{{Tsujimoto}(2021)}]{Tsujimoto.ea_2021.ApJL_two.proc.sites}
{Tsujimoto}, T. 2021, \apjl, 920, L32

\bibitem[{{van de Voort} {et~al.}(2020){van de Voort}, {Pakmor}, {Grand},
  {Springel}, {G{\'o}mez}, \&
  {Marinacci}}]{vandeVoort.ea_2020_rproc.enrich.nsm.rare.ccsne}
{van de Voort}, F., {Pakmor}, R., {Grand}, R. J.~J., {et~al.} 2020, \mnras,
  494, 4867

\bibitem[{{Villar} {et~al.}(2018){Villar}, {Cowperthwaite}, {Berger},
  {Blanchard}, {Gomez}, {Alexander}, {Margutti}, {Chornock}, {Eftekhari},
  {Fazio}, {Guillochon}, {Hora}, {Nicholl}, \&
  {Williams}}]{Villar.ea_2018.ApJ_kn170817.mir.spitzer}
{Villar}, V.~A., {Cowperthwaite}, P.~S., {Berger}, E., {et~al.} 2018, \apjl,
  862, L11

\bibitem[{{Vlasov} {et~al.}(2017){Vlasov}, {Metzger}, {Lippuner}, {Roberts}, \&
  {Thompson}}]{Vlasov.ea_mnras.2017_weak.rproc.ms.magnetar.vwinds}
{Vlasov}, A.~D., {Metzger}, B.~D., {Lippuner}, J., {Roberts}, L.~F., \&
  {Thompson}, T.~A. 2017, \mnras, 468, 1522

\bibitem[{{Waxman} {et~al.}(2018){Waxman}, {Ofek}, {Kushnir}, \&
  {Gal-Yam}}]{Waxman.ea_2018.mnras_gw170817.emcp.ejecta.model}
{Waxman}, E., {Ofek}, E.~O., {Kushnir}, D., \& {Gal-Yam}, A. 2018, \mnras, 481,
  3423

\bibitem[{{Winteler} {et~al.}(2012){Winteler}, {K{\"a}ppeli}, {Perego},
  {Arcones}, {Vasset}, {Nishimura}, {Liebend{\"o}rfer}, \&
  {Thielemann}}]{Winteler.ea_2012.ApJL_magnetorot.sne.rproc}
{Winteler}, C., {K{\"a}ppeli}, R., {Perego}, A., {et~al.} 2012, \apjl, 750, L22

\bibitem[{{Woosley} \& {Bloom}(2006)}]{WoosleyBloom06_snegrbRev}
{Woosley}, S.~E., \& {Bloom}, J.~S. 2006, \araa, 44, 507

\bibitem[{{Yoon} {et~al.}(2019){Yoon}, {Chun}, {Tolstov}, {Blinnikov}, \&
  {Dessart}}]{Yoon.ea_2019.ApJ_ni56.mix.color.ev}
{Yoon}, S.-C., {Chun}, W., {Tolstov}, A., {Blinnikov}, S., \& {Dessart}, L.
  2019, \apj, 872, 174

\bibitem[{{Yoshii} {et~al.}(2003){Yoshii}, {Tomita}, {Kobayashi}, {Deng},
  {Maeda}, {Nomoto}, {Mazzali}, {Umeda}, {Aoki}, {Doi}, {Enya}, {Minezaki},
  {Suganuma}, \& {Peterson}}]{Yoshii.ea_2003.ApJ_opt.nir.phot.sn2002ap}
{Yoshii}, Y., {Tomita}, H., {Kobayashi}, Y., {et~al.} 2003, \apj, 592, 467

\bibitem[{{Zevin} {et~al.}(2019){Zevin}, {Kremer}, {Siegel}, {Coughlin},
  {Tsang}, {Berry}, \& {Kalogera}}]{Zevin.ea_2019.ApJ_nsm.rproc.glob.cluster}
{Zevin}, M., {Kremer}, K., {Siegel}, D.~M., {et~al.} 2019, \apj, 886, 4

\end{thebibliography}

\end{document}